%% file: ms.tex
\documentclass[iop,numberedappendix]{emulateapj}

\usepackage{amsmath}

\newcommand{\OVI}{\ion{O}{6}}

\newcommand{\hst}{{\sl HST}}
\newcommand{\kms}{\ensuremath{{\rm km\,s}^{-1}}}
\newcommand{\lya}{\ensuremath{{\rm Ly}\alpha}}
\newcommand{\dNdz}{\ensuremath{d\mathcal{N}/dz}}

\begin{document}

\title{Absorption-line Detections of 10$^{5-6}$ K Gas in Spiral-Rich Groups of Galaxies
\footnote{Based on observations with the
NASA/ESA {\sl Hubble Space Telescope}, obtained at the Space Telescope
Science Institute, which is operated by AURA, Inc., under NASA contract NAS
5-26555.}}

\author{John T. Stocke, Brian A. Keeney, Charles W. Danforth, David Syphers, H. Yamamoto, J. Michael
Shull, James C. Green \& Cynthia Froning\footnote{Current address: Department of Astronomy, University of Texas, Austin, TX
78712}}
\affil{Center for Astrophysics and Space Astronomy, Department of
Astrophysical and Planetary Sciences, University of Colorado, 389 UCB,
Boulder, CO 80309, USA; john.stocke@colorado.edu}

\author{Blair D. Savage, Bart Wakker \& Tae-Sun Kim\footnote{Current address: Osservatorio Astronomico di Trieste,
 Via G. B. Tiepolo 11, 34131 Trieste, Italy}
}
\affil{Department of Astronomy, U of Wisconsin, Madison, WI 53706}


\author{Emma V. Ryan-Weber \& Glenn G. Kacprzak\footnote{Australian Research Council Super Science Fellow}}
\affil{Centre for Astrophysics \& Supercomputing, Swinburne University of Technology, VIC 3122, Australia}

\shorttitle{Spiral Group Medium}
\shortauthors{Stocke et~al.}

\begin{abstract}

Using the Cosmic Origins Spectrograph (COS) on the {\it Hubble Space Telescope} (HST) the COS Science Team has conducted a
high signal-to-noise survey of 14 bright QSOs. In a previous paper (Savage et al.\ 2014) these
far-UV spectra were used to discover 14 ``warm'' (T $\geq$ 10$^5$ K) absorbers using a combination of broad Ly$\alpha$ and broad
O~VI absorptions. A reanalysis of a few
of this new class of absorbers using slightly relaxed fitting criteria
finds as many as 20 warm absorbers {\it could be} present in this sample. 
A shallow, wide spectroscopic galaxy redshift survey has been conducted around these sightlines to investigate the warm absorber environment, which
is found to be spiral-rich groups or cluster outskirts with radial
velocity dispersions $\sigma$ = 250-750 km s$^{-1}$. While 2$\sigma$ evidence is presented favoring the hypothesis that these 
absorptions are associated with the galaxy groups and not with the individual, nearest galaxies, this evidence has considerable systematic
uncertainties and is based on a small sample size so it is not entirely conclusive. 
If the associations are with galaxy groups, the observed frequency of warm absorbers (\dNdz\ = 3.5--5 per unit
redshift) requires them to be very extended as an ensemble on the sky ($\sim$ 1 Mpc in radius at high covering factor). Most likely these warm absorbers
are interface gas clouds whose presence implies the existence of a hotter (T$\sim$ 10$^{6.5}$ K), diffuse and probably very massive 
($>$ 10$^{11}$ M$_{\odot}$) intra-group medium which has yet to be detected directly. 

\end {abstract}


\section{Introduction}
\label{intro}

On the basis of theoretical and observational considerations \citep* [e.g.,][] {spitzer56, mcgaugh00, pagel08} 
and numerical simulations \citep{klypin01} a massive halo of gas is expected around every luminous
late-type galaxy, otherwise these systems would be  baryon deficient relative 
to dark matter \citep{klypin01, stocke13}. If present in spiral galaxy halos, these as yet undetected baryons
could add considerably to the overall baryon inventory at $z \approx$ 0, which has not yet been finalized
\citep*{fukugita98, bregman07, shull12}. 
Additionally, while cosmological simulations find that the amount of gas accretion, both
cold and hot, has declined by an order of magnitude since redshifts of 2--4 \citep{keres09}, there
still needs to be sufficient infall in recent times to satisfy the constraints of the ``G-dwarf problem'' 
\citep* {larson72, pagel08} and to maintain the high star-formation-rates (SFR) in current day, 
massive spiral galaxies \citep{binney87, chomiuk11}. 

On a larger scale, our Milky Way and virtually all large spirals are
members of small groups of galaxies which may or may not be bound entities \citep{pisani03, berlind06}. 
Detailed studies of the Local Supercluster
by \citet{tully09} concluded that there are virtually no isolated galaxies although more recently a few
relatively isolated dwarfs have been located \citep{karachentseva10}. 
Almost all galaxies in the Local Supercluster can be identified as members of bound groups and clusters or possibly unbound
associations. Therefore, it is likely that the distinction between individual galaxy halos and a larger, intra-group
medium in spiral-rich groups may be largely semantic; i.e., individual galaxy halos may merge to form a
larger gas reservoir that could fill much or all of the volume of the group. Numerical simulations \citep* {cen99, dave99}
suggest that metal-enriched gas escaping from galaxies by supernova- and AGN-driven winds reaches distances of $\geq$
1 Mpc and observations find that O~VI absorbers (the most sensitive tracer of metal-enriched gas at low-$z$) extend
$\sim$ 1 Mpc from the nearest L$^*$ galaxy \citep* {tumfang05, stocke06}. However, the ultimate source, physical conditions and extent of this
low-metallicity, halo/intra-group gas has not yet been identified and characterized. 

On the other hand, large reservoirs of low-metallicity ($[Z]\approx$ -0.5 solar) hot gas already are
detected via thermal bremsstrahlung X-ray emission in groups and clusters of early-type galaxies
\citep{sarazin88, mulchaey00}. The intra-cluster and intra-group gas contains a greater number
of baryons than in the member galaxies and raises the baryon-to-dark matter ratio in groups and
clusters to the cosmic mean \citep [1:5;][] {white93}. This suggests that these systems are ``closed
boxes'' from an evolutionary perspective. However, only those clusters and groups sufficiently evolved
dynamically to contain giant elliptical galaxies at or near their centers are detected in soft X-ray
emission \citep{mulchaey00}. Since no spiral-rich groups were detected in their large {\it ROSAT} survey
of nearby groups of galaxies, \citet {mulchaey96} suggested
that the intra-group gas in these systems is too cool to emit X-rays in sufficient numbers and energy
to be detectable using current instruments. Using the observed scaling-relation between the X-ray 
determined gas temperature T$_x$ and the group
velocity dispersion $\sigma_v$, a factor of 5--30 cooler gas temperatures are predicted for spiral-rich 
groups based on their lower observed galaxy radial velocity dispersions of $\sigma_v$=100-300 km
s$^{-1}$ \citep* {zabludoff98, mulchaey98}. Indeed, some of these groups may not even be fully
gravitationally-bound much less virialized. Instead, \citet {mulchaey96} suggested that this $T\approx$ 10$^6$ K gas
would be most easily detected using absorption-line spectroscopy of background UV-bright sources.
Given the expected temperature range for spiral group gas, the UV absorption doublet of O~VI 1032,
1038 \AA\ would be the most sensitive indicator. However, the fraction of oxygen which is in the
quintuply-ionized state is small ($\leq$10\%) in T $\geq$ 10$^6$ K gas with most of the oxygen being
in more highly-ionized states (O~VII and O~VIII), which are detectable in soft X-ray absorption
lines. If imaging such sources of diffuse, very soft X-ray emitting gas were possible,
then their 
physical extents, covering factors, densities, metallicities and temperatures could be measured.
But this is not currently possible. Following \citet [][Paper 1 hereafter]{savage14} we will term
this potential reservoir ``warm gas'' (T$\sim$ 10$^{5-6}$ K), in contrast to cool (T$\sim$ 10$^4$ K),
photo-ionized gas and hot (T$\geq$ 10$^6$ K) intra-group and cluster gas.

Since the original
proposal of \citet{bahcall69}, it has been suspected that many of the multitude of
absorption lines seen in the spectra of high-$z$ QSOs are due to galaxy halos. The discovery of the
low-$z$ Ly$\alpha$ ``forest'' \citep{morris91, bahcall91} with the {\it Hubble Space
Telescope} (HST) offers us an opportunity to test these theoretical predictions for spiral galaxies and
their environments. An FUV QSO absorption-line method for detecting this warm gas is  
feasible:  Broad Ly$\alpha$ Absorbers (BLAs) with b-values $\geq$ 40 km s$^{-1}$ corresponding 
to $T\geq$ 10$^{5.0}$ K \citep* [][and Paper 1]{richter04, lehner07, danforth10, narayanan09, savage10} and the O~VI
1032,1038 \AA\ absorption doublet associated with these BLAs. But BLAs are very difficult to detect in low or even modest signal-to-noise
(S/N) spectra (e.g., S/N $\leq$ 15) obtained with the Goddard High Resolution Spectrograph (GHRS) or
the Space Telescope Imaging Spectrograph; \citep* [STIS;][]{lehner07, danforth10, tilton12}. Currently,
HST's Cosmic Origins Spectrograph \citep* [COS;][]{green12, osterman11} is routinely
returning S/N $\geq$ 20 $R\approx$ 18,000 far-ultraviolet (FUV) spectra in the 1150-1750 \AA\ band. 

The COS Science Team has obtained S/N$\geq$ 20 FUV spectra of 14 bright QSOs for the purpose
of conducting a sensitive spectroscopic survey, including searching for O~VI absorption and BLAs
\citep* [Paper 1 and][] {savage10, savage11a, savage11b} that could be associated with warm gas. 
Starting with 54 detected O~VI absorption systems containing 85 individual O~VI components,
Paper 1 identified 54 possibly aligned O~VI and Ly$\alpha$ components, out of which temperature measurements
were possible for 45 of the 54. Of these forty-five, 31 (69\%)
show association with cool, photo-ionized gas; the remaining 14 aligned components have observed $b$-values which are
indicative of T$>$10$^5$ K gas and are most likely collisionally-ionized. 
The O~VI in the misaligned O~VI components could be produced in warm
collisionally-ionized or in cool photoionized-ionized gas.
HST/COS is our first opportunity to study these unusual absorbers in some detail; for the most part 
the warm absorbers studied here are detected at too low a contrast to have
been discovered and studied previously \citep* [but see][for early reports from this survey] 
{narayanan10, savage10, savage11a, savage11b}. 
Despite very high S/N ground-based optical spectra,
similar BLA detections have not been made at $z\geq$2 owing to the density of the photo-ionized
Ly$\alpha$ forest absorptions at those redshifts.  

HST/STIS and COS already have provided the first detailed look at the cooler, metal-enriched clouds in
galaxy halos at low redshift using H~I absorption with log N$_{HI} \geq$ 14.0 cm$^{-2}$ (all column
densities herein will be quoted in cm$^{-2}$) and b-values of 20-30 km s$^{-1}$ (all b-values will be
quoted in km s$^{-1}$) and metal ion absorption in species such as C~II, Si~II, Si~III, C~III,
Si~IV, C~IV, N~V and O~VI. Targeting specific galaxy halos at $z\sim$ 0.2 with COS, \citet{tumlinson11}
discovered log N$_{O~VI} \geq$ 14.3 O~VI absorption out to 150 kpc radius around late-type galaxies at
high covering factor.  While ion states lower than O~VI are well-modeled as being in photo-ionization
equilibrium \citep [e.g.,][]{werk14}, the ionization source for the O~VI absorption associated with these cool clouds remains
controversial \citep*{tripp08, danforth08, dave07}. 
\citet{stocke06}, \citet{wakker09} and \citet* {prochaska11a, prochaska11b} used STIS archival spectra to
make galaxy halo detections serendipitously. These studies showed that O~VI absorption extends much
further away from galaxies than the lower ions, as much as $\sim$ 1 Mpc away from $L \geq L^*$
galaxies, much farther than the virial radius of an L$^*$ galaxy (see Stocke et al. 2013 for the luminosity-determined
virial radius values used herein). These stronger O~VI systems are  more readily detected in lower S/N COS spectra \citep [e.g.,][] {tumlinson11}
than the broad, shallow O~VI associated with BLAs in the warm absorbers studied here and in Paper 1 in higher-S/N spectra. Often the
much stronger, photo-ionized absorbers are present in the same absorber complex with BLAs and aligned
O~VI. It is the presence of the cooler clouds which ultimately sets the detectability of the warm
absorbers in many cases rather than the raw S/N of the spectra.

\citet*{werk13, werk14, stocke13, keeney13, keeney14} and others have studied these cooler
clouds using simple photo-ionization modeling to determine approximate physical conditions in cool
halo clouds:  $T\sim$ 10$^4$ K; log n$_{HI}$ = -3 to -4 cm$^{-3}$ and metallicities ranging from $<$ 1\%
solar to slightly super solar. These cloud conditions are inferred by assuming single phase, uniform
density clouds in photo-ionization equilibrium with the extragalactic ionizing radiation field. Given
the very high covering factors observed for the H~I and metal-line absorption \citep*[$\sim$ 100\% inside the
virial radius of late-type, $L\geq$ 0.1L$^*$ galaxies;][]{prochaska11a, stocke13},
these clouds must be quite numerous with a very large total mass. Stocke et al. estimate a total mass
in these clouds approaching 10$^{10}$ M$_{\odot}$ for L$^*$ galaxies, comparable to the mass in the
parent galaxy disk. \citet {werk14} suggest an even larger mass in the photo-ionized clouds.

The O~VI absorption lines usually are not included as constraints in this modeling because the O~VI
may not be photo-ionized in all cases \citep*{dave07, tripp08, danforth08, smith11, shull12}. If the
ubiquitous O~VI absorption discovered by \citet{tumlinson11} is a shock-heated interface between
these cool clouds and a hotter medium, then what is the nature of the hotter medium, its physical
conditions, extent, total mass, metallicity and source? Based on the few, early warm absorber
detections and the properties of the cool clouds, \citet{stocke13} hypothesized that this warm
reservoir of halo gas could contain $\geq$ 10$^{11}$ M$_{\odot}$ of gas. If this is correct, then, as with
more virialized groups and clusters of galaxies, this warm halo gas could contain most of the baryons in
the system. 

Following the lead of Paper 1 \citep [see also] [] {savage10, 
savage11a, savage11b} we use BLA plus O~VI absorption to
investigate this warm halo gas in the light of the group gas hypothesis. In Section 2 and Appendix A we
reanalyze a small fraction of the O~VI plus H~I absorbers presented in Paper 1 to determine a plausible
{\it maximum} number of warm absorbers. The detailed component model fits for these few warm absorber candidates
are shown in the Appendix A, for which a summary of results is provided at the end of Section 2. 

In Section 3 the galaxy environment of the low-$z$ warm absorbers is presented. 
Since the data in-hand comes from wide-field multi-object spectroscopy (MOS) on 4m-class
telescopes, it is not exceptionally deep, limiting discussions of warm absorber environments to
$z <$ 0.15. A comparison between warm absorber environments and those in which cooler, photo-ionized absorbers
are found also is presented. Sections 3.4 and 3.5 provide a summary of the inferences from the galaxy group analysis.
Section 4 describes the major pieces of evidence that favor the hypothesis that these warm absorbers are the detection of a massive reservoir
of gas in spiral galaxy groups. Section 5 summarizes the main results of this study. Throughout this paper we use the
WMAP-9 cosmology with H$_0$= 70 km s$^{-1}$ Mpc$^{-1}$ \citep {hinshaw13}.

\section{Broad Lyman Alpha in Warm Gas at $T\geq 10^{5}$ K}

Paper 1 presented an investigation of all 14 high-S/N COS FUV spectra obtained by the COS Science Team's 
Guaranteed Time Observations (GTO). These 14 sightlines were chosen to provide the highest quality FUV spectra in the
shortest observing time. In these spectra Paper 1 identified 14 potential warm absorbers defined as having 
inferred temperatures T $\geq$ 10$^{5}$
K based on their Ly$\alpha$ and O~VI thermal line widths in systems for which the H~I and O~VI
absorptions are aligned in velocity. The thermal (T) and non-thermal (NT) widths of these absorption
lines were determined from the observed H~I and O~VI line widths by assuming that the NT width is independent of
species while the T width is inversely proportional to the square root of atomic weight. Also these
different species must arise in the same parcel of gas, an assumption allowed for but not demanded by
their alignment in velocity. For this type of analysis to be robust, several species, not just
H~I and O~VI, are required, but are generally not available. 

One physical source of line-broadening not specifically addressed in Paper 1 is Hubble flow broadening. Because later in
this paper we will propose that these warm absorbers might be associated with groups of galaxies requiring their 
physical extents to be $\sim$ 1 Mpc, significant Hubble flow broadening could be present if the associated galaxy groups
are not gravitationally-bound. As with other NT broadening mechanisms, if Hubble flow were important it would require 
near equality of H~I and O~VI b-values (as well as for other species detected) and so would be taken into account in the
analysis of Paper 1. A few of the identified ``warm absorber'' systems do have significant $b_{NT}$ values, which could be
Hubble broadening but this does not alter the temperature determinations found herein. 
  
Paper 1 also identifies two broad O~VI absorbers without associated Ly$\alpha$ absorption,
PKS~0405-123/0.16716 \citep{savage10} and  PHL~1811/0.13280 (Paper 1; Figure 10d) which are likely
warm absorbers (throughout this paper we identify a particular absorber by its sightline/redshift).
Although they do not meet the selection criteria adopted in Paper 1 because a BLA is
not detected, their inferred T $\geq $10$^5$ K makes these probable warm absorbers.   
Except for the few potential warm absorbers described here and reanalyzed in Appendix A, we adopt the classification of
the other $\sim$ 30 H~I $+$ O~VI absorbers from Paper 1 as ``cool'' photo-ionized absorbers and ``misaligned'' 
H~I Ly$\alpha$/O~VI absorbers.

\begin{deluxetable*}{lcccl}
  \tabletypesize{\footnotesize}
  \tablecolumns{5}
  \tablewidth{0pt}
  \tablecaption{O\,VI Absorbers Reanalyzed Here}
\tablehead{\colhead{Sight Line}    &
           \colhead{$z_{\rm abs}$} & 
           \colhead{aligned?} &
           \colhead{new fit} &
           \colhead{Comments}}
 \startdata
  PKS\,2155$-$304 & 0.05722 & NO  & NO     & Ly$\alpha$ complex; possible O\,VIII
absorption \\
  3C\,273         & 0.09010 & NO  & Fig.~12 & probable red wing to Ly$\alpha$ \\
  PG\,1116$+$215  & 0.13850 & YES & Fig.~13 & $b_{NT} \gg b_{thermal}$ suggested by
Paper 1  \\
  PHL\,1811       & 0.17651 & NO  & NO     &                           \\
  H\,1821$+$643   & 0.17036 & NO  & NO, Fig.~15 & O\,VI absorption very weak \\
  Ton\,236        & 0.19457 & YES & Fig.~16 & broad absorber location uncertain \\
  HE\,0153$-$4520 & 0.40052 & YES & NO     &                            \\
  PKS\,0405$-$123 & 0.09657 & YES & NO     & O~VI association with broad or narrow H~I ? \\
  PKS\,0405$-$123 & 0.29770 & YES & Fig.~17 &                            \\
  3C\,263         & 0.11389 & NO  & NO     & BLAs without associated O~VI \\
\enddata
\end{deluxetable*}

Here we adopt a complementary tactic in reanalyzing a {\it few} of the Paper 1 potential warm absorbers by starting with the O~VI
line profile and width, which is often less confused by lower temperature (i.e., cool, photo-ionized)
absorption than the BLA. If the O~VI data comes from COS, then the S/N is usually sufficient to
determine if the profile is symmetrical enough to be well-fit with a single Gaussian profile. In some cases
where H~I Ly$\alpha$ is detected by COS at low-redshift ($z <$ 0.12), 
the O~VI lines are in the spectroscopic band covered by the {\it Far Ultra-violet
Spectroscopic Explorer} (FUSE) satellite; for these 14 very bright targets  
FUSE spectra often possess the quality to permit such an assessment as well. We do {\bf not} reanalyze
those absorbers with complex O~VI line profiles, arguing that in these cases there is ample evidence for
considerable turbulent or component motion. However, for those O~VI absorbers well-fit with a single, 
smooth Gaussian component, we assume that the observed O~VI width is almost entirely thermal. While
this is not guaranteed, it would be quite unlikely (but
not impossible) for turbulent or component motion to mimic a single Gaussian profile. We then
determine whether the Ly$\alpha$ profile is consistent with a broader, thermal Gaussian profile at a velocity
consistent with the O~VI thermal width. An alignment between the BLA and the O~VI absorption and log T(K) $\geq$ 5.0 are
required for an absorber identification as ``warm''. In some cases the S/N is insufficient to make
this determination while in others a symmetric Gaussian is not a good fit to the O~VI profile or the
O~VI does not align with the BLA. In these cases we do not attempt a reanalysis and we adopt the
spectral fits presented in Paper 1 as final. In fact, all line-fits derived in Paper 1 are fully acceptable
but, owing to blends of broad, shallow components and saturated cool components, the Ly$\alpha$ line widths and
velocity centroids are uncertain in some cases.

We emphasize that the Ly$\alpha$ line fits, whether in Paper 1 or herein, are inherently uncertain due to
the complex superposition of ``warm'' and ``cool'' components to the absorption. And in some cases the uncertainties 
in the temperature estimates are large, although almost all absorbers classified as ``warm'' are unambiguously at log T(K) $\geq$ 5.0. 
 
Scrutiny of the O~VI absorbers in Paper 1 finds only seven cases where a reanalysis is suggested by
the O~VI line profile (see Table 1). In these cases the O~VI absorption profile is symmetric but Paper
1 identified a large proportion of the line width as NT or did not find a corresponding Ly$\alpha$
component aligned with the O~VI. Table 1 contains the identification of these seven absorbers by
sightline and redshift, listing b-values, whether the Ly$\alpha$ and O~VI components are aligned in
velocity as determined in Paper 1, whether a new fit is presented herein and whether a new warm
absorber is proposed. The intent of these reanalyses is to determine if more warm absorbers {\it could}
be present in this sample if a different approach to the line-fitting is performed. These reanalyses
were performed by an independent set of co-authors (JTS, CWD) from the analysis in Paper 1.

To this list we have added three absorbers which were not classified as broad, symmetric and aligned
by Paper 1, PKS~2155-304/0.05722, PG~1116+215/0.13850 and 3C~263/0.11389. 
The former
absorber has a tentative O~VIII absorption detection in a {\it Chandra ACIS/LETG} grating spectrum
\citep*{fang07} indicative of warm gas if the O~VIII absorption is real. The second absorber
possesses a very symmetric O~VI line profile and weak broad wings to Ly$\alpha$, which were ascribed
to non-thermal motion in Paper 1. The third absorber has a BLA that appears to be consistent in its
large $b$-value between Ly$\alpha$ and Ly$\beta$ but has no associated O VI. The detailed reanalysis 
of each of these absorbers can be found in Appendix A, from which we now summarize the results. 

Our reanalysis found that in six cases there was no reasonable alternative to the model solutions found by Paper 
1, so that for these six absorbers and all others not reanalyzed, we adopt the model solutions presented in
Paper 1 (see Table 2 and Paper 1, Table 5 for a list of the warm absorbers in order of their quality). 
An updated solution is suggested only for four absorbers, which are presented in Table 3 and summarized below.  

\noindent (1) In the case of 3C~273/0.09010 a fourth Ly$\alpha$ component aligned with the O~VI was added but
with poorly constrained $b$-values. For our best-fit result, the $b$-values derived suggest that a
warm absorber probably is present.  \\
(2) For PG~1116+215/0.13850 our reanalysis finds that a much broader fit to
Ly$\alpha$ is plausible, which is a better match to the broad, symmetric O~VI profile. This new
solution obviates the need for an absorber model which is dominated by NT
motions despite having O~VI 1032, 1038 \AA line profiles which are demonstrably smooth and symmetric to
the limits of the COS signal-to-noise. In this new model the broad absorber has line profiles which
are dominated by thermal broadening and a high value of log T(K)=5.6. \\ 
(3) For Ton~236/0.19457 a viable, alternative fit is suggested which does not differ
significantly from the results of Paper 1 and the presence of a single ``warm absorber'' is confirmed
but with a hotter temperature (log T(K)=5.8) than the Paper 1 solution. \\
(4) For PKS~0405-123/0.29770 we have found that a very broad Ly$\alpha$ absorber aligned with the
broad O~VI absorption can be fit to the data. Thus, we consider the existence of this warm absorber 
at log T(K)=5.7 as probable.

\begin{deluxetable*}{llccccc}
\tabletypesize{\scriptsize}
\tablecolumns{7}
\tablewidth{0pt}
\tablecaption{Properties of Collisionally-Ionized ``Warm Absorbers''}
\tablehead{\colhead{Sight Line}   &
        \colhead{$z_{\rm abs}$} &
        \colhead{$\log N_{\rm HI}$} &
        \colhead{$b_{\rm HI}$} &
        \colhead{$\log N_{\rm OVI}$} &
        \colhead{$b_{\rm OVI}$} &
        \colhead{$\log T$} \\
        \colhead{} &
        \colhead{} &
        \colhead{(cm$^{-2}$)} &
        \colhead{(km s$^{-1}$)} &
        \colhead{(cm$^{-2}$)} &
        \colhead{(km s$^{-1}$)} &
        \colhead{(K)} }
\startdata
HE\,0153$-$4520 & 0.22600 &$ 13.58\pm0.05 $&$ 151\pm15 $&$ 14.23\pm0.01 $&$ 37\pm 1  $&$ 6.14^{+0.08}_{-0.10} $ \\
Mrk\,290        & 0.01027 &$ 14.38\pm0.01 $&$  53\pm 2 $&$ 13.65\pm0.10 $&$ 31\pm10  $&$ 5.07^{+0.13}_{-0.19} $ \\
H1821$+$643     & 0.26656 &$ 13.64\pm0.02 $&$  46\pm 2 $&$ 13.61\pm0.03 $&$ 24\pm 2  $&$ 4.99^{+0.05}_{-0.06} $ \\
3C\,263         & 0.14072 &$ 13.47\pm0.10 $&$  87\pm15 $&$ 13.60\pm0.09 $&$ 33\pm12  $&$ 5.62^{+0.15}_{-0.24} $ \\
PKS\,0405$-$123 & 0.49507 &$ 14.14\pm0.03 $&$  51\pm 5 $&$ 14.31\pm0.02 $&$ 32\pm 2  $&$ 5.00^{+0.12}_{-0.18} $ \\
Mrk\,876        & 0.00315 &$ 14.00\pm0.02 $&$  58\pm 2 $&$ 13.38:       $&$ 35:      $&$ 5.14^{+0.16}_{-0.25} $ \\
3C\,273         & 0.00336 &$ 13.54\pm0.10 $&$  64\pm 7 $&$ 13.39\pm0.08 $&$ 43\pm10  $&$ 5.16^{+0.20}_{-0.38} $ \\
PHL\,1811       & 0.15785 &$ 13.25\pm0.04 $&$  97\pm10 $&$ 13.68\pm0.03 $&$ 31\pm 3  $&$ 5.73^{+0.09}_{-0.11} $ \\
HE\,0226$-$4110 & 0.20701 &$ 13.45\pm0.16 $&$ 100\pm25 $&$ 14.37\pm0.01 $&$ 36\pm 1  $&$ 5.75^{+0.20}_{-0.37} $ \\
H\,1821$+$643   & 0.22497 &$ 13.94\pm0.17 $&$  95\pm11 $&$ 14.25\pm0.01 $&$ 45\pm 1  $&$ 5.65^{+0.11}_{-0.15} $ \\
H\,1821$+$643   & 0.12141 &$ 13.80\pm0.05 $&$  82\pm 7 $&$ 13.69\pm0.04 $&$ 58\pm 7  $&$ 5.33^{+0.15}_{-0.24} $ \\
PHL\,1811       & 0.07773 &$ 13.60\pm0.07 $&$  82\pm10 $&$ 13.53\pm0.08 $&$ 43\pm 7  $&$ 5.49^{+0.13}_{-0.19} $ \\
HE\,0238$-$1904 & 0.47204 &$ 13.83\pm0.03 $&$  47\pm 4 $&$ 14.16\pm0.01 $&$ 20\pm 1  $&$ 5.06^{+0.08}_{-0.10} $ \\
H\,1821$+$643   & 0.17035 &$ 13.65\pm0.01 $&$  54\pm 2 $&$ 13.20\pm0.22 $&$ 31\pm 2  $&$ 5.10^{+0.14}_{-0.22} $ \\
PKS\,0405$-$123 & 0.16716 & \nodata\tablenotemark{a} & \nodata    &$ 13.85\pm0.01 $&$ 56\pm 2  $&$>6.1                  $ \\
PHL\,1811       & 0.13280 & \nodata\tablenotemark{b} & \nodata    &$ 13.57\pm0.03 $&$ 25\pm 2  $&$\sim5.7               $ \\
\enddata

\tablenotetext{a}{HI non-detection}
\tablenotetext{b}{Marginally detected}
\end{deluxetable*}

\begin{deluxetable*}{llccccc}
\tabletypesize{\scriptsize}
\tablecolumns{7}
\tablewidth{0pt}
\tablecaption{Properties of Probable Collisionally-Ionized ``Warm Absorbers''}
\tablehead{\colhead{Sight Line}   &
        \colhead{$z_{\rm abs}$} &
        \colhead{$\log N_{\rm HI}$} &
        \colhead{$b_{\rm HI}$} &
        \colhead{$\log N_{\rm OVI}$} &
        \colhead{$b_{\rm OVI}$} &
        \colhead{$\log T$} \\
        \colhead{} &
        \colhead{} &
        \colhead{(cm$^{-2}$)} &
        \colhead{(km s$^{-1}$)} &
        \colhead{(cm$^{-2}$)} &
        \colhead{(km s$^{-1}$)} &
        \colhead{(K)} }
\startdata
3C\,273         & 0.09010 &$ 13.23\pm 0.02 $&$ 49\pm3 $&$ 13.0 \pm0.3 $&$ 20:    $&$ \approx5.1 $ \\
PG\,1116$+$215  & 0.13850 &$ 13.6 \pm 0.1  $&$ 86\pm11$&$ 13.8 \pm0.1 $&$ 35\pm5 $&$ 5.6^{+0.1}_{-0.2}$ \\
Ton\,236        & 0.19457 &$ 13.7 \pm 0.1  $&$ 90\pm9 $&$ 13.6 \pm0.2 $&$ 40\pm17$&$ 5.8^{+0.2}_{-0.3}$ \\
PKS\,0405$-$123 & 0.29770 &$ 13.15\pm 0.10 $&$100\pm10$&$ 13.55\pm0.10$&$ 57\pm4 $&$ 5.7^{+0.1}_{-0.2}$ \\
\enddata
\end{deluxetable*}

\begin{deluxetable*}{llccccc}
\tabletypesize{\scriptsize}
\tablecolumns{7}
\tablewidth{0pt}
\tablecaption{Properties of Photo-Ionized ``Cool Absorbers''}
\tablehead{\colhead{Sight Line}   &
        \colhead{$z_{\rm abs}$} &
        \colhead{$\log N_{\rm HI}$} &
        \colhead{$b_{\rm HI}$} &
        \colhead{$\log N_{\rm OVI}$} &
        \colhead{$b_{\rm OVI}$} &
        \colhead{$\log T$} \\
        \colhead{} &
        \colhead{} &
        \colhead{(cm$^{-2}$)} &
        \colhead{(km s$^{-1}$)} &
        \colhead{(cm$^{-2}$)} &
        \colhead{(km s$^{-1}$)} &
        \colhead{(K)} }
\startdata
PG\,0953+414    & 0.00212 &$ 13.15\pm 0.03 $&$ 38\pm3 $&$ 13.60\pm0.06$&$ 43\pm8 $&$ < 4.5            $ \\
HE\,0226-4110   & 0.01750 &$ 13.20\pm 0.04 $&$ 27\pm4 $&$ 13.91\pm0.16$&$ 10\pm6 $&$ 4.61^{+0.14}_{-0.22}$ \\
H\,1821+643     & 0.02443 &$ 14.24\pm 0.01 $&$ 29\pm1 $&$ 13.42\pm0.09$&$ 23\pm7 $&$ 4.30^{+0.31}_{-0.30}$ \\
PG\,1259+593    & 0.04625 &$ 15.30\pm 0.20 $&$ 30\pm1 $&$ 13.90\pm0.10$&$ 40\pm13$&$ < 4.80              $ \\
PG\,1259+593    & 0.04625 &$ 14.70\pm 0.20 $&$ 25\pm10$&$ 13.70\pm0.15$&$ 17\pm8 $&$ 4.50^{+0.50}_{-0.30}$ \\
PG\,1116+215    & 0.05897 &$ 13.57\pm 0.01 $&$ 33\pm2 $&$ 13.49\pm0.05$&$ 27\pm16$&$ 4.36^{+0.54}_{-0.36}$ \\
PG\,0953+414    & 0.06808 &$ 14.34\pm 0.05 $&$ 19\pm1 $&$ 14.29\pm0.03$&$ 12\pm1 $&$ 4.14^{+0.11}_{-0.14}$ \\
3C\,273         & 0.12007 &$ 13.50\pm 0.01 $&$ 23\pm1 $&$ 13.38\pm0.02$&$ 10\pm2 $&$ 4.44^{+0.06}_{-0.07}$ \\
PG\,0953+414    & 0.14231 &$ 13.57\pm 0.01 $&$ 26\pm1 $&$ 14.09\pm0.01$&$ 19\pm1 $&$ 4.21^{+0.08}_{-0.10}$ \\
PG\,0953+414    & 0.14213 &$ 13.47\pm 0.01 $&$ 29\pm1 $&$ 13.60\pm0.03$&$ 29\pm4 $&$ < 4.25              $ \\
HE\,0153-4520   & 0.14887 &$ 13.26\pm 0.03 $&$ 34\pm4 $&$ 14.02\pm0.01$&$ 25\pm1 $&$ 4.53^{+0.18}_{-0.32}$ \\
\enddata
\end{deluxetable*}

In addition, for the 3C~263/0.11389 absorber we confirm the detailed fits in Paper 1 but also find that a warm
absorber is possibly present based on a smooth, symmetric BLA detected consistently in Ly$\beta$ (i.e., both
line-fits yield consistent column densities and $b$-values). There is no detection of O~VI absorption
coincident with this BLA.
 
With these revisions there is a net gain of three warm absorbers to the BLA $+$ broad O~VI list. By
including the two O~VI-only absorbers from Paper 1 (PHL~1811/0.13280 and PKS~0405-123/0.16716) as also
being probably warm, a total of as many as 20 warm absorbers can be present in this sample. (The
possible warm absorber, 3C~263/0.11389, is not included in this tally). The
distribution of temperature values for the 20 members of the COS/GTO warm absorber sample is shown in
Figure 1. Unlike the initial discovery of the O~VI-only system PKS~0405-123/0.16716 \citep {savage10}, most
of the derived temperatures are below T$\sim$10$^6$ K. Tables 2 and 3 provide the basic observational
data for all of the warm absorbers in this sample. Table 2 compiles the spectral data from 
Paper 1 for the warm absorbers for which we have adopted the Paper 1 line-fits. Table 3 lists the same information
for the warm absorbers for which this paper supplies revised line-fits (see Appendix A). Table 4 provides the
same observational data for the sample of low-$z$ ``cool'', photo-ionized absorbers which are used for statistical
purposes herein (11 from Paper 1 and two, PG~1259+593/0.04625a and b, added from another high-S/N COS spectrum
and which is described in \citet {danforth14}).

\begin{figure}
\epsscale{1}
\plotone{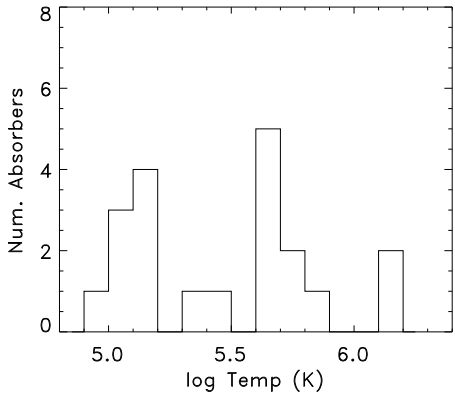}
\caption{The Inferred Temperatures for the 20 Warm Absorbers in the COS High-S/N Survey. The errors bars 
on the log T values are typically 0.1-0.3 dex (see Tables 2 and 3).}
\end{figure}

Due largely to the presence of narrow Ly$\alpha$ and O~VI components that are plausibly cool
halo clouds and their shocked interfaces \citep* {tumlinson11, stocke13} and due somewhat
to the limited S/N of the COS spectra, we cannot dismiss the possibility that very low
contrast, very broad ($b>$ 100 km s$^{-1}$) Ly$\alpha$ lines are present in some of these absorber
complexes. We can state that their presence is neither required by the constraints of the O~VI and
Ly$\alpha$ line fits of Paper 1 nor by the reanalysis in the Appendix. But as the very high S/N COS
spectra of 3C~273, PKS~2155-304 and PKS~0405-123 show, it is the presence of the cooler component
absorption, not the S/N, which is the major hindrance in discovering more warm absorber systems.
These limitations are difficult to quantify, so we make no attempt to do so. Rather we determine
the total pathlength for discovering warm absorbers using a simplified calculation set by the S/N of
the COS spectra alone. 

The pathlength ($\delta$z) for warm absorber discovery is estimated using the entire
redshift path of the 14 QSOs observed in this program minus the obscuration  created by the Galactic
damped Ly $\alpha$ line plus the high and low ionization metal lines that arise in the Galactic disk
or halo. In this case we obtain a $\delta$z= 4.0 \citep[see][] {danforth14} and the number of warm absorbers per unit 
redshift, \dNdz\ = 3.5--5 based on the range in the number of warm absorbers in these 14 sightlines: 
14 proposed by Paper 1; 20 by this
paper. This is approximately 4 times the \dNdz\ for Mg~II/Lyman limit system absorbers, which are
thought to be optically-thick clouds in extended galaxy halos of radius $\sim$ 100 kpc with high
covering factors \citep*{steidel95, churchill00, kacprzak10, kacprzak11}. If the warm absorbers
also are associated only with luminous galaxies similar to the Mg~II absorbers, their numbers suggest an
absorber size of $\sim$ 200 kpc, comparable to the virial radius of an L$^*$ galaxy. As with the class
of Mg~II absorbers this inference needs to be confirmed by detecting galaxies associated with these
warm absorbers. Luckily the combination of HST/COS and {\it FUSE} UV spectroscopy provides some
detections at very low redshift by historical absorption line standards (i.e., $z\leq$ 0.15) whose
fields can be searched for associated galaxies. 

\section{Galaxy Groups Associated with \OVI\ Absorbers}
\label{groups}

In order to determine the galaxy environment of the warm and cool absorbers (see
Tables 2, 3 \& 4 for a summary of observed properties of the warm and cool absorbers
considered here), we
have used a combination of large-area redshift surveys like the Sloan
Digital Sky Survey (SDSS) and the 2dF survey done at the
Anglo-Australian Telescope (AAT) and our own multi-object spectroscopy
(MOS) program centered on the sightlines observed by the COS Science
Team. \citet* {stocke06, stocke13} contains a description of our use of
the large-area sky survey redshift catalog, which includes targeted observations
made by others of specific sightlines in this survey \citep [e.g.,][] {morris93, tripp98, chen09, prochaska11b,
johnson13}. Our new observations include all but one of the 14 sightlines in Paper 1 and extend $\sim2$--3
magnitudes deeper than the SDSS \citep{keeney14}; the H~1821+643
field has not yet been observed by our survey because the target was
observed with COS only recently (the only galaxy redshift observations
in this region were reported by \citet{tripp98} and have a
somewhat uncertain depth and breadth reported). For these 14 sightlines spectra
and redshifts were obtained mostly at the Wisconsin-Indiana-Yale-NOAO
(WIYN) 3.5-m telescope using the HYDRA MOS.  Additional observations of
southern targets were obtained at the Cerro-Tololo Inter-American
Observatory's (CTIO) 4m Blanco Telescope using the version of HYDRA
available there. A few additional southern targets were observed at
the AAT using the AAOmega MOS system. Occasional long-slit spectra
of galaxies nearest to absorbers were obtained at the Apache Point
Observatory (APO) 3.5m telescope using the Dual-Imaging Spectrograph
(DIS). Individual galaxy redshifts from this survey have typical velocity errors of
$\pm$ 30 km s$^{-1}$.

\input tab5.tex
\input tab6.tex
\input tab7.tex

The MOS grating setup was selected to obtain good throughput just
long-ward of 4000~\AA\ so that \ion{Ca}{2} H \& K might be detectable
even for faint and diffuse targets at low redshift. At WIYN the
$60\arcmin$ field-of-view (FoV) corresponds to 6.7 Mpc at $z=0.1$ and
so is excellent for determining the large-scale environment of low-$z$
absorbers. The FoVs at the CTIO 4m ($40\arcmin = 4.5$ Mpc at $z=0.1$)
and at the AAT ($120\arcmin = 13.4$ Mpc at $z=0.1$) are also good
matches to this goal. The $>90$\% completeness magnitude for our
observations at WIYN/HYDRA (in approximately two-hour exposure times)
is $g=20$. The 90\% completeness limit at AAT/AAOmega is $g=20.5$, but
CTIO's HYDRA system is less efficient than its counterpart at WIYN,
with the CTIO observations being 90\% complete to $g=19$. A $g=20$ mag limit
corresponds to $0.16\,L^*$ at $z=0.1$ and $1.2\,L^*$ at $z=0.25$. Since it is
necessary to find other associated galaxies at least 2 magnitudes
fainter than $L^*$ in order to determine with some confidence if a
small galaxy group is present or not, this 4m
MOS survey is excellent for determining details of the galaxy
environments only at $z\leq0.10$. And it will provide some preliminary
information for $0.10 \leq z \leq 0.15$, but little detail in most
cases at $z>0.15$. Details of the wide-angle 4-m galaxy survey work
can be found in \citet{keeney14}.

\subsection{Procedure for Defining Groups of Galaxies Associated with Absorbers}

In this discussion and throughout this paper we use the term ``galaxy group'' to describe the large-scale
(2-5 Mpc) galaxy environment of the absorbers studied here. We have defined these groups entirely by the
observational properties provided by individual galaxy redshifts and sky locations 
using a ``friends-of-friends'' (FoF) approach similar to that employed by \citet*{berlind06, berlind08}
using the SDSS data; i.e., we do not require detections of soft X-ray bremsstrahlung, which occurs only
in elliptical-dominated groups \citep{mulchaey00}, nor a theoretically-inferred dark matter halo mass threshold. 
Indeed, most of the groups we have found do not contain early-type galaxies nor are they known sources of 
diffuse, soft X-ray emission but rather these groups are similar in galaxy content to spiral-rich groups
found in the Local Supercluster (e.g., the M101/M51 group or the M96 group; see Section 3.2). Our use of the term 
``galaxy group'' does not necessarily imply that these physical entities are either bound or virialized. 
Using standard crossing time arguments  \citep*{pisani03, berlind06}, the velocity dispersions measured for these groups 
are sufficient for only one or two crossing times in a Hubble time. Further, many of these groups do not have
Gaussian-distributed velocity distributions and physical locations required for such arguments to be robust (see
the velocity distributions and galaxy locations for these groups in Figures 2, 3 \& 4). 
Therefore, by using the term ``galaxy group'', we imply only that
these absorbers are surrounded by large and small galaxies in their vicinity, for which
we have measured velocity distributions whose dispersions provide characteristic kinetic
energies to which the gas kinetic energies can be compared using the COS-observed line-widths.
Deeper spectroscopic observations for many of these groups are underway to further characterize
the group memberships and velocity distributions; present observations limit a detailed discussion
of group properties to all absorbers at $z\leq 0.10$ and a few others out to $z\leq$0.15, based
on the depth to which galaxy surveys have been completed in those regions.

Tables 5, 6 \& 7 list the basic characteristics of nearest galaxies and galaxy groups found around
the QSO sightlines at the redshifts of the absorbers. For each absorber in these tables the galaxy group was 
identified and its membership defined using an FoF analysis \citep [e.g.,][] {berlind06}. 
Table 5 lists the galaxy environment data for the warm absorbers at
$z\leq 0.15$. Similar data are shown in Table 6 for absorbers judged
to be in photo-ionization equilibrium (i.e., ``cool'' absorbers at $\log{T(K)} < 5.0$) and
Table 7 for absorbers whose \lya\ and \OVI\ absorption lines are
``misaligned'' so that no unambiguous model can be constructed for these absorbers (see Paper
1 for details). Information in these Tables includes (in columns 1 \&
2) the sightline and absorber redshift; (3) the nearest galaxy
luminosity (L$_{ng}$) in L$^*$ units; (4) the impact
parameter ($\rho$) to the nearest galaxy in kpc; (5) the impact parameter
as a fraction of the virial radius of the nearest
galaxy (R$_{vir}$). The virial radii for the nearest galaxies
are calculated using the ``halo matching'' scaling between luminosity and virial
radius due to \citet {moster13} as shown in Figure 1 of \citet {stocke13}; 
(6) the velocity difference between the absorber and the
nearest galaxy ($\Delta$v$_{ng}$) in km s$^{-1}$ ($\pm$ 30 km s$^{-1}$);
(7) the integrated group luminosity using all group galaxies observed
(L$_{grp}$) in L$^*$ units; (8) the number of galaxies in the group
(N$_{grp}$); (9) the impact parameter ($\rho_{grp}$) to the group centroid in kpc; (10) the
group impact parameter in units of the virial radius (R$_{vir}$). The virial radii 
for the galaxy groups were calculated using $M_{\rm grp}/L_{\rm grp} = 500\,(M/L)_{500}$
 in solar units, which is close to the mean value for groups of the richness found (see discussion below);
(11) the velocity difference between the
absorber and the group mean velocity ($\Delta$v$_{grp}$)
in km s$^{-1}$ ($\pm$ 30 km s$^{-1}$); (12) the group velocity dispersion
($\sigma_{grp}$ and its associated error; see detailed discussion below) in km s$^{-1}$. Values and errors are listed
for all groups with $N_{grp} \geq$ 8. Single valued estimates are made for groups with $N_{grp} \geq$ 3 (see detailed
discussion below);
(13) a second estimate of the group velocity dispersion assuming that the observed group is virialized using
the procedure of \citet* [][see detailed discussion below] {bryan98}; 
and (14) the galaxy survey $>$90\% completeness limit (L$_{comp}$) in L$^*$ units at the absorber redshift out to 2 Mpc
radius from the sightline.

The velocity differences between absorbers and galaxies (column 6) or groups (column 11) are computed as follows:
$\Delta v = c (z - z_{abs}) / (1 + z_{abs})$. Due to the range in redshifts and the variation in galaxy
survey work along these sightlines, the completeness limits (column 14) can vary substantially.
In some cases the completeness luminosity is a function of
impact parameter; e.g., the H~1821+643/0.12141 field has deeper MMT/MOS
close to the sightline as reported by \citet {tripp98}. In
these cases the completeness luminosity is listed with its associated
impact parameter (in kpc) in parentheses and the narrow/deep survey limit separated from the broad/shallow
survey limit by a ``/'' in column (14).

Our group finding algorithm has several steps. In each case we start with a catalog of all galaxies within $\pm1000$~\kms\ of the absorber redshift
with impact parameters $\leq2$~Mpc from the QSO sight line. Then we use an FoF algorithm to search for
galaxies associated with the closest galaxy to the QSO sight line. In our FoF algorithm each galaxy has individualized linking
lengths equal to $5R_{\rm vir}$ on the sky and five times larger than that in velocity space to match the relative scalings 
between the plane-of-the-sky linking length and the redshift linking length ($l_{\rm sky}$ and $l_z$) from \citet{berlind06}.
In order to make sure that we have investigated a large enough volume around the absorber to locate all group members, we proceed as follows.
We assume that all of our potential groups have $M_{\rm grp}/L_{\rm grp} = 500\,(M/L)_{500}$ in solar units, which provides
a scaling relation between a group's luminosity and its radius and velocity dispersion (this mass-to-light ratio is close to the median
value for the groups we have found \citep* [see analysis associated with Figure 1 in][] {stocke13}). Using the preliminary group luminosity, 
L$_{grp}$, we then estimate the ``virialized'' group radius and velocity dispersion using:

\begin{align}
\label{eqn:Rgrp}
R_{\rm grp} &=  \left(\frac{3 M_{\rm grp}}{4\pi\Omega_{\rm m}\Delta_{\rm vir}^0\rho_{\rm crit}}\right)^{1/3} = 446\,\left(\frac{M}{L}\right)_{500}^{1/3}\,\left(\frac{L_{\rm
grp}}{L^*}\right)^{1/3}~{\rm kpc} \\
\label{eqn:sigma}
\sigma_{\rm vir} &= \sqrt{\frac{GM_{\rm grp}}{2R_{\rm grp}}} = 155\,\left(\frac{M}{L}\right)_{500}^{1/3}\,\left(\frac{L_{\rm grp}}{L^*}\right)^{1/3}~\kms,
\end{align}

\noindent where we have assumed $\Omega_m = 0.282$ \citep{hinshaw13}, $\Delta_{\rm vir}^0 \equiv \Delta_{\rm vir}(z=0) = 350$ \citep{bryan98},
and $\rho_{\rm crit} = 9.205\times10^{-30}\,h_{70}^2~{\rm g\,cm^{-3}}$ \citep*{shull12}. Since all of our groups have $z_{\rm grp}<0.15$,
we ignore any evolution of $\Omega_{\rm m}$ or $\Delta_{\rm vir}$ with redshift. While we make no correction in this calculation to account for galaxies 
with luminosities below our completeness limit, such a correction is small in all cases for which there are sufficient group galaxies 
identified to quote reliable velocity disperisons (N$_{grp} \geq$ 8). Values calculated using equation (2) above are listed in column (13) of Tables 5, 6 and 7.
After the values of $R_{\rm grp}$ and $\sigma_{\rm vir}$ are calculated, we check to ensure that our search volume is large enough to include
$\pm5\sigma_{\rm vir}$ from the group redshift and $2.5R_{\rm grp}$ from the group center. If it is not, then we expand the search volume,
determine a new FoF group with updated values of $R_{\rm grp}$ and $\sigma_{\rm vir}$, and iterate until our search volume satisfies these criteria.

Once this process is complete we use the FoF group as the first step in an iterative procedure within the final search volume to define the final group membership.
At each step we determine the group's center on the sky and in redshift space and identify all galaxies (i.e., not just those identified as
members in the previous step) within $\pm3\sigma_{\rm vir}$ of the group redshift and $1.5R_{\rm grp}$ of the group center as being group members.
We then repeat this process until the group membership does not change from one iteration to the next. Usually the group membership converges in $\leq3$ iterations.

We have utilized the robust statistics described in \citet*{beers90} to minimize the effect of potentially spurious group members on the physical
quantities of interest for our groups. The geometric center of the group on the sky and the group redshift are determined using the bi-weight
location estimator, and the group's velocity dispersion ($\sigma_{\rm grp}$) is calculated using the ``gapper'' scale estimator \citep*{beers90}, 
which involves ordering the recession velocities $v_i = cz_i/(1+z_{\rm grp})$ from smallest to largest and defining weights, $w_i$, and gaps, 
$g_i$, such that:

\begin{align}
\label{eqn:wg}
w_i &= i\,(N_{\rm grp}-i) \\
g_i &= v_{i+1} - v_i, \qquad i=1,2,...,N_{\rm grp}-1. \nonumber
\end{align}

The rest-frame velocity dispersion of the group is then:
\begin{equation}
\label{eqn:gapper}
\sigma_{\rm grp} = \frac{\sqrt{\pi}}{N_{\rm grp}(N_{\rm grp}-1)} \sum^{N_{\rm grp}-1}_{i=1} w_i g_i.
\end{equation}

\noindent These values are listed in column (12) in Tables 5, 6 and 7. 

Our group finding procedure must indicate that a putative group has more than 
two members for us to quote any group properties. For groups with 3--7 members we list tentative group properties with approximate velocity
dispersions (see column 12 for listings without formal error estimates). 
When a group has $N_{\rm grp} \geq 8$ members we estimate the 90\% confidence interval of the velocity dispersion with an ``$m$ out of $n$''
bootstrap method using 10,000 resamplings of size $8 \leq m \leq N_{\rm grp}$, where $m$ is chosen using the prescription of \citet{bickel08}.
When a group has $N_{\rm grp} < 8$ members we quote the dispersion returned by the gapper estimator without error bars because we have too few
samples of the parent distribution to generate reliable confidence intervals. These 90\% confidence errors are much greater than the 
typical uncertainty of our galaxy redshifts ($\sim30$~\kms).
The measured velocity dispersion, $\sigma_{\rm grp}$, usually agrees remarkably well with that predicted solely from the group's luminosity,
$\sigma_{\rm vir}$ (see Equation~\ref{eqn:sigma}), and for groups with $N_{\rm grp} \geq 8$ members $\sigma_{\rm vir}$ is within the 90\%
confidence interval of $\sigma_{\rm grp}$ for all but one case, PG~1116+215/0.05897, whose $\sigma$ values differ by an amount just greater than
the estimated 90\% confidence interval. We take the general agreement between these two values as an indication that our
group identifications are relatively secure and that most of these systems are close to being virialized.

Associated groups with $N_{\rm grp} <$ 8 members may have significantly larger dispersions than we have calculated 
\citep* {beers90, zabludoff98, osmond04, helsdon05}; for this reason we have not quoted errors on these values.
Deeper observations are required to characterize these groups adequately since both their group boundaries and memberships and also their velocity dispersions
can change appreciably with the discovery of a few new group members. 
Basic galaxy identifiers, sky positions and redshifts for all group members shown in Figures 2, 3 and 4 ($N_{grp} \geq$ 8) can be found in
Appendix B. Full galaxy redshift results from our on-going survey along these and other COS sightlines can be
found in \citet{keeney14}.

\begin{figure*}[!ht]
\begin{center}
\epsscale{1}
\plottwo{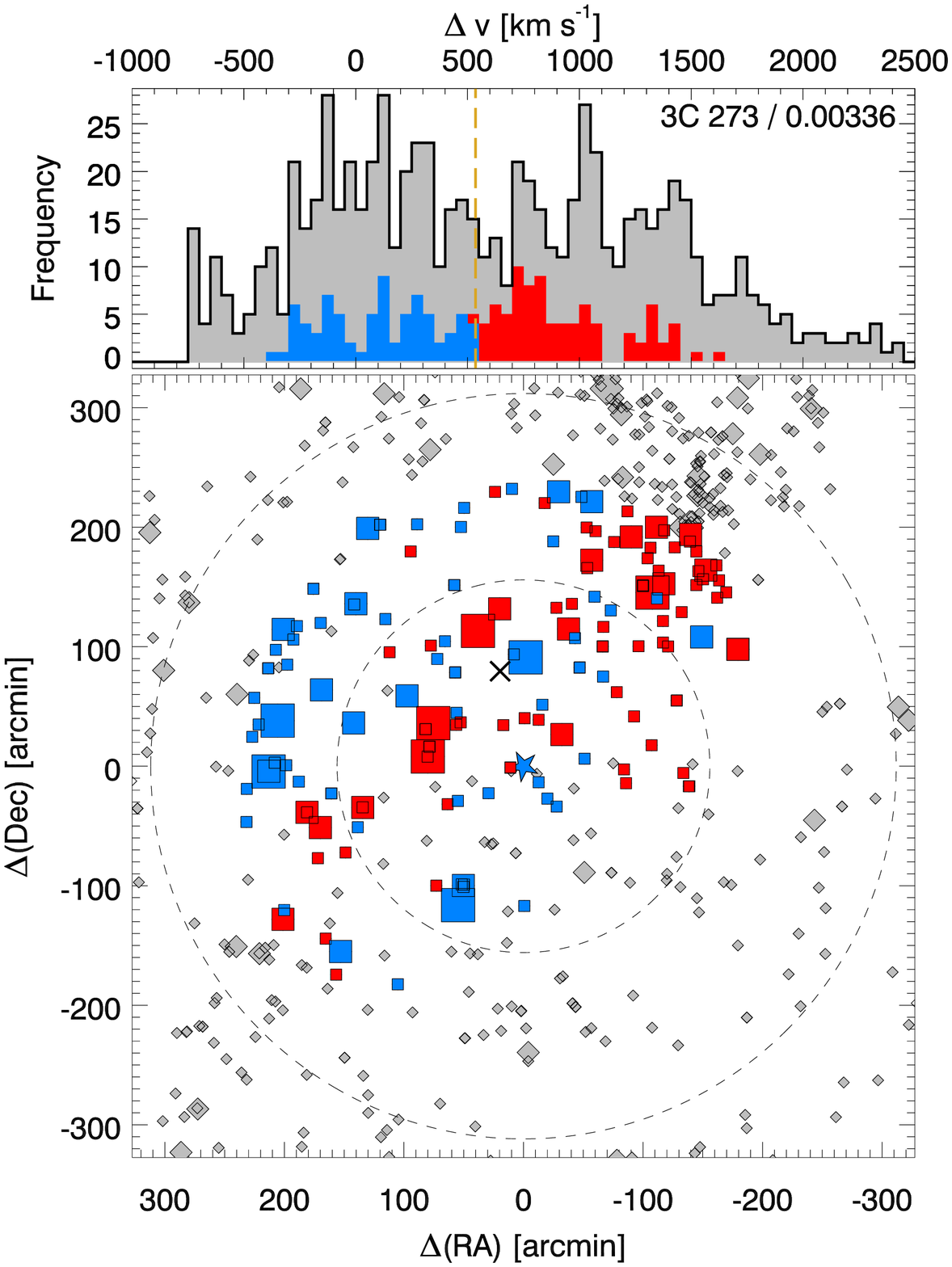}{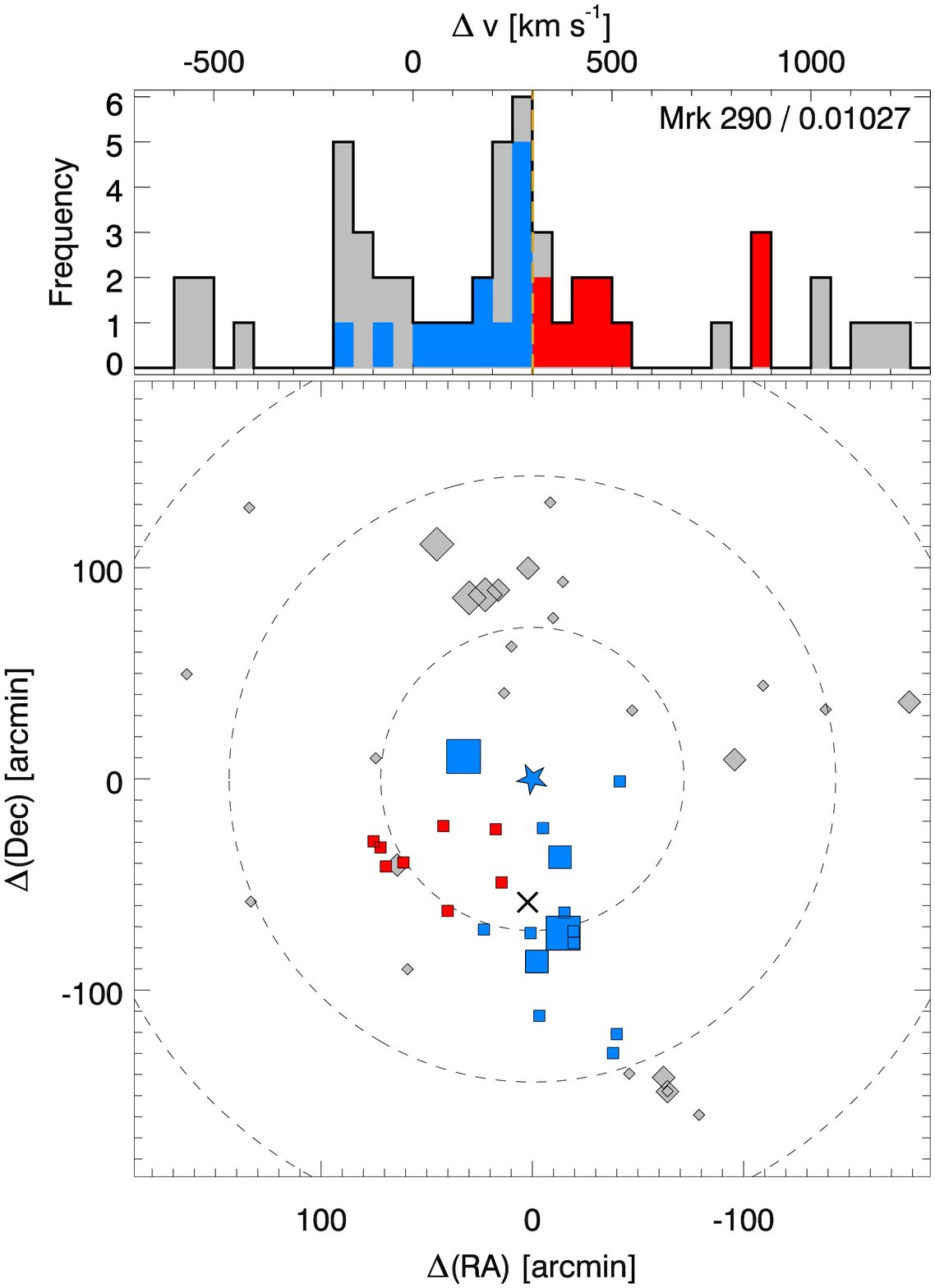}
\plottwo{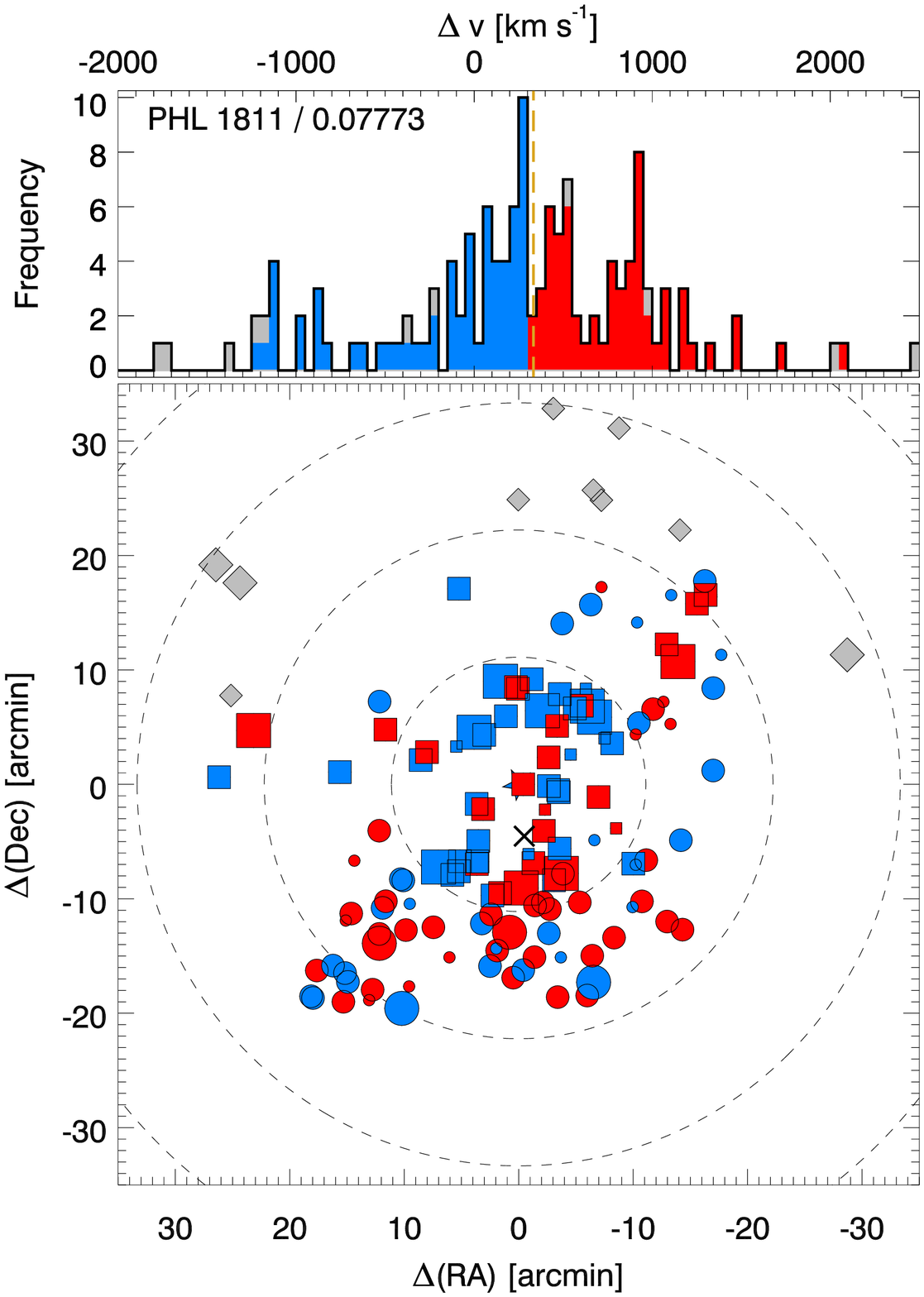}{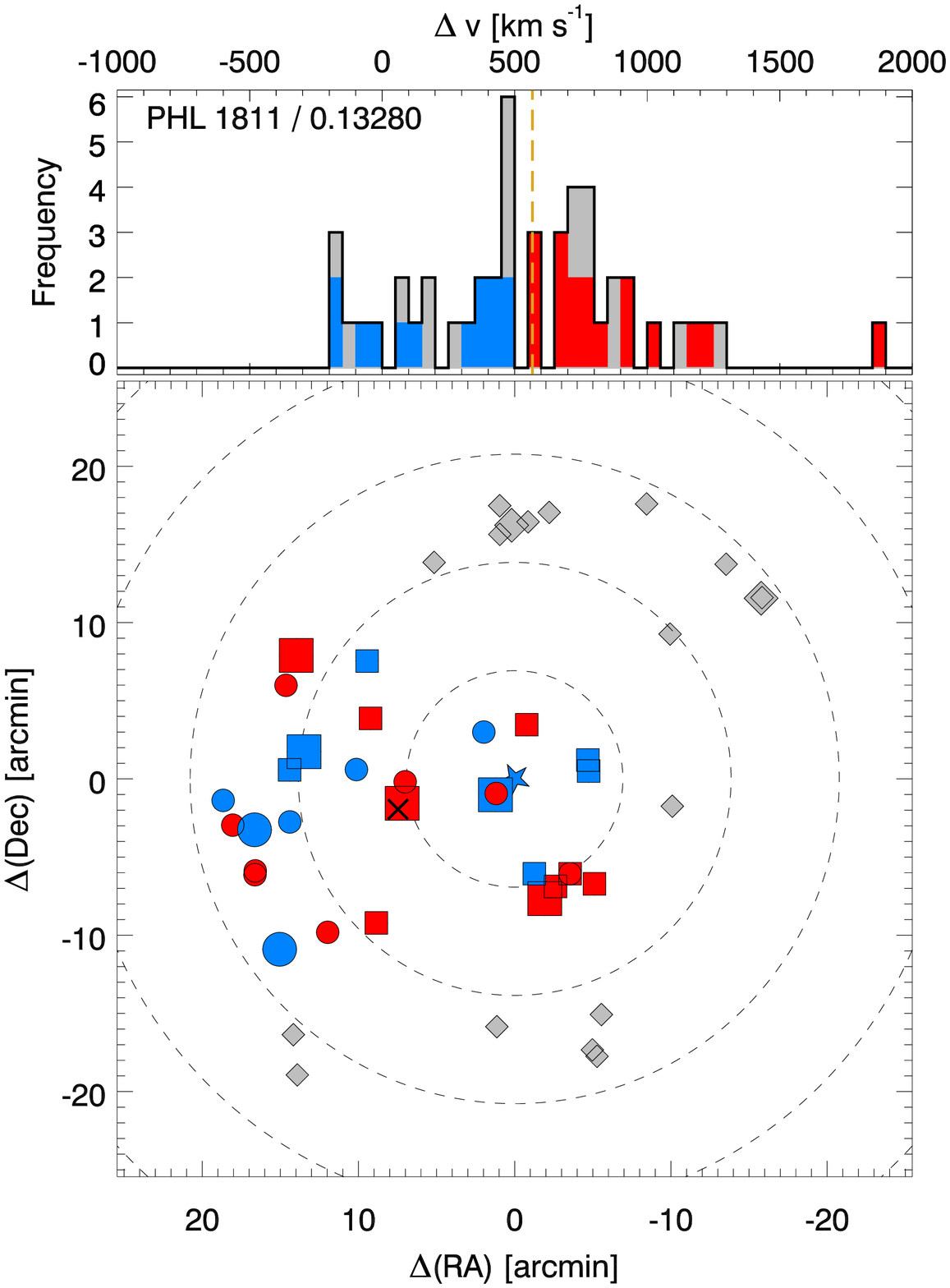}
\caption{A montage of four galaxy groups associated with warm absorbers (see Table 5). At the top of each absorber plot is a
histogram of galaxy radial velocities in $\Delta v$ relative to the absorber. The colored bins
are the group members, blue and red bins are blueshifted and redshifted relative to the group centroid; 
the grey bins are the galaxies excluded by the group finding algorithm. 
The dashed yellow vertical line is the group velocity centroid. At bottom is the spatial
distribution of galaxies around the absorber on the sky. The coordinates
are in arcminutes relative to the QSO sightline which is indicated by the large star at the origin.
The dashed circles show impact parameters in 1 Mpc units increasing outwards.
The positions of group galaxies are marked with squares for redshift data from SDSS or our nearby
galaxy catalog \citep [see][and Appendix B]{stocke06} or circles for
redshift data from our own 4m MOS survey \citep [see] [and Appendix B]{keeney14};  
the grey diamonds are non-group members \citep [see][]{keeney14}. 
The symbol size indicates the galaxy luminosity: large = super-L$^*$, medium = sub-L$^*$, and small = dwarfs. 
Blue symbols mark group galaxies with $z < z_{grp}$; red symbols are group galaxies with $z > z_{grp}$. 
The large ``X'' is the group centroid on the plane of the sky.}
\end{center}
\end{figure*}

\begin{figure*}[!ht]
\begin{center}
\epsscale{1}
\plottwo{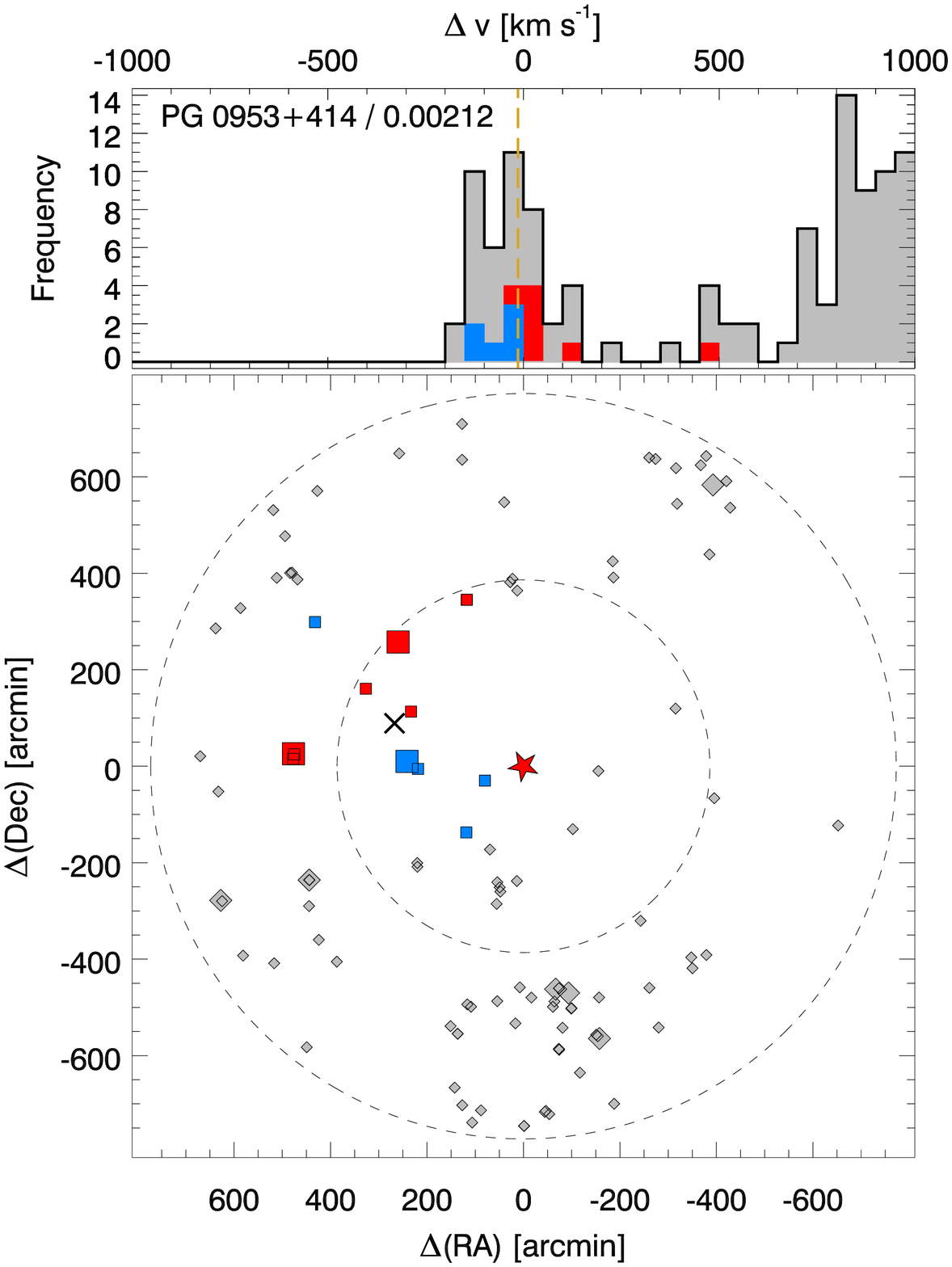}{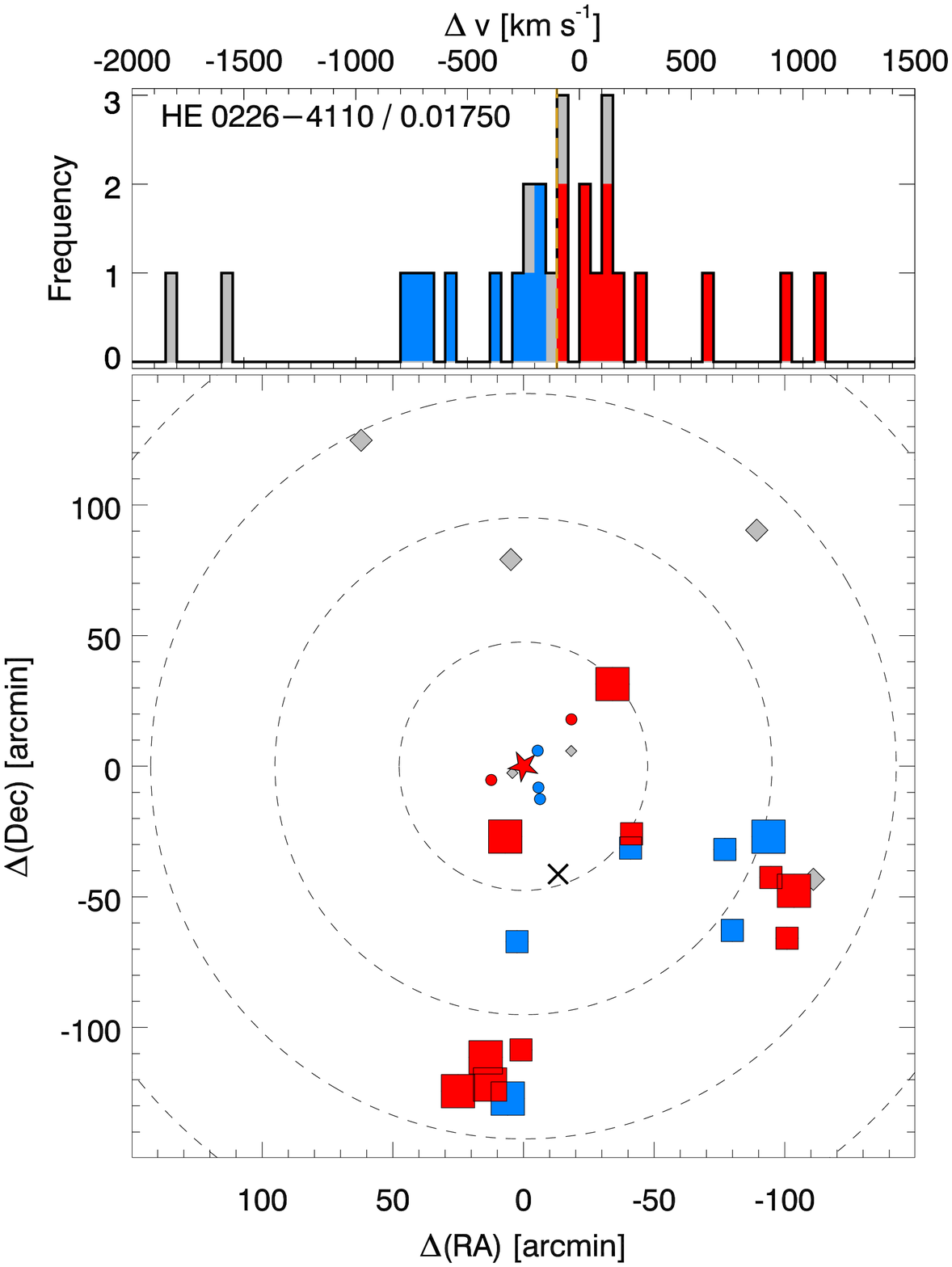}
\plottwo{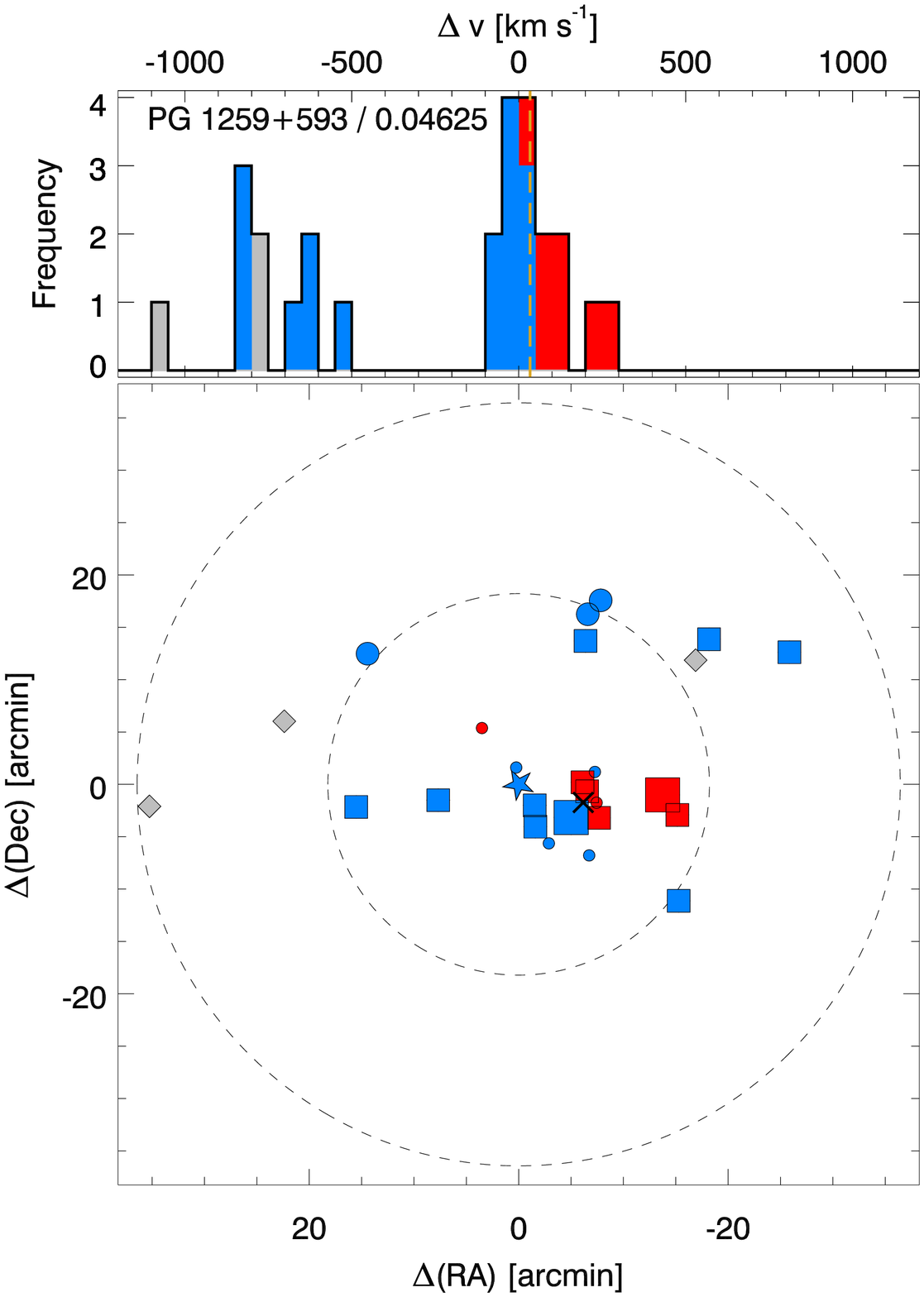}{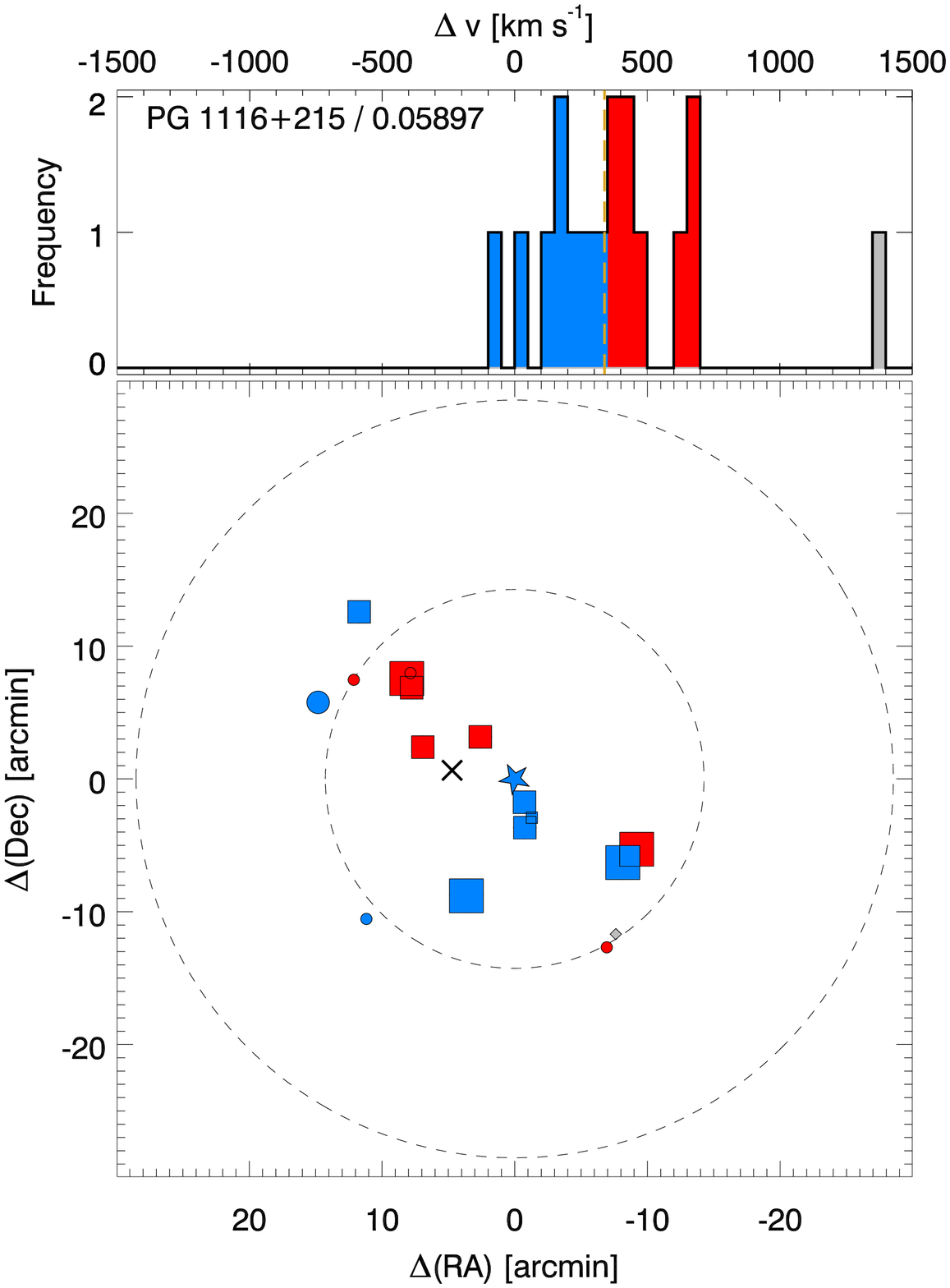}
\caption{A montage of five galaxy groups associated with cool absorbers (see Table 6).
Coordinates and symbology are described in the caption to Figure 2. See the text for a
detailed discussion of the
absorber groups shown in this Figure.}
\end{center}
\end{figure*}

\begin{figure}[!ht]
\begin{center}
\figurenum{3}
\epsscale{1}
\plotone{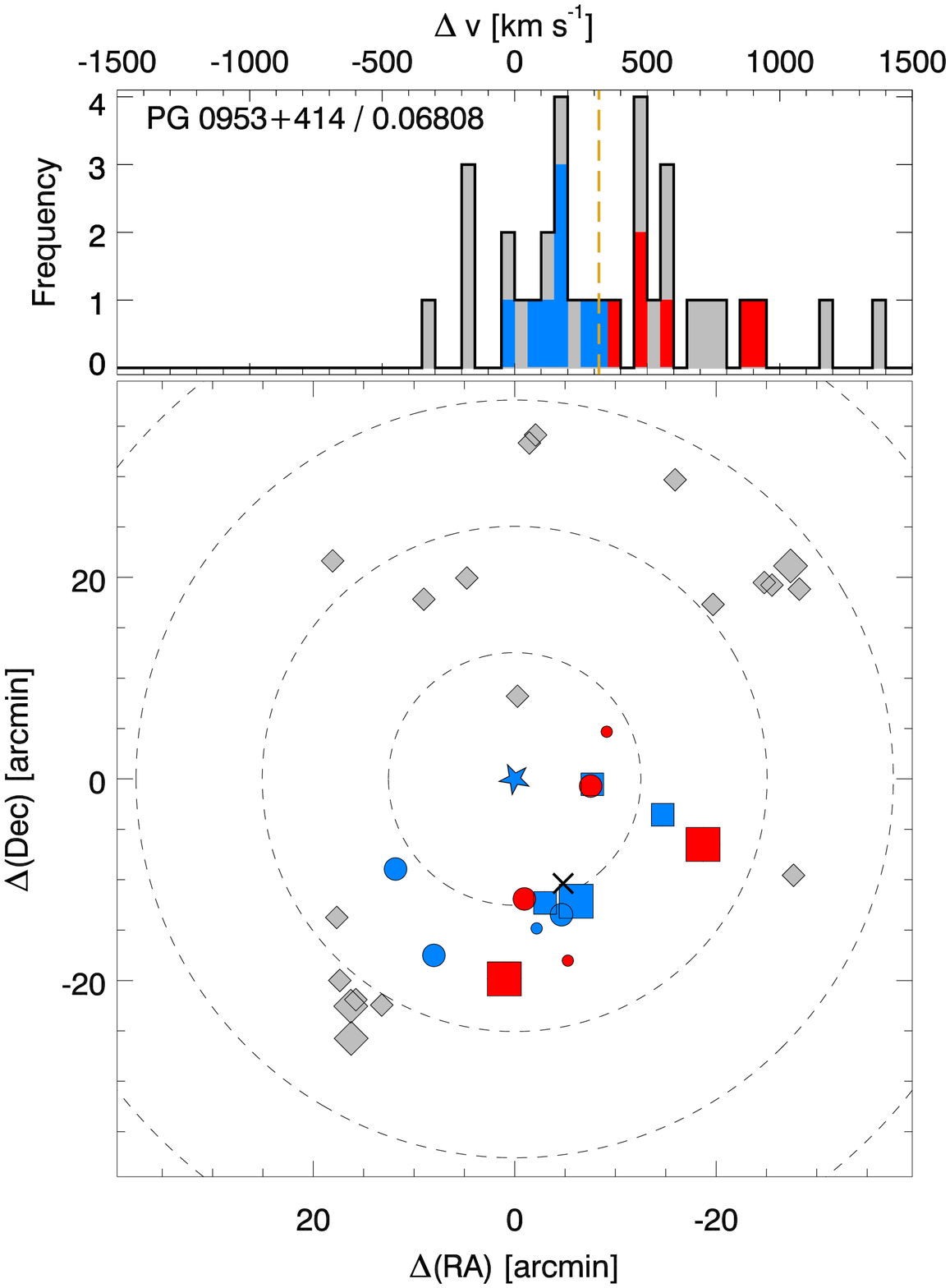}
\caption{(continued)}
\end{center}
\end{figure}

\begin{figure*}[!ht]
\begin{center}
\epsscale{1}
\plottwo{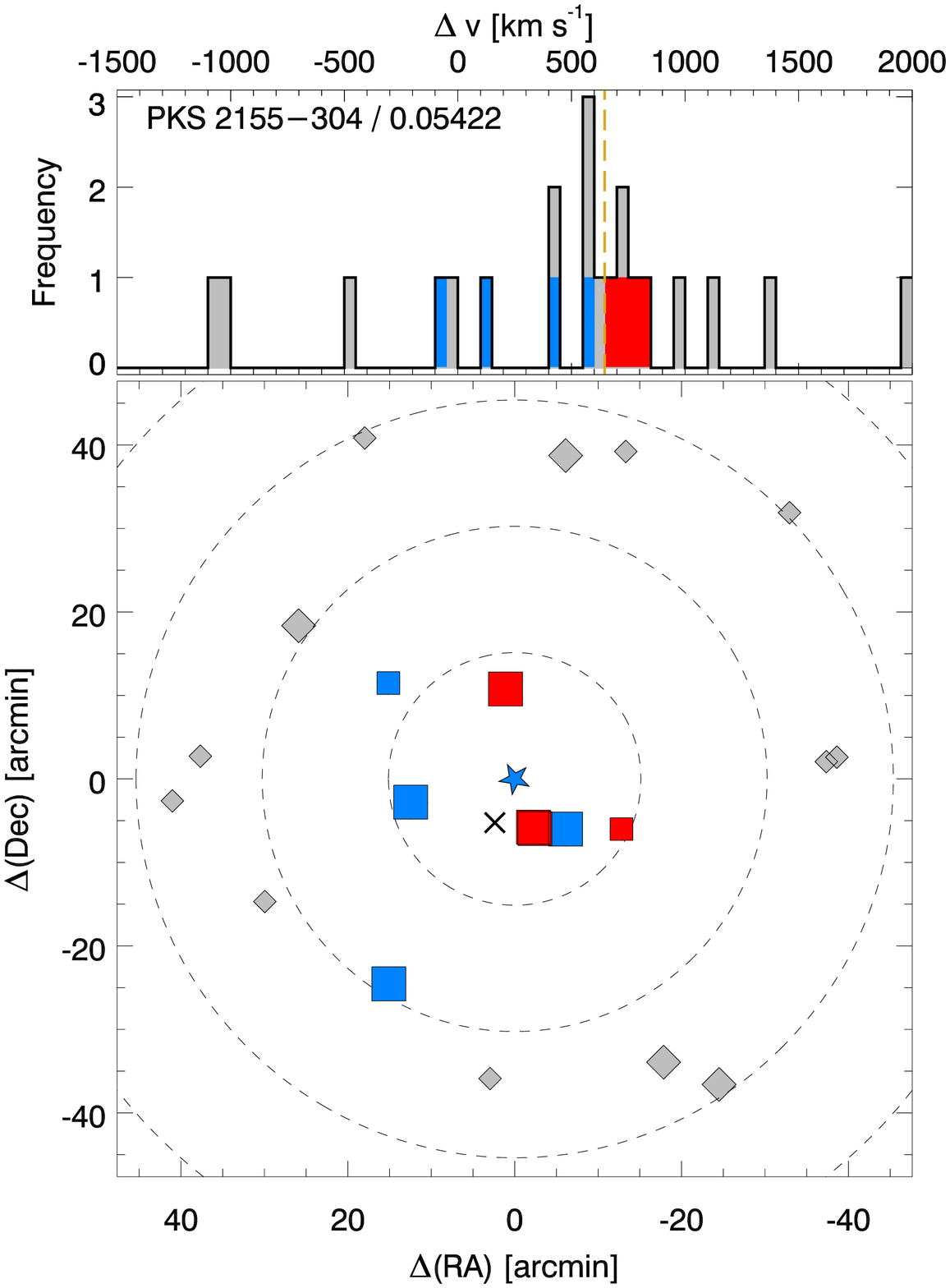}{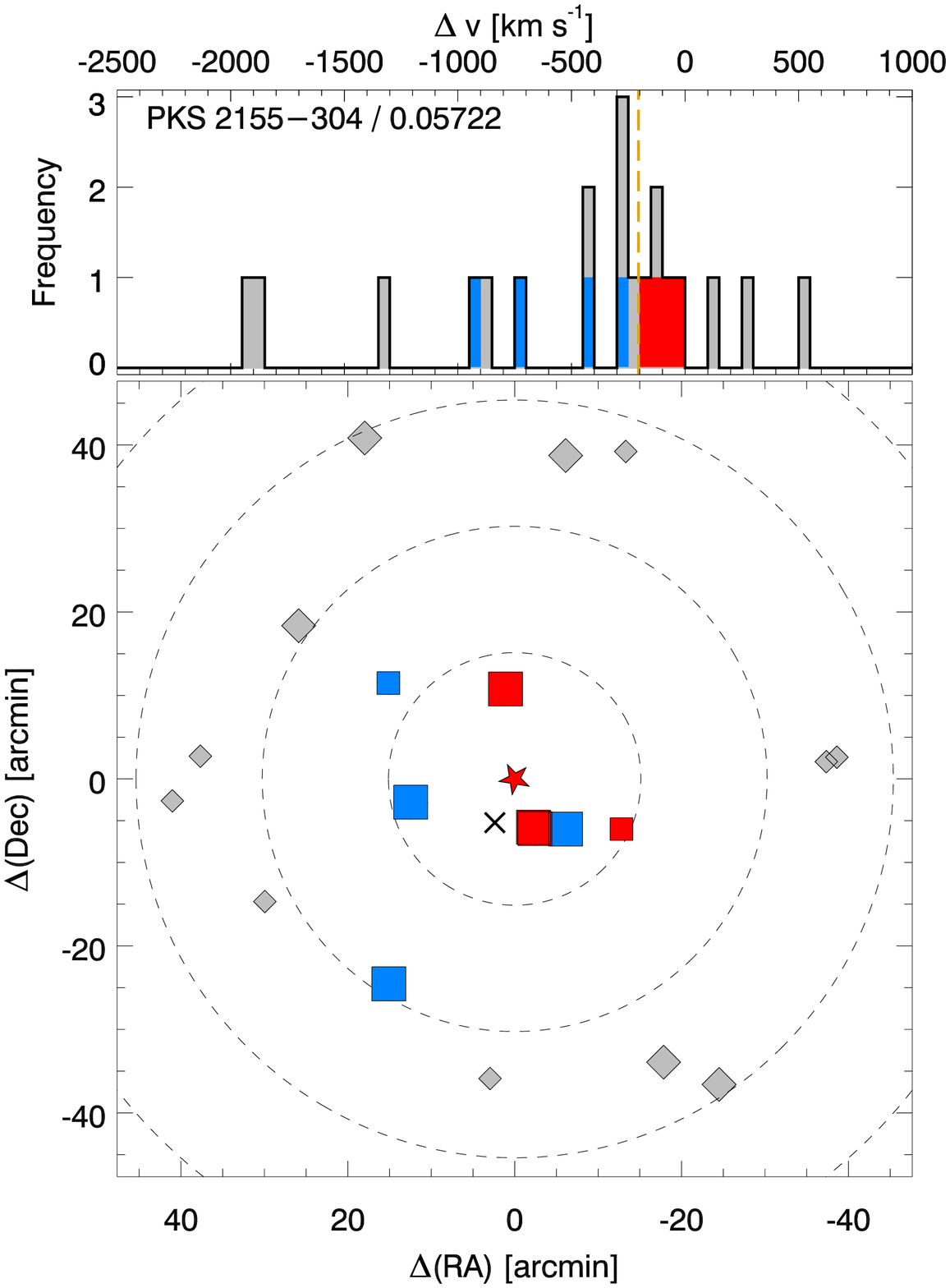}
\plottwo{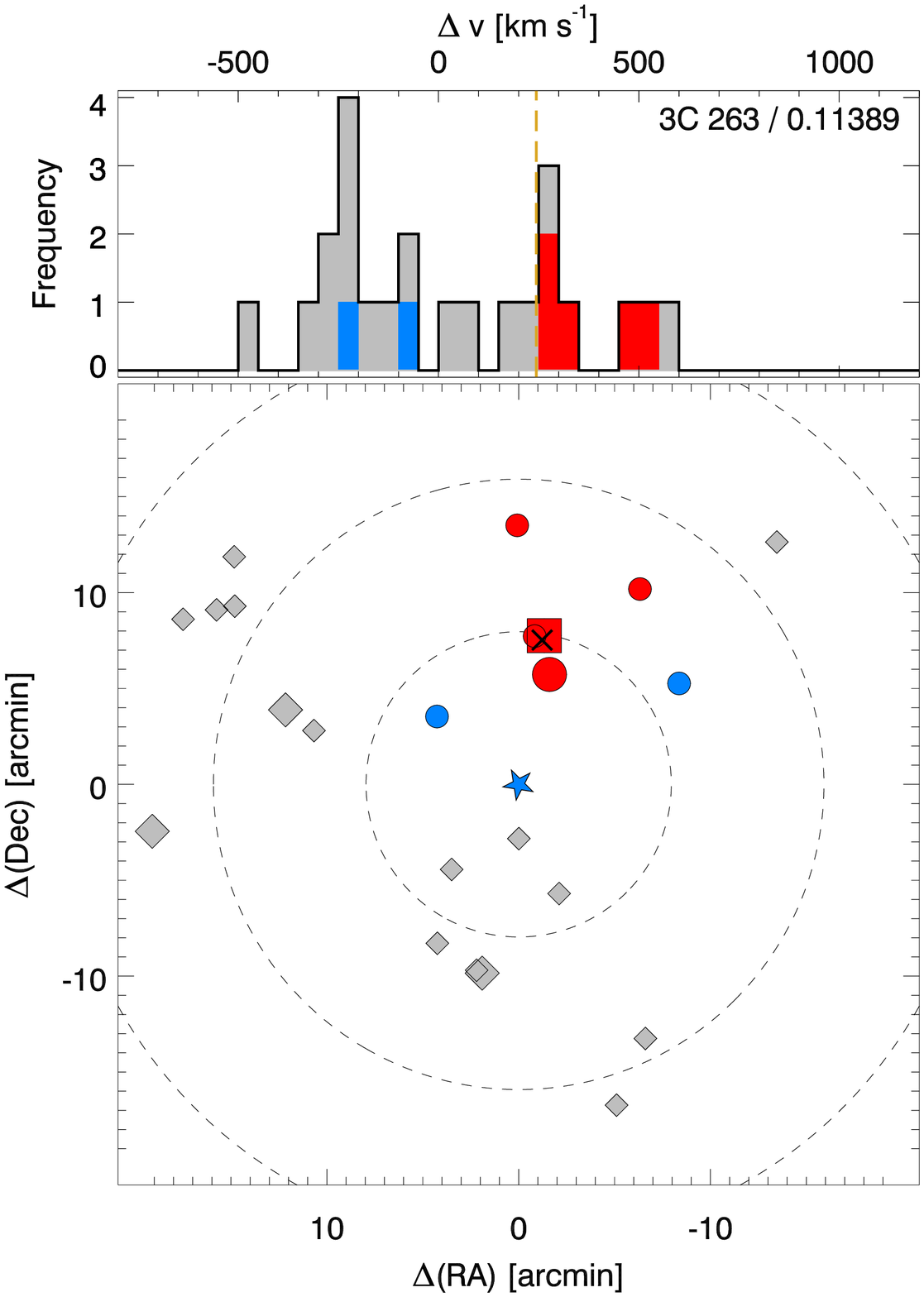}{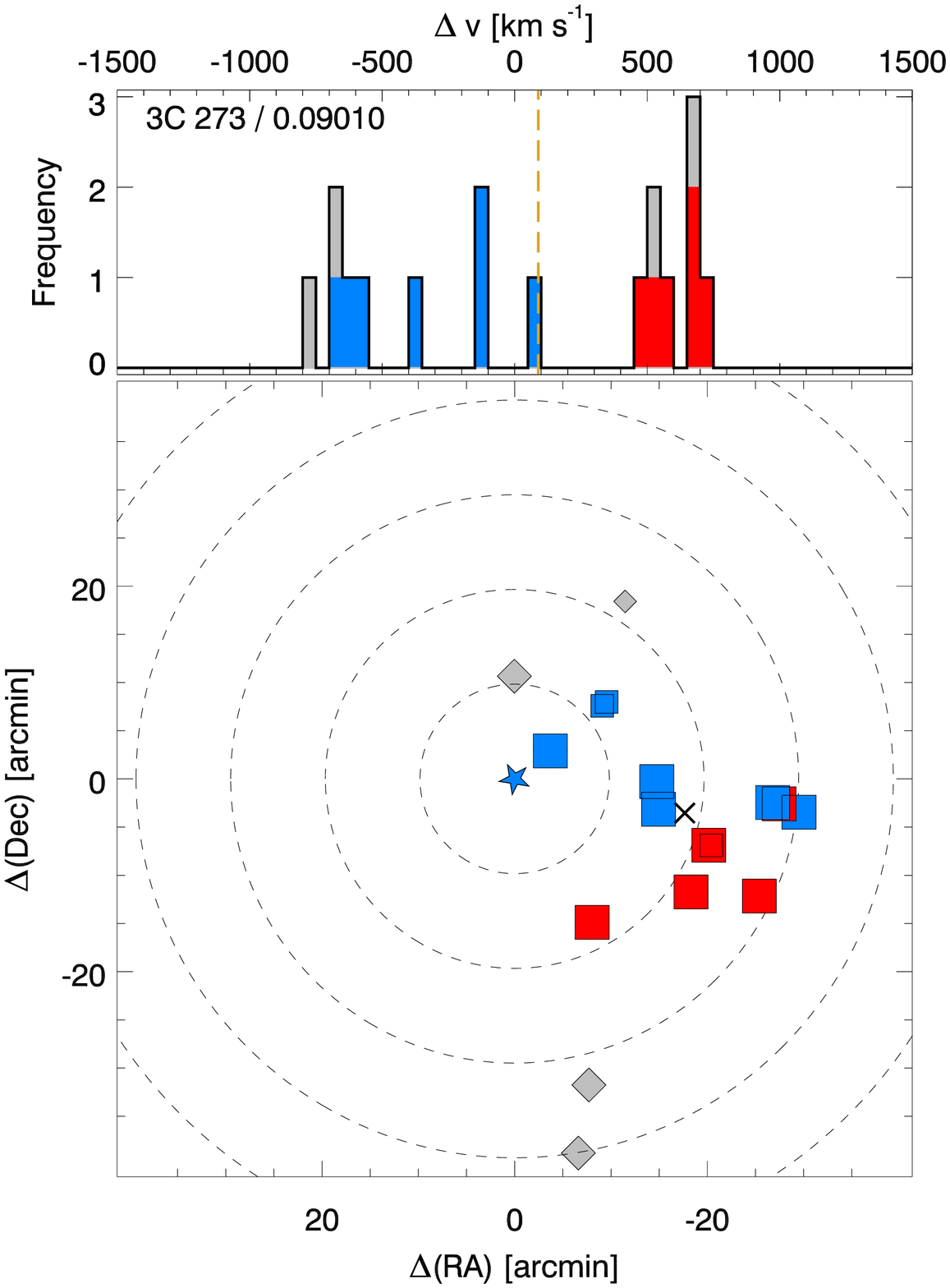}
\caption{A montage of four galaxy groups which are associated with ``misaligned'' absorbers (Table 7).
Coordinates and symbology are described in the caption to Figure 2. The two PKS~2155-304 O~VI + H~I
absorber groups at top are the same physical group since these absorbers bracket the redshifts
of this group \citep{shull98}.}
\end{center}
\end{figure*}

Figures 2 (warm absorbers), 3 (cool absorbers) and 4 (misaligned absorbers) show all groups in this study for which 
$\geq$ 8 group members were identified by the group finding analysis described above. 
These figures show group members as colored circles or squares with non-group members in the same regions as grey triangles. 
The presence of grey triangle galaxies within the search region for virtually all these groups indicates that the group memberships
are limited by the linking length criteria not the imposed boundaries of the search.
However, two of these ``groups'' may extend beyond the initial search regions; these two groups are also much richer 
than the others in this study and contain luminous elliptical galaxies:

1.  3C273/0.00336 \citep*{tripp02, rosenberg03, stocke04, yoon12} is in the southern outskirts of the
Virgo Cluster (see Figure 2). When left unrestricted the FoF technique includes nearly 1000 galaxies in this group, including
luminous ellipticals near the Virgo Cluster center $>$ 5 Mpc away. However, in an effort to better characterize the more local 
environment of this absorber, we have decreased the linking lengths by a factor of two and restricted the search region to $\leq$ 
1 Mpc from the absorber. The resulting values are still unusually large among warm absorber associations both 
in galaxy numbers ($\sim$ 150) and also in total group luminosity ($\sim$ 32 L$^*$). While the FoF algorithm identifies only 
a fraction of the galaxies in this restricted region as group members (colored vs. grey symbols in the 3C~273/0.00336 panel 
in Figure 2), inclusion of all of the grey-symboled galaxies (mostly dwarfs) as group members does not change the calculated
velocity dispersion by an amount more than the quoted 90\% confidence error range. The agreement between  $\sigma_{\rm grp}$ and  $\sigma_{\rm vir}$
in this case (see Table 5) are suggestive that our restrictive approach has defined a warm absorber group within a larger supercluster region.  

2. The other very rich group found around warm absorber locations (PHL~1811/0.07773) is between two rich clusters, 
Abell 2402 and SDSS-C4 2012. However, the algorithm does
identify a group confined within the search region with a few non-group members present within the region (see Figure 2). We conclude that
this is a well-defined group with a well-defined velocity dispersion (see the approximately Gaussian velocity distribution in Figure 2)
within a larger supercluster region.

Three other fields deserve some brief comments with regard to the FoF group membership algorithm. In the case of the cool absorber group 
PG~0953+414/0.00212 (see Figure 3) the FoF analysis finds a fairly isolated group even though a higher redshift group 
($\Delta v \geq$ 400 km s$^{-1}$) is also present between 1 and 2 Mpc to the south (concentration of grey points
to the south of the absorber and at higher radial velocities in the $\Delta$v histogram). However, many small galaxies
which might be included in the cool absorber group are excluded by the FoF analysis (grey symbols in the spatial plot and histogram mixed in with the
identified group members). If those excluded galaxies are instead included, this group increases from 1.5L$^*$ to
3.0L$^*$ in total luminosity but the velocity dispersion remains nearly identical to the value reported in Table 6. Therefore, despite potential differences in
group membership accounting, the characteristics of this group are fairly robust.

The warm absorber group Mrk~290/0.01027 shown in Figure 2 has been pointed out previously by \citet {narayanan10}, who describe this group as being in a 
``filament'' of galaxies. Indeed, the FoF analysis links to galaxies out to $\sim$ 3 Mpc from the absorber but we truncate our analysis at $\pm$ 1.5$R_{grp}$
and 3$\sigma_{vir}$. While either the group or filament descriptions
seem valid, we retain the values shown in Table 2 for this warm absorber group for this analysis. 

There are two anomalies created by the procedure of group identification above (see table notes to Table 5). For the PHL~1811/ 0.07773
absorber, a $>$ 100 member group is identified which is well-centered around the absorber on the plane of the sky but offset
in velocity ($\Delta$v$_{grp}$ = 330 km s$^{-1}$) and with a very large velocity dispersion ($\sigma$ = 630 km s$^{-1}$).
In this case there is an 0.7 L$^*$ group member which is closer on the sky to the absorber (0.22 R$_{vir}$ away)
than a similarly bright galaxy, which is 1.75 R$_{vir}$ away (as listed in Table 5). However, this proximate galaxy has a $\Delta$v = 904 km s$^{-1}$
relative to the absorber, which greatly exceeds our adopted maximum velocity difference of 400 km s$^{-1}$ to be associated with the absorber. 
This same galaxy has been identified as the one associated with a slightly higher redshift LLS absorber in the PHL~1811 FUV spectrum \citep {jenkins03}. 
Similarly, in the 3C~273/0.00336 absorber region there is an extreme dwarf ($<$ 1\% L$^*$) quite
close to the sightline ($\rho$ = 46 kpc) but at substantial $\Delta$v = 610 km s$^{-1}$; this dwarf is associated with a slightly higher redshift
absorber in this same sightline \citep {stocke04}.

\subsection{Spiral-Rich Galaxy Groups Associated with Warm Absorbers at $z\leq$ 0.15}
\label{groups:warm}

For the four warm absorbers at $z \leq$ 0.10 our 4m survey is more than
adequate to supply a good description of their associated galaxy
groups; all but one (Mrk~876/0.00315) have $>$ 10 group members for galaxy survey completeness
levels deeper than 0.1 L$^*$. A fifth absorber, PHL~1811/0.13280, has also been surveyed deeply
enough to establish the presence of a rich group. These four have
velocity dispersions $\sigma_{\rm grp} \approx$ 250--650~\kms\  and total
group luminosities of $\sim$ 5--50 L$^*$. These numbers correspond to significant (4--100$^+$)
overdensities in galaxies.

While the richness, total luminosity and velocity dispersion of the Mrk876/0.00315 group 
would be much larger if many of the dwarfs in its region were included in this group by the FoF
analysis, its listed (Table 5) values are less than the Local Group, dominated by a single 0.25L$^*$
galaxy. If we were to include
all the 22 galaxies in the original search volume as group members, this ``group'' would have
a total group luminosity of $\sim$ 2L$^*$ and a velocity dispersion $\sigma \sim$ 250 km s$^{-1}$, which
would make it comparable to the Local Group and only slightly poorer but with
a velocity dispersion comparable to the Mrk~290/0.01027 group. The lack of sufficient group members to
provide a robust velocity dispersion leads us
to discard this group from the statistical analysis despite having a warm absorber at log T(K) = 5.14 (Paper 1). 

The two richest absorber environments have velocity dispersions comparable to many elliptical-dominated groups
of galaxies \citep* [see Table 2 and][]{zabludoff98, mulchaey00}. 
The richest ``group'', associated with 3C~273/0.00336, is in the southern outskirts of the Virgo Cluster with
a couple of luminous early-type galaxies (NGC~4636, a 2L$^*$ E0 and NGC~4643, a 1.6L$^*$ SB0) $\sim$ 0.8 Mpc away 
from the sightline. The 5L$^*$ Sbc spiral M61 is the most luminous galaxy in the region.
Similarly the PHL~1811/0.07773 group is also in the vicinity of richer galaxy regions and has a 
total group luminosity (55 L$^*$) and velocity dispersion of $\sim$ 600 km s$^{-1}$, very large  
for an isolated spiral-rich group \citep* [see Table 5, Figure 2 and][]{zabludoff98, mulchaey00}. This group also contains several
early-type galaxies including 2MASX J21551727-0917517 \citep[see also][]{jenkins03}, a 2L$^*$ elliptical 0.6 Mpc from the 
sightline. Given its large
velocity dispersion and approximately Gaussian-distributed positions and velocities
(comparable to many poor clusters of galaxies), this group should be searched for diffuse soft X-ray emission.
Evidently, warm absorbers can be found either in small spiral-rich groups, in richer galaxy groups within filaments or 
in regions on the outskirts of clusters of galaxies. 

A fifth nearby group is associated with the {\it possible} warm
absorber, 3C~263/0.11389 (listed as a ``misaligned absorber'' by Paper 1 but see 
Appendix A for a reanalysis of this possible warm absorber). This absorber region has been
surveyed for galaxies to $0.2\,L^*$, has 7 identified
group members and a velocity dispersion of $\geq$270~\kms. The three other low-$z$ warm absorbers in Table 5 have regions surveyed
to 0.2--0.3L$^*$ and only 4 group members identified. While an extrapolation from these few galaxy identifications
using a standard luminosity function \citep{montero09} suggests the presence of groups comparable in richness to Mrk~290/0.01027,
deeper MOS is required to characterize them more precisely. Therefore, based
on the eight absorbers in Table 5 and a ninth, possible warm absorber
in Table 7 (3C~263/0.11389), the warm absorber environment is usually spiral-rich galaxy groups with total
luminosities of 2--20L$^*$ (assuming the larger value for the Mrk~876/0.00315 group) 
and velocity dispersions of 250--450 km s$^{-1}$. A couple of absorbers are found 
within richer groups and in cluster outskirts with larger total luminosities and velocity dispersions. These
conclusions are all based on a small size sample and need to be confirmed by discovering many more warm absorbers at
very low redshifts, where galaxy groups can be characterized in detail. It will be particularly 
important to target groups with well-defined properties, selected consistently from large galaxy databases \citep [e.g., SDSS][]
{berlind08} to verify the diversity of groups associated with these ``warm absorbers''.

\subsection{Galaxy Groups Associated with Cool and Misaligned Absorbers}
\label{groups:cool}

Table 6 lists all the lower temperature \OVI\ +
\lya\ absorbers at z $\leq$ 0.15 from Paper 1. Generally the cool absorbers are in
groups of galaxies with slightly lower total luminosities ($L_{\rm grp}$ = 1.5--25 L$^*$) 
and velocity dispersions ($\sim$ 130--500~\kms) than the warm absorber groups. 
These small differences between warm and cool absorber
environments are not due to these fields having a brighter completeness
luminosity as all but three absorbers (3C~273/0.12007; PG~0953+414/0.14231
and HE~0153-4520/0.14887; i.e., the three highest redshift absorbers in Table 6)
have been surveyed well below $0.1\,L^*$. Due to the dearth of cool absorbers 
with good galaxy data we have added
PG~1259+593/0.04625 to Table 6. This QSO has a high-S/N (36:1) COS spectrum and has been
well-surveyed for galaxies using the same galaxy redshift database and WIYN/HYDRA MOS described in Section 3.1
despite not being a sightline included in Paper 1. A detailed analysis of this absorber complex is presented in
Appendix A. See Figure 3 for the galaxy distributions around
cool absorbers with $\geq$ 8 group members identified.

Interestingly, despite having similarly faint completeness limits, galaxy
redshift surveys fail to find many galaxies around the misaligned
absorbers (see Table 7). Only three misaligned absorber groups have sufficient numbers of identified 
group galaxies to quote a firm velocity dispersion (i.e., N$_{grp} \geq$ 8). The three well-sampled groups
associated with misaligned absorbers are shown in Figure 4 along with the N$_{grp}$ = 7 group associated
with 3C~263/0.11389, which may contain a ``warm absorber'' (see Appendix A).

\subsection{Statistical Properties of Galaxy Groups Associated with Warm Absorbers}
\label{groups:summary}

In this Sub-section we explore the possible differences between the nearest galaxies and the galaxy groups
surrounding warm and cool absorbers. Since the COS spectroscopy provides only single probes through this
gas, it is not obvious whether these absorbers are small ($\leq$ 20 kpc) and more closely related to an
individual galaxy or are large ($\geq$ 100 kpc) and are associated with an
entire group of galaxies. We perform three observational tests to 
determine which of these two possibilities is more likely: (1). Do nearest galaxy
properties or galaxy group properties differ between warm and cool absorbers? (2). Do the
observed absorber line-widths and inferred gas temperature values correlate with the kinetic energy in the group
as measured by the group velocity dispersion? and (3). Which potential association, nearest galaxy or galaxy group,
provides an absorber size which better matches the observed impact parameters shown in columns (4) and (9) of Table 5? 
This third test will be presented in the next sub-Section.

Each of these tests is limited both by the small sample size and by the statistical and systematic errors in
the absorber and group measurements. While the O~VI line widths offer the most unambiguous measurement of
the temperature of the warm or cool gas, the O~VI line profiles also may include some contribution from unresolved components, 
turbulence or Hubble flow motions (see Paper 1 and Appendix A). For example, the PG~0953+414/0.14231 absorber includes two
O~VI components with $b$=19$\pm$7 and 29$\pm$4 km s$^{-1}$, but from which similar gas temperatures (log T(K)$<$ 4.25 and
log T(K) = 4.21) are derived when the O~VI and H~I $b$-values are used together. Secondly, the
Ly$\alpha$ line-widths are derived from 
best-fits to multiple components within a complicated line profile (see Appendix A and Paper 1). This makes the Ly$\alpha$ $b$-values
less reliable than b$_{O~VI}$ due to the presence of strong, photo-ionized absorption in Ly$\alpha$, which creates both 
statistical and systematic errors. Finally, gas temperatures are derived from a 
combination of the measured O~VI and Ly$\alpha$ $b$-values and so have the greatest statistical uncertainties. But this 
procedure does allow for the removal of other contributors to the line widths that otherwise would create large systematic errors in log T.
This makes the temperatures the most reliable of these three measures since the systematical uncertainties are minimized, so that the statistical
errors are a good measurement of the total uncertainties in log T. Further, the number of group members defined by the FoF 
algorithm is modest in most cases analyzed, making both the statistical and the systematic errors in the velocity dispersion values
potentially quite large. Of particular concern are the non-Gaussian distributed 
velocities and sky locations of the group galaxies (see Figures 2, 3 and 4). At issue is not only the detailed methodology for computing
velocity dispersions for small numbers of galaxies \citep{beers90} but also the robustness of the group membership algorithms \citep{berlind06},
including our choice of ``linking lengths''. However, our tests in Section 3.1 show that the velocity dispersions computed from group
membership agree to within the errors with the group velocity dispersion computed from total group galaxy luminosities under the assumption 
that the group is virialized. This agreement provides some confirmation for the group definition process we have used, but it may be that a
combination of systematic and statistical errors in measuring absorber and group properties (T$_{abs}$, $\sigma_{grp}$, etc) may actually 
obscure any real correlations that are present.

\begin{figure}[!t]
  \epsscale{1}
  \plotone{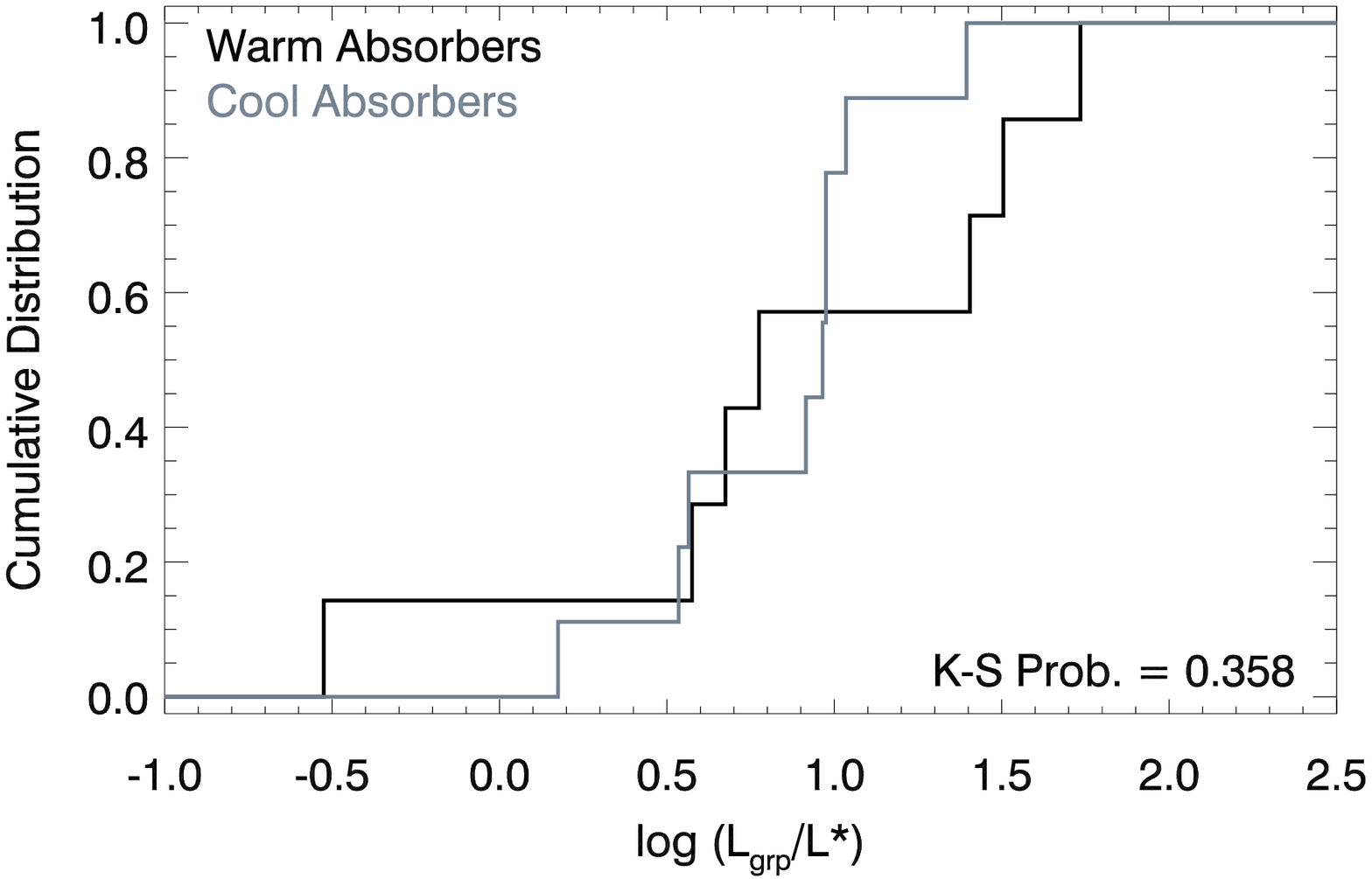}
  \caption{Cumulative distribution functions of total group luminosity, $L_{\rm grp}$.
  \label{fig:cdf_Lgrp}}
\end{figure}

\begin{figure}[!t]
  \epsscale{1}
  \plotone{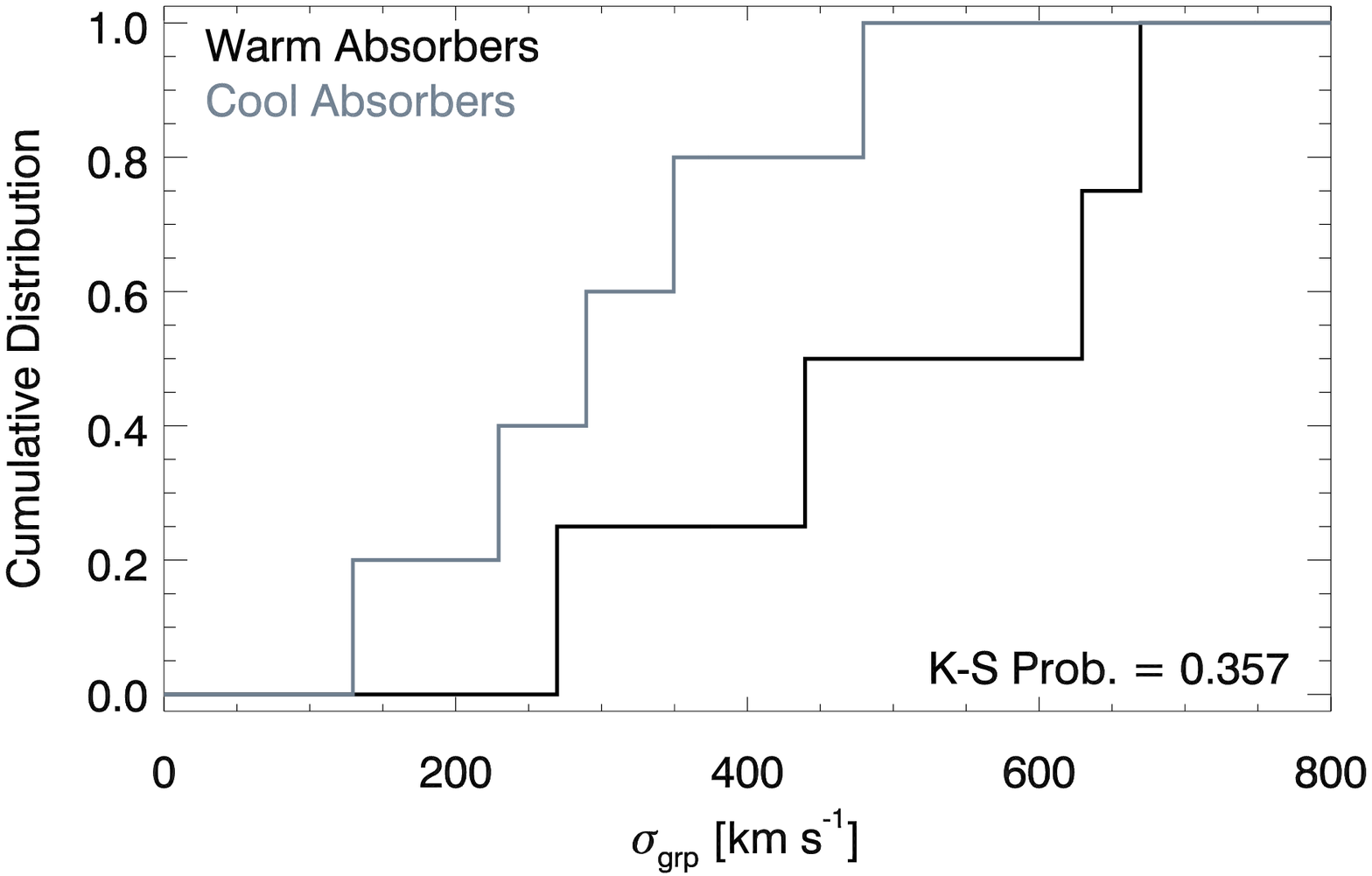}
  \caption{Cumulative distribution functions of group velocity dispersion, 
           $\sigma_{\rm grp}$, for groups with $N_{\rm grp} > 8$~members.
  \label{fig:cdf_sgrp_rich}}
\end{figure}

With these caveats in mind, the very slight systematic differences in galaxy group environments between the
warm and cool absorbers in this sample are shown not to be significant in Figures 5 and 6.
The warm absorbers are found in galaxy groups which are both slightly richer
(marginally greater total luminosity; K-S test probability of chance occurrence
$\sim$ 36\%) and which possess systematically larger velocity
dispersions (K-S test probability of chance occurrence $\sim$ 36\%). 
For Figure 5 we have used all warm absorbers in Table 5 (plus the
possible warm absorber 3C~263/0.11389 from Table 7) and cool
absorbers in Table 6, judging that all of these fields have been
surveyed deeply enough to find most of the total luminosity of the groups.
For Figure 6 we have used only those groups in Tables 5 and
6 which have $\geq8$ members surveyed, so that a velocity dispersion with estimated errors is quoted. 
Including all groups with N$\geq$ 3 (i.e., all values in Tables 5, 6 \& 7 where a $\sigma_{grp}$ value is quoted)
decreases the K-S test probability for chance occurrence to
$\sim$ 10\%, but this comparison simply requires more groups before it can be judged to show a significant difference
or not. Thus, using the most secure velocity dispersion data, 
we find only a very slight difference ($<$1$\sigma$ level) that the warm absorber
groups have a larger velocity dispersion and higher total luminosities. 

\begin{figure}[!t]
  \epsscale{1}
  \plotone{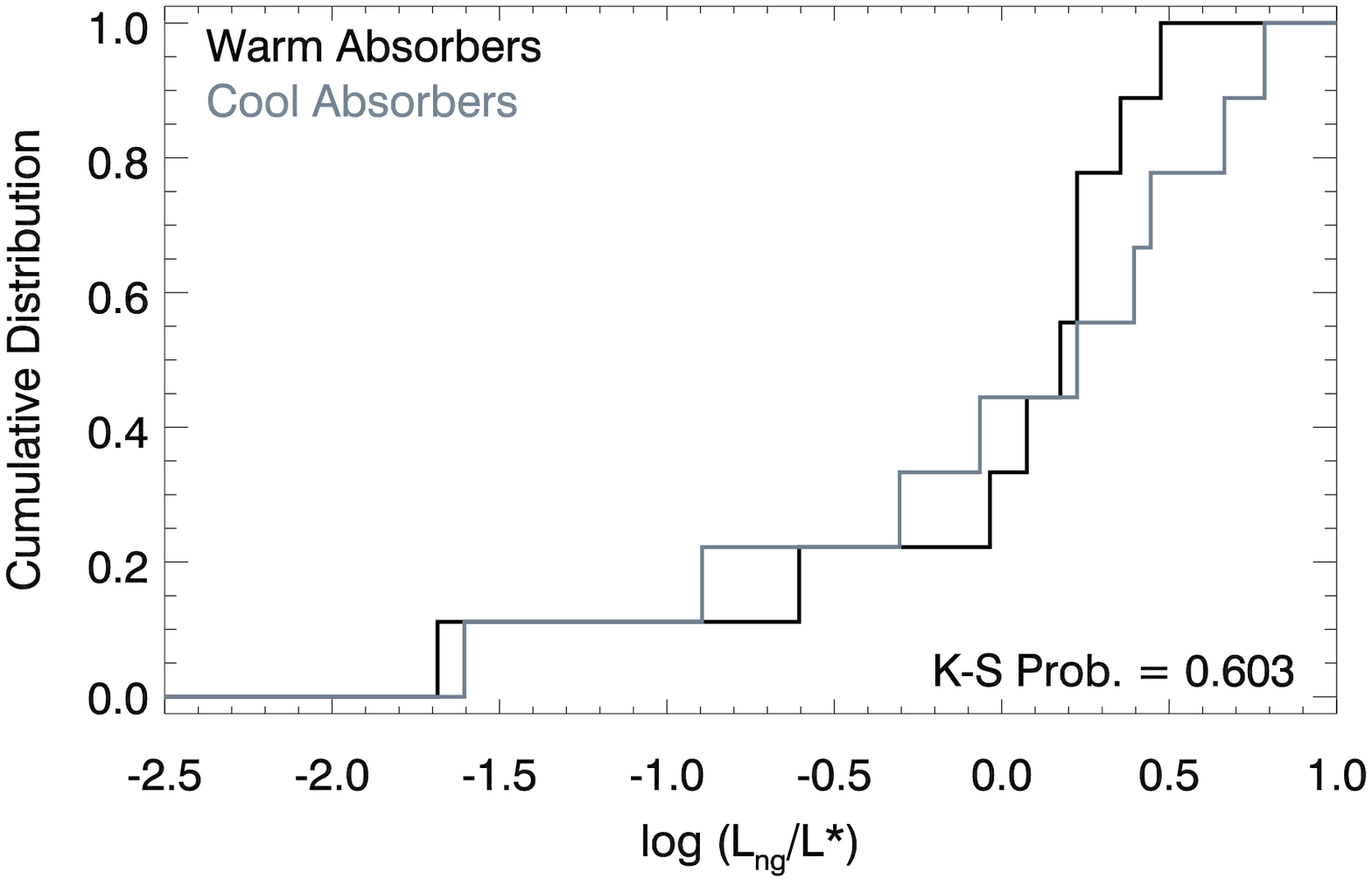}
  \caption{Cumulative distribution functions of nearest galaxy luminosity, 
           $L_{\rm ng}$.
  \label{fig:cdf_Lng}}
\end{figure}

Figure 7 shows that there is no dichotomy at all between the warm and cool
absorbers if the luminosity of the nearest associated galaxy is
plotted. In this case we have used all the absorbers
in Tables 5 and 6 because the two samples have a comparable range of 
completeness luminosities close to their
sightlines. A K-S Test yields a probability of $\sim$60\% that these
two distributions have been drawn from the same parent
population. These comparisons offer no strong support for or against an association between these
absorbers and either their nearest galaxy neighbors or the groups in which they are imbedded.

Dividing the full galaxy groups sample by absorber temperature may actually obscure a more conclusive result
for test (1) above because the statistical uncertainties in the measured temperature for some absorbers are so large. Therefore,
in test (2) we use the raw line-width data as well as the derived temperature values, but taking the uncertainties
explicitly into account. In Figure 8 we show a plot of O~VI $b$-values versus 
$\sigma_{\rm grp}$. No correlation between these quantities is present. A weak correlation ($\sim$ 50\% probability
of chance occurrence) may be present
between the Ly$\alpha$ line width and $\sigma_{\rm grp}$ (Figure 9) but its existence depends entirely on the one point associated
with PHL~1811/0.07773. While any real correlation between these quantities could be masked by the potentially large 
systematic errors in the line-fitting, these plots offer no substantial support for or against a group origin for this gas.

\begin{figure}[!t]
  \epsscale{1}
  \plotone{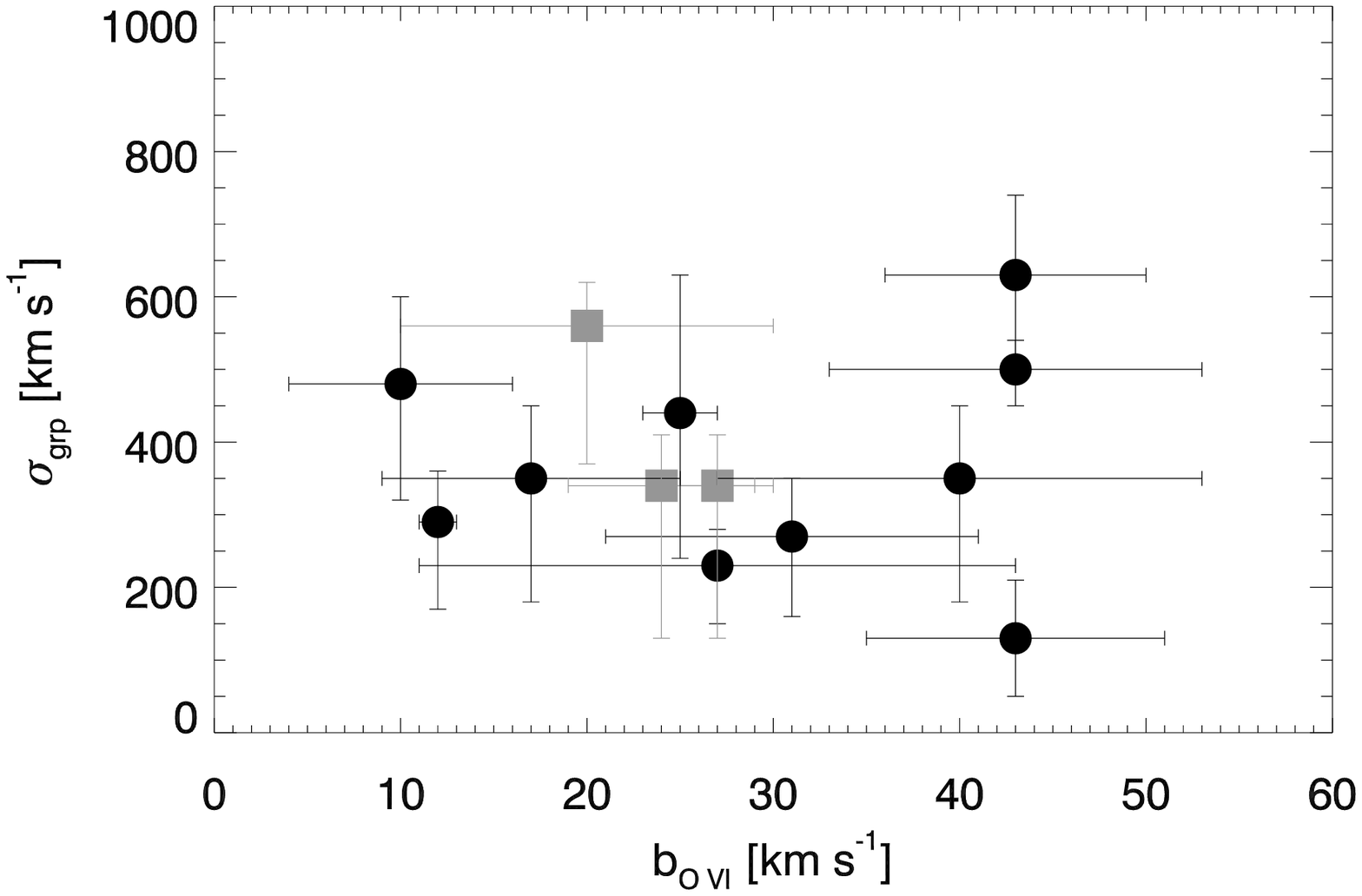}
  \caption{The relationship between the line widths of the O~VI absorption ($b_{OVI}$) and the group velocity dispersion
($\sigma_{grp}$). The black circles are from Tables 5 and 6 (warm and cool absorbers); the grey squares are from 
Table 7 (misaligned absorbers). There is no correlation present either using the misaligned absorbers or not. 
  \label{fig:bOVI_sigma.eps}}
\end{figure}

\begin{figure}[!t]
  \epsscale{1}
  \plotone{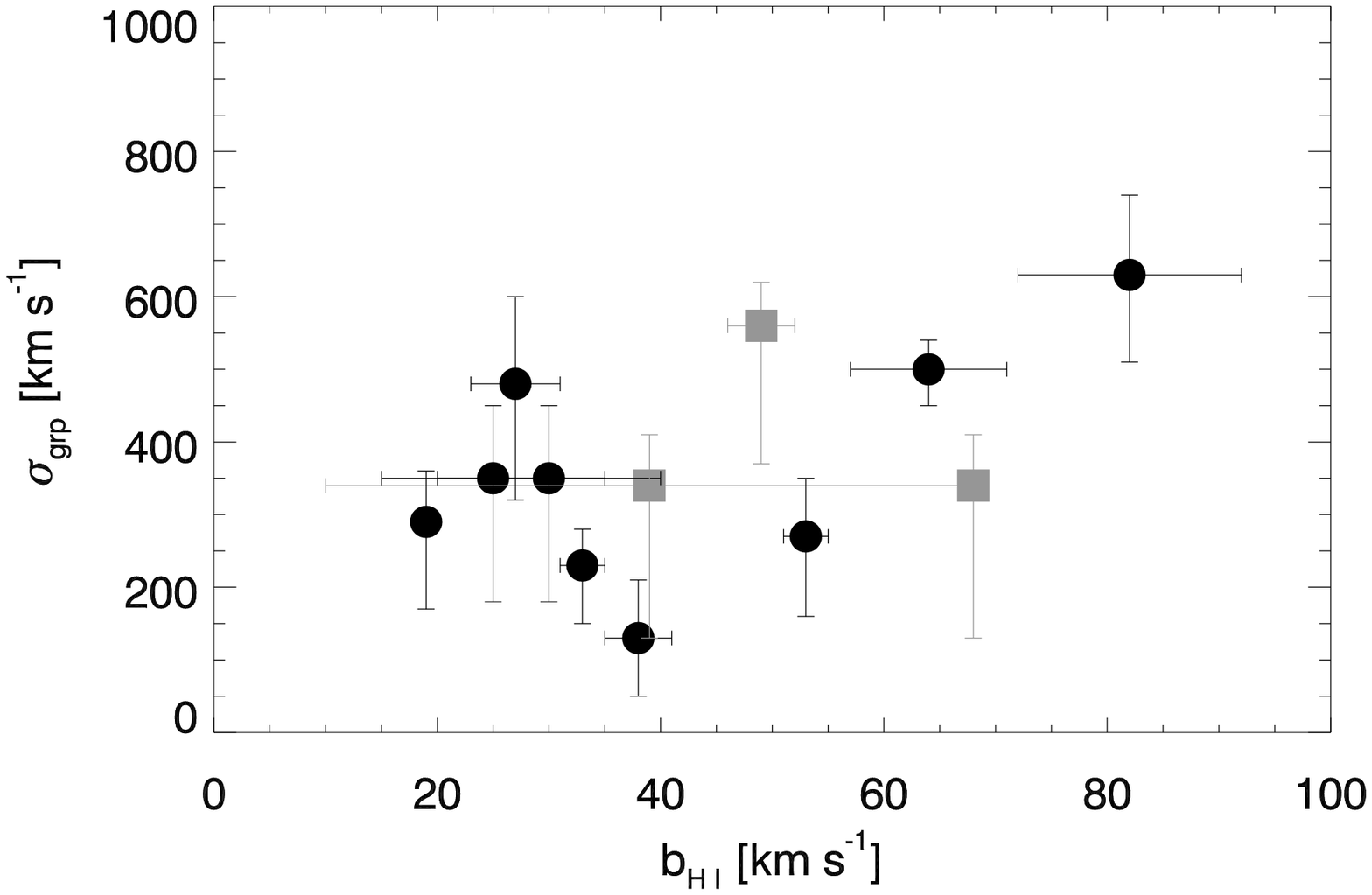}
  \caption{The relationship between the line widths of the H~I Ly$\alpha$ absorption ($b_{H~I}$) and the group velocity dispersion
($\sigma_{grp}$). The black circles are from Tables 5 and 6 (warm and cool absorbers); the grey squares are from Table 7 (misaligned absorbers). 
There is a very slight (50\% probability) correlation present either using the misaligned absorbers or not. 
However, if one point, PHL~1811/0.07773, was removed no correlation is present.
  \label{fig:bHI_sigma.eps}}
\end{figure}

However, some support for a group origin for the warm and cool O~VI gas detected spectroscopically by COS
is shown in Figures 10 \& 11 (n.b., these plots are not fully independent since group luminosities and velocity dispersions
correlate). These plots use the inferred absorber gas temperatures (Tables 2, 3 \& 4) and the group
velocity dispersions and luminosities (Tables 5, 6 \& 7) to determine whether there is a relationship between the
energy in the gas and in the galaxies as would be expected if they trace the same
gravitational potential. All absorbers with reasonably robust velocity dispersions ($N_{grp} \geq$ 8 members) 
have been included in Figures 10 \& 11 as filled black symbols.
These data result in a correlation between
these two variables at the 95\% level ($\sim$ 2$\sigma$) using various statistical tests (e.g., Kendall and Spearman rank correlation
coefficients). These possible correlations are far from compelling due to the small sample size 
and requires new observations to be confirmed (e.g., two groups, H~1821+643/0.12141 and 3C~263/0.14072, which
do not appear in Figures 10 \& 11 and which have log T (K) $>$ 5.5 are in regions not yet surveyed deeply for galaxies). However,
the derived temperatures may provide a better indication of the specific energy in the gas compared to the raw
line widths due to the removal of non-thermal contributions to $b_{O~VI}$ and $b_{H~I}$. 

Also, as a follow-up to the possible correlation between group luminosity and absorber temperature, it will
be important to conduct a quantitative comparison between the galaxy morphologies in warm and cool groups. While
such an analysis is beyond the scope of this paper, it may provide important insights into the dynamical evolution of these groups
and will be presented at a later time \citep {keeney14}.

\begin{figure}[!t]
  \epsscale{1}
  \plotone{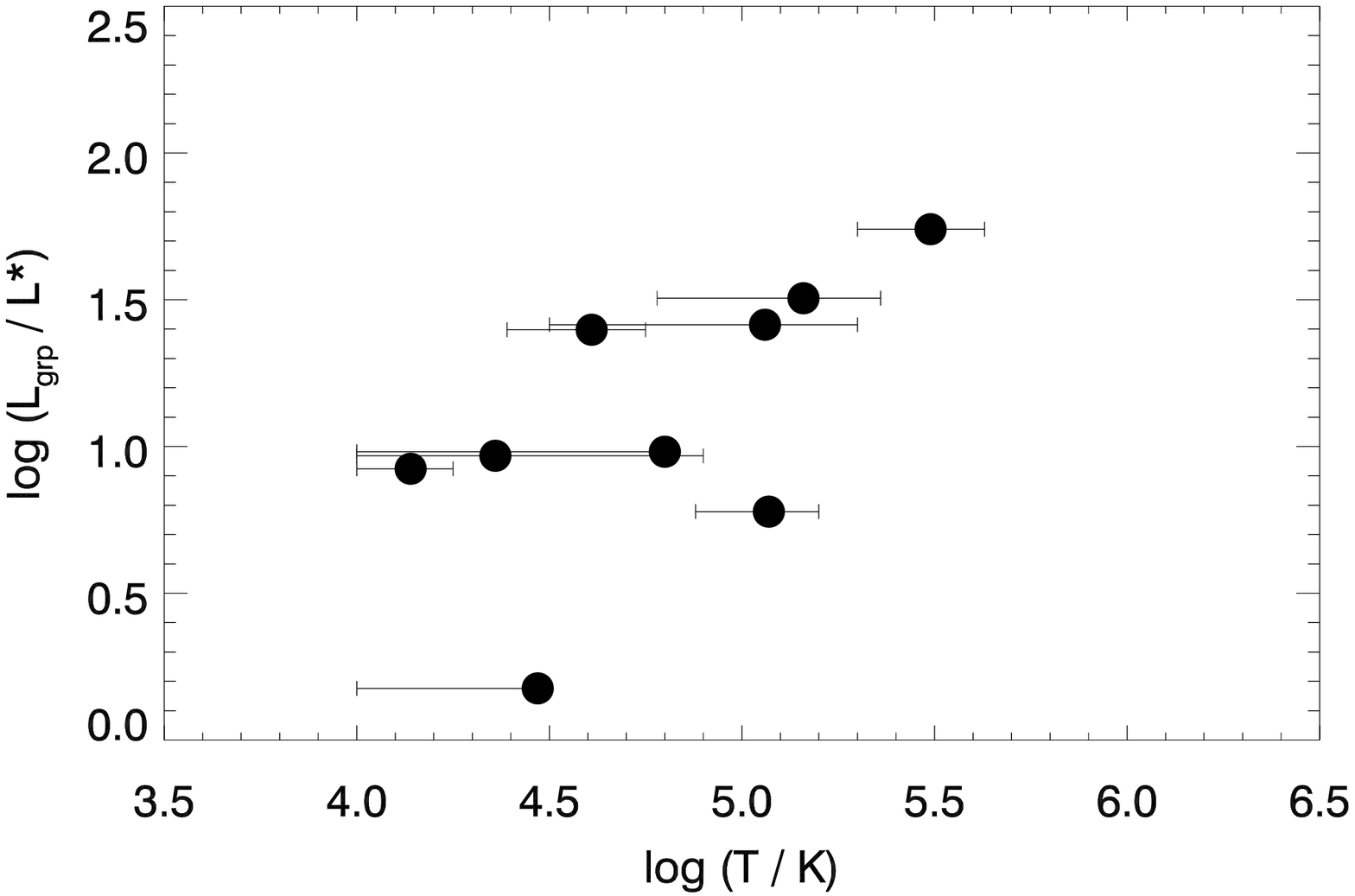}
  \caption{Plot of $\log{T}$ versus $L_{\rm grp}$. The black
symbols are those groups with $N_{\rm grp} \geq$ 8 members. These data result in a 95\% confidence 
(2$\sigma$) level correlation.
  \label{fig:logT_Lgrp}}
\end{figure}

\begin{figure}[!t]
  \epsscale{1}
  \plotone{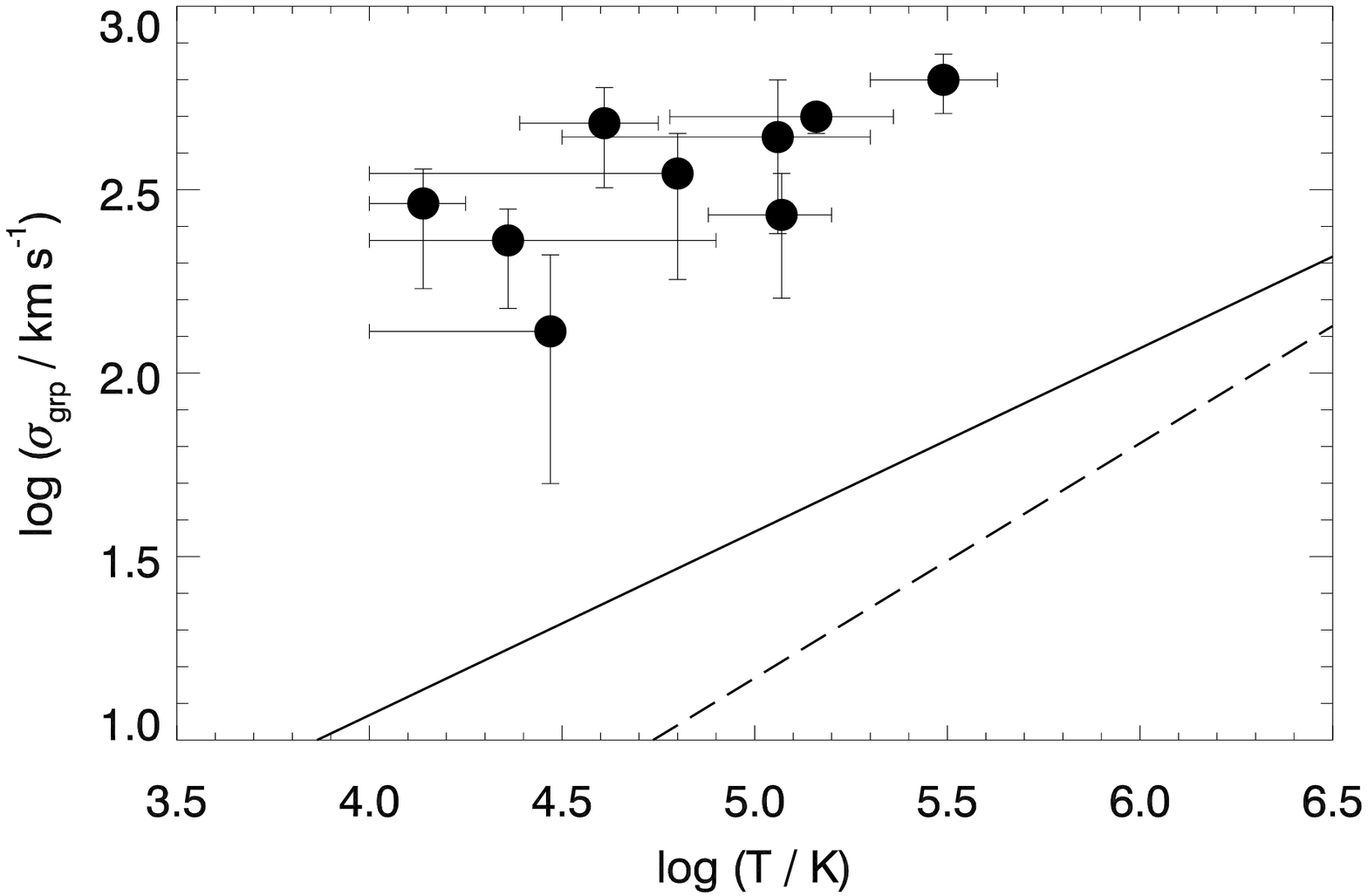}
  \caption{Plot of $\log {T_{gas}}$ versus $\sigma_{\rm grp}$ for the ``cool''' and ``warm'' groups listed in 
Tables 5 \& 6. This plot includes only those absorber groups with robust velocity dispersions;
i.e., $\geq$ 8 members with measured redshifts. 
For a few of the photo-ionized absorbers, we have truncated the error bars on temperature
from Paper 1 at log T (K)= 4.0 to be consistent with standard photo-ionization models. These data result
in a 95\% confidence (2$\sigma$) level correlation between these two variables. The solid line is the
extrapolation of $\beta_{spec}$ = 1 (energy equipartition between galaxies
and gas) from the clusters and
groups data found in \citet {osmond04}; see their Figure 16. The dashed
line is an extrapolation of the best-fit power-law
for clusters and groups from \citet {wu99}. See Section 4 for a discussion of the spiral-group data in
comparison to the extrapolations of \citet {osmond04} and \citet {wu99}. 
\label{fig:logT_sigma}}
\end{figure}

\subsection{Inferred Sizes of Warm Absorbers}

The third test for discriminating between nearest galaxy or galaxy group
associations with warm absorbers uses the number density
\dNdz = 3.5--5 per unit redshift as a discriminator (see Section
2). The bounds on the \dNdz\ value above refer to the number of warm absorbers found in Paper
1 (14 plus 2 O~VI-only absorbers) and herein (20 in total, including those same two
O~VI-only absorbers). We use \dNdz = 4$\pm$1 per unit
redshift for this analysis as a compromise value between the results of Paper 1 and the 
results in Section 2 and Appendix A herein. Note that this value is several times less than
the \dNdz\ for all O~VI absorbers \citep{danforth08}. Here, by using a mean space density of
associated objects (galaxies of varying luminosities and galaxy groups) we can use the observed
\dNdz\ value to compute a characteristic size for the absorber
assuming an 100\% covering fraction. These characteristic sizes can then be
compared with the observed impact parameters to determine if an
association with each object class is viable. 

Previously this type of analysis had been conducted for photo-ionized absorbers, which have been found to be associated
with individual galaxy halos at $\rho \leq R_{vir}$ at high covering factors
\citep* {tumlinson11, prochaska11a, stocke13}. These studies find near unity covering fractions
around $L\geq0.1\,L^*$ galaxies at  $\rho \leq R_{vir}$. \citet{stocke13} also finds 
substantial covering fractions around dwarfs at  $\rho \leq R_{vir}$. Among the warm absorption complexes studied in detail here (i.e.,
those in Table 5), only two absorbers (PHL~1811/0.07773 and PG~1116+215/0.13850) 
have low metal ions (e.g., Si~II, Si~III, etc) present, which are associated with the narrow Ly$\alpha$ component
in the absorption complex. These are photo-ionized clouds that may or may not be associated physically with the warm absorber. 
As has been found by the three studies referenced above, these two are at low impact parameter (0.22 and 0.67 R$_{vir}$) from the 
nearest galaxy in the group. The remainder in Table 5 have only H~I + O~VI present and are found at larger impact parameters 
(typically at $\rho > R_{vir}$) from the nearest galaxy in the groups.

\begin{deluxetable*}{llccccc}
\tabletypesize{\scriptsize}
\tablecolumns{5}
\tablewidth{0pt}
\tablecaption{Collisionally-Ionized ``Warm Absorber'' Size Estimates}
\tablehead{\colhead{Galaxy Luminosity Range}   &
        \colhead{$\phi(L_{low}$)} &
        \colhead{Required Absorber Radius} &
        \colhead{Max Observed Impact} &
        \colhead{Sample Size} & 
        \colhead{Percent with} \\
        \colhead{($L_{low}/L^*$)} &
        \colhead{($\times$ 10$^{-3}$ Mpc$^{-3}$)} &
        \colhead{R$_{abs}$ (Mpc)} &
        \colhead{Parameter $\rho$ (Mpc)} & 
        \colhead{} &
        \colhead{$\rho < R_{abs}$}} 
\startdata
$\geq$ 1.0      & $2.8\pm 0.2$ & $0.33\pm0.18$ & $>$ 2.0 & 9 & 45\% \\
$\geq$ 0.1      & $4.9\pm 0.6$ & $0.25\pm0.15$ & 0.62    &  4 &  50\% \\
$\geq$ 0.03     & $7.7\pm 1.4$ & $0.20\pm0.10$ & 0.43    &  3 & 33\% \\
$\geq$ 0.01     & $12.1\pm 2.8$ & $0.16\pm0.08$ & 0.43    &  2 & 50\% \\
Galaxy Groups   & 0.30        & $1.0 \pm0.55 $ & 1.11    & 10 & 90\% \\
\enddata
\tablenotetext{a}{calculations assume \dNdz\ = 4$\pm$1 and unity covering factor}
\end{deluxetable*}

Here we use the SDSS luminosity function of \citet {montero09} 
to determine the space density of galaxies using our assumed $H_0=70~\kms\,{\rm
Mpc}^{-1}$ value (see Table 8).
Recently \citet* {berlind06, berlind08} have undertaken a census of nearby
galaxy groups using the SDSS. Using their median catalog of groups
converted to $H_0=70~\kms\,{\rm Mpc}^{-1}$ yields $\phi=3\times
10^{-4}~{\rm Mpc}^{-3}$. \citet {pisani03} used a different 
approach which nets approximately the same space density for a group velocity
dispersion of $\sigma \geq$ 200 km s$^{-1}$, which appears to be a reasonable
lower limit to use in this estimate based on the warm absorber group values in Table 5.
An earlier study by \citet {girardi00} found a slightly larger value for the
space density of groups, 4.5 to 6.5 $\times$ 10$^{-4}$ Mpc$^{-3}$, but included
lower mass groups that are not as good a match to the current sample of
warm absorbers groups. 

The Berlind et al. value ($\phi=3\times 10^{-4}~{\rm Mpc}^{-3}$), which will
be used here for the space density of galaxy groups (see Table 8), is ten times less 
than the space density of $L^*$ galaxies, $\phi(L \geq L^*) =
2.8\times10^{-3}~{\rm Mpc}^{-3}$. The Berlind et al. catalog of galaxy groups are
dominated by small (3--5 L$>$ L$^*$, members), spiral-rich groups,
$\geq80$\% of which are found to be virialized on the basis of crossing
times $<$ 0.3 Hubble times. Almost all of these groups
are significantly richer and with higher velocity dispersions than the Local Group 
and so are similar to the warm absorber groups found here.  

Using the observed \dNdz = 4$\pm$1 (Poisson error) of warm absorbers and assuming a unity covering factor, 
the required absorber sizes listed in Table 8 (column 3) were computed using the \citet{montero09}
luminosity function integrated from the lower bound (L$_{low}$) in column (1) to infinity. The quoted errors
in column (3) are the quadrature addition of the \dNdz\ Poisson error and the errors in $\phi$ in column (2). 
The inferred sizes obtained by this calculation are half-again larger than the virial radii of 
L$_{low}$ galaxies but much less than the maximum impact parameter observed between the warm absorbers
and the nearest galaxies at L $\geq$ L$_{low}$. If the 
covering factor of these absorbers is $<$ 1 then the required absorbers sizes are larger still and so
must be much larger than the virial radii of the nearest galaxies to provide a plausible association. For the 8 warm absorbers in our small 
sample (9 including the possible warm absorber 3C~263/0.11389 in Table 7), we find a maximum 
impact parameter to the nearest $L \geq L^*$ galaxy of $>$ 2 Mpc and a range of nearest L$> L^*$ galaxy neighbors of
140 kpc to $>$ 2 Mpc. Lower luminosity bins have similar discrepancies. The last column in Table 8 shows that for all
galaxy luminosities the percentage of cases where the observed impact parameter ($\rho$) is less than the required
maximum absorber size (R$_{abs}$) is small. 
Unless the covering factors of these absorbers are $\ll$ 1 and their gaseous extents are $\gg$ R$_{vir}$,
an association between warm absorbers and individual galaxies appears unlikely.
However, caution involving this conclusion is required with respect to a possible association between warm absorbers
and dwarf galaxies due to the very small numbers of absorbers (3 or 2) which have been surveyed to 0.03 or 0.01L$^*$ completeness levels. 
Deeper observations of many of these groups are required.
 
On the other hand, the largest observed impact parameter between the
warm absorbers and their associated galaxy group's centroid is 1.11
Mpc (PHL~1811/0.13280), with a spread in $\rho/R_{vir}$ values of 0.24--1.27 (see column 10 in Table 5 and
the last row of values in Table 7). Unlike for the individual galaxies, these values fit closely with the 1 Mpc 
required absorber radius in Table 8. Additionally, these numbers suggest that the warm absorbers are found approximately
within the virial radius of the galaxy group.
If only a fraction of spiral-galaxy groups contain warm
absorbers and/or if the covering factor of these absorbers is
significantly less than unity, an even larger size than 1 Mpc is required
for the gas in these groups to account for the observed \dNdz\ value. Note well, that this
size is a statistically-determined extent of the ensemble of these absorbers at high covering factor, not the physical size of any one absorbing cloud.

In summary, small galaxy groups or groups in cluster outskirts with velocity
dispersions of $\sigma \sim$ 250-650 km s$^{-1}$ are typical warm
absorber environments. While simple comparisons between warm absorber groups and groups associated
with cooler, photo-ionized O~VI absorbers find little difference between these two samples, the sample sizes are small 
and systematics in the observables can be large. However, using the absorber temperature rather than the O~VI and H~I
line widths results in modest (95\% confidence, 2$\sigma$) correlations between log T and group luminosity and log T and 
group velocity dispersion. These marginal differences are suggestive, but not conclusive, that the warm absorbers found in 
high-S/N COS spectra are detections of diffuse gas in small, spiral-rich galaxy groups. But, using the observed
absorber \dNdz\ and the impact parameters found for these absorbers relative to the galaxy group centroid finds support
for a group association. The rather substantial \dNdz\ $\approx$ 4 per unit
redshift requires these absorbers to be extended over $\sim$ 1 Mpc in radius at high
covering factor. This large extent is consistent with the observed impact 
parameters between the QSO sightlines and the foreground group centroids. 

\section{Discussion}

The existence of a massive circum-galactic medium (CGM) or gaseous halo around individual late-type galaxies is now well-established. 
Several recent studies \citep* {prochaska11a, tumlinson11, stocke13, tumlinson13, werk14} 
have discovered and investigated cool, photo-ionized galaxy halo clouds found within the virial radii of large and small galaxies. These studies used
STIS and COS spectroscopy to detect photo-ionized Ly$\alpha$ and metal lines at very high covering factor out to at least the virial radius 
from the nearest, usually star-forming galaxy. The presence of similar clouds around early-type galaxies has been proposed but is still
controversial \citep{thom12}. Where sufficient ions were available to
provide good constraints on photo-ionization modeling, single-phase, homogeneous models were used to
calculate cloud temperatures and densities from which sizes and masses were inferred \citep* {stocke13, werk14,
keeney14}. The resulting sizes (diameters of 0.1 - 30 kpc) and masses 
(10$^{1-8}$ M$_{\odot}$) of these halo clouds are consistent with estimates for high-velocity-clouds (HVCs) 
in the Milky Way \citep* [e.g.,][] {wakker07, wakker08}. Estimates of the ensemble mass of the cool, photo-ionized
clouds around a super-L$^*$ galaxy exceeds 10$^{11}$ M$_{\odot}$ \citep*{stocke13, werk14}. Computing internal pressures for these clouds 
based on the derived temperatures and densities from the photo-ionization modeling finds
$\langle P/k \rangle \approx$ 10 cm$^{-3}$ K ($\pm$ 0.5 dex) \citep{stocke13}.

Surprisingly, \citet{stocke13} found no evidence for a declining pressure while \citet{werk14} finds
2.3$\sigma$ evidence for a shallow decrease in density $n \sim$ ($\rho$/R$_{vir}$)$^{-0.8}$. 
In either case the density decline in the halo is unexpectedly slight. It is important to be cautious in this interpretation
since it rests on data from single sightline probes past many galaxies as indicative of the CGM around any single galaxy in the sample. 
This conclusion also assumes that the gaseous halo is centered on the nearest galaxy, which would not necessarily be the case if the
confining medium were related to an entire group of galaxies. If this result is interpreted in the context of pressure equilibrium 
between these clouds and a hotter, more diffuse medium,
the absence of a rapid decrease in cool cloud pressure with increasing impact parameter is unexpected if these clouds (and any
surrounding intra-cloud medium) are related exclusively to an individual galaxy \citep* {anderson10, anderson11}. 

The discovery of the 20 warm absorbers studied here may be related to the existence of the hotter gas whose presence is suggested by
the cool, photoionized cloud pressures. It is important to recall that the absorbers found by Paper 1 and studied herein are too shallow and broad to have been 
detected in the lower S/N COS and STIS spectra published earlier. The strong O~VI absorbers previously studied by \citet* {tumlinson11, prochaska11a, werk13,
stocke13, werk14} are found well within $R_{vir}$ of the nearest galaxy's halo and appears associated in some way with the cool,
photoionized clouds. As found in the previous section, these ``warm absorbers'' are located on average much farther from their nearest galaxy neighbor
(i.e., typically at $\rho > R_{vir}$) but within the confines of a galaxy group containing that nearest galaxy. Most of these warm absorbers do 
not contain lower ions within the absorption complex which differs from the higher column density O~VI systems found by \citet {tumlinson11}. 

While simple correlation studies fail to show strong evidence 
for this proposed absorber/group association, modest ($\sim$ 2$\sigma$) correlations between absorber temperature and group luminosity and
velocity dispersion were found (n.b., these are not independent correlations). Also, the observed impact parameters to the nearest galaxy 
appear much too large (i.e., $\rho \gg R_{vir}$
in most cases) to explain an individual galaxy association with these absorbers unless the gas is extremely patchy. An association with group gas is more likely based on 
a match between the observed absorber impact parameters from group centroids and the required cloud sizes. While suggestive, the current data
are too sparse to warrant a claim that the warm absorbers are the first solid detections of an intra-group medium in spiral-rich galaxy
groups. While this is a possibility, the 10$^{5-6}$ K temperatures of these absorbers are an order of magnitude too low to be the intra-group
medium {\bf if} these groups are virialized. Most likely these absorbers mark the transition regions between the hot, 
intra-group gas and cooler, photo-ionized absorbers \citep{smith11, shull12, cen13}.

Figure 11 in the previous Section showed the T-$\sigma$ absorber/group data with extrapolations from hotter temperature groups and 
clusters shown as solid and dashed lines.
The dashed line indicates the extrapolation of the best-fit power-law from \citet [][as shown in Mulchaey 2000, Figure 4] {wu99} 
for clusters and groups into the T-$\sigma$ domain of the spiral groups studied here. It is on the basis of an extrapolation like this
that M96 predicted that a typical spiral-rich group with $\sigma$=100 km s$^{-1}$ would possess an intra-group gas 
with log T(K)=6.3. When \citet {savage10} discovered a broad O~VI-only absorber in the COS spectrum of PKS~0405-123 
whose $b$-value and absence of associated H~I Ly$\alpha$ strongly suggested log T(K)=6.1, it was natural to associate this 
absorption line detection of warm gas with the prediction of M96. However, as shown in Figures 1 \& 11 the observed 
(absorption-line determined) temperatures in spiral-rich groups presented herein are at least
an order of magnitude lower than as predicted by the velocity dispersions of spiral-rich groups {\it if} an
extrapolation of the \citet {wu99} rich cluster and group data is appropriate.  

More recently \citet {osmond04} constructed a large group catalog using the {\it ROSAT} all-sky X-ray survey and a
number of nearby group catalogs. The resulting log T vs log $\sigma$ 
correlation found for these groups compared to a sample of rich clusters is shown in \citet [][see their Figure
16] {osmond04}. What is most apparent is that the 
groups have a much larger spread in velocity dispersion at a given log T$_x$ than the rich clusters. It is controversial
whether a single power-law fits both the cluster and group data or whether the slope steepens in the group regime \citep* [e.g.,][]
{mulchaey00, osmond04, helsdon05}. But neither the cluster plus group best-fit slope (0.71), nor the group-only best-fit slope (1.15), 
predicts the gas temperatures that we observe for the spiral-rich groups. 
An extrapolation of a single power-law which fits both the clusters and the most X-ray luminous, hottest groups (but not
the lower L$_x$ E-dominated groups) and is theoretically motivated by equipartition of energy between gas and galaxies; i.e.,
$\beta_{spec}$ = 1 \citep {osmond04} is shown in Figure 11 as a solid line. 

While the solid-line extrapolation comes closer to predicting the temperatures for the warm absorbers, 
even the $\beta_{spec}$ = 1 extrapolation falls well below the spiral data points from this study, indicating that 
there is proportionately more energy in the galaxies than in the warm absorber gas. If a group is still in the process of virializing, it is 
{\it possible} that the galaxy motions have not had the time required to heat the gas to a virial equilibrium temperature in the case that the gas has been ejected 
from the galaxies only recently.  But it is also
possible that the O~VI + H~I warm absorbers are sensitive only to that portion of the intra-group gas closest to the temperature where the fraction of oxygen in
the O~VI ion maximizes (log T(K) $\approx$ 5.5) and that these warm absorbers indirectly trace the presence of even hotter gas, to which the HST/COS
spectra are not sensitive. At T $>$ 10$^6$ K the neutral fraction of hydrogen is {\it f$_{HI} \sim$ 10$^{-7}$} \citep{danforth10} and the fraction of oxygen which is
in the O~VI ion is $<$ 10\% \citep {bregman07}. If the bulk of the intra-group gas is at temperatures much hotter than 10$^6$ K, 
the warm absorbers could be interfaces with or cooler portions of this hotter gas and the bulk of the intra-group gas has not been detected directly as yet.  

It is also reasonable to consider the possibility that the warm absorbers are a diffuse, high filling-factor gas in these groups as a limiting case; 
i.e., while not yet in equilibrium with the energy in the galaxies, these warm absorbers are nevertheless the bulk of the intra-group medium. 
To illustrate the consequences of this possibility, we assume pressure balance between the cool,
photo-ionized clouds ($\langle P/k \rangle \approx$ 10 cm$^{-3}$ K) and a ``warm'' diffuse medium with T$\sim$ 10$^{5.5}$ K based on our warm absorber detections
(see Figure 1). This constrains the density of this diffuse gas to be $\sim$ 10$^{-4.5}$ cm$^{-3}$ over a region with a minimum radius of $\geq$ 300 kpc.
These density and minimum size estimates yield a {\it minimum} warm mass reservoir of $\geq$ 10$^{11}$ M$_{\odot}$. 
An even larger gas extent at high covering factor is required by the \dNdz\ of the warm absorbers ($\sim$ 1 Mpc in radius; 
see Section 3.5). Assuming a standard $\beta$-model \citep {mulchaey98} with $\beta$ = 0.67 for the warm gas density profile 
with a 300 kpc core radius, a maximum gas radius of 1 Mpc and a central gas density of 3 $\times$ 10$^{-5}$ cm$^{-3}$ yields a total warm gas mass of
$\sim$ 10$^{12}$ $M_{\odot}$. However, the large extent of this
diffuse gas would create unacceptably large line widths for the O~VI and
Ly$\alpha$ absorption. We estimate that the required velocity dispersions in such large
structures would create line widths at least 2--3 times larger than what we find for these
absorbers (see Table 2) either from Hubble flow broadening (unbound case) or from the substantial velocity
dispersions of these groups (bound case). Also gas at the temperatures of these warm absorbers has a very short cooling time and so is unlikely to be the dominant 
gas reservoir around late-type galaxies and in their groups \citep{shull12, cen13}.
\citet{smith11} find that in the present epoch, most of the gas at log T(K) = 5.0-6.0 has 
recently cooled from hotter gas that is the dominant WHIM reservoir at $z\sim$0.

In order to satisfy the observations of nearly constant pressure over a minimum impact parameter of $\sim$ 300 kpc as well as the likely, but not firm, association
of the warm absorbers with galaxy groups rather than individual galaxies, a diffuse hot medium is still required. Using Figure 11 as a guide, a reasonable temperature
for the intra-group medium in groups with $\sigma_{grp}$ = 250-450 km s$^{-1}$ is T$\sim$ 10$^{6.5}$ K. Pressure balance then requires a central gas density of 3 $\times$ 10$^{-6}$
cm$^{-3}$ and a total hot gas mass of $\sim$ 10$^{11}$ $M_{\odot}$ (see previous paragraph). 
The core radius (300 kpc) and full extent (1 Mpc) we assume are similar to what is found in 
elliptical-dominated groups \citep {mulchaey00} but the total mass is significantly less. While we have assumed
a value of $\beta$ comparable to that found for clusters of galaxies, \citet {osmond04} find
evidence for a somewhat smaller value of $\beta$ in groups which leads to a shallower density profile and a greater mass. Truncating the mass
distribution at 1 Mpc may also be artificial and a substantial gas reservoir could be present farther away than 1 Mpc ($>$ R$_{vir}$)
as in the rich clusters \citep{simion11}. Therefore, the bulk of the evidence presented herein suggests that the warm absorbers are an {\it indirect, not a direct} 
detection of a massive baryon reservoir of hot gas in small, spiral-rich groups of galaxies at T$\sim$ 10$^{6.5}$ K.

\section{Summary of Conclusions}

The Cosmic Origin Spectrograph (COS) Science Team has conducted a high-S/N ($\geq$20:1), high resolution (R $\approx$ 18,000) far-UV
spectroscopic survey of 14 bright QSOs. Paper 1 reported some of the results from this survey concentrating on the
absorbers detected in O~VI. Specifically, Paper 1 identified a new class of broad, shallow O~VI and Ly$\alpha$ absorbers which are
demonstrably hot enough to be fully in the collisionally-ionized regime; we term these ``warm absorbers'' in Paper 1 and herein (see Figure 1).
As a class these warm absorbers have not been detected or studied heretofore in any detail due to the absence of sufficiently high-S/N FUV spectra,
which the COS GTO observations have now provided.

Additional absorption systems with and without metals from these spectra are presented in \citet {danforth14}.
Most of the O~VI absorbers were also detected in H~I Ly$\alpha$ with COS but two O~VI absorbers
were not detected in H~I at all and a few with broad Ly$\alpha$ (BLA) detections had O~VI detections in the {\it FUSE} band. One {\it possible}
warm absorber contains a BLA in Ly$\alpha$, which is confirmed to be warm using the observed Ly$\beta$ width but has no corresponding O~VI.
Fourteen broad O~VI + BLA detections in the Science Team's survey have absorption in these two species aligned in velocity with $b$-values requiring
collisionally-ionized gas at T $\geq$ 10$^5$ K, so called, ``warm gas''. Two other absorbers found in the spectra analyzed in Paper 1 possess
broad O~VI absorption with no corresponding H~I Ly$\alpha$ absorption, also requiring the presence of warm gas. In this paper the galaxy environments of these 
warm absorbers are investigated in order to determine the most likely type of galaxy or galaxy group with which they 
are physically associated. Given an association with either individual galaxies or with the entire galaxy group,
the spatial extent and gas mass of these warm absorbers is also estimated. We find the following results:  

1. We have reviewed the results of Paper 1 concentrating on those few broad, symmetric O~VI absorbers suggestive of warm
gas which were not identified as such by Paper 1 (see Section 2). While almost all of the model fits made by Paper 1 
were confirmed, we suggest alternative fits for 4 absorbers only (see
Appendix A). These reevaluations, plus two O~VI-only absorbers that
lack associated Ly$\alpha$, suggest that as many as 20 warm absorbers may be present in this sample compared to 14 
(plus 2 O~VI-only) warm absorbers identified in Paper 1. Using the 
pathlength probed by these 14 high-S/N spectra of $\delta z $ = 4.0, we find a \dNdz\ for warm absorbers of 3.5-5 per unit 
redshift, several times the number density of Mg~II/LLS absorbers and about one-quarter of the \dNdz\ value for all previously-detected 
O~VI absorbers \citep{danforth08}.   

2. Spiral-rich galaxy groups are found around the location of most of these warm absorbers with total
luminosities of $\sim$2--26L$^*$ and velocity dispersions, $\sigma$ = 250--450 km s$^{-1}$. These groups have 2--5 
bright (L$\geq$L$^*$) group members that are usually luminous early-type spirals. However, two 
warm absorbers have galaxy environments more appropriately described as cluster outskirts or as a galaxy group within a filament between clusters. 
These two warm absorber groups have velocity dispersions of $\geq$ 500 km s$^{-1}$, total luminosities of $\sim$30-60 L$^*$ and 
brightest galaxies that are 2-4 L$^*$ early-types.

3. Evidence is presented in Section 3 that the warm absorbers are more likely
associated with the galaxy group than the nearest galaxy, although this
evidence is not entirely conclusive due to the small sample size. 
Group properties associated with either warm or cool absorbers do not show significant differences. 
Modest (95\% probability) 2$\sigma$ correlations are found between both the total group luminosity and 
also the gas temperature in comparison with the group velocity dispersion (Figures 10 and 11). On the other hand we find no evidence favoring an 
association with the nearest galaxy neighbors to these absorbers. 

4. The large \dNdz $\approx$ 4 per unit redshift favors an association with galaxy groups rather than individual galaxies (see Table 8 in Section 3.5). 
Regardless of their luminosity, most of the nearest galaxies are systematically too far away from the warm absorbers to be viably associated.
On the other hand, if the associations are with galaxy groups, then the impact parameters match the absorber sizes of $\sim$
1 Mpc radius required in the case of high covering factors. One Mpc is also the approximate size of the virial radius
of these groups, if they are in fact gravitationally bound. 
But in this case the rather narrow O~VI and Ly$\alpha$ line widths
observed are
explicable only if these absorbers are individually small with respect to
the galaxy group but which in ensemble cover the entire group at very high
covering factor.

5. While the T-$\sigma$ relation for spiral groups in this sample (Figure 11) lies well above an extrapolation of the relation seen for
clusters and elliptical dominated groups, the data for spiral-rich groups lie much closer to an extrapolation of the cluster $+$ group X-ray data if
equipartition of energy between gas and galaxies ($\beta_{spec}$ = 1) is assumed. However, the temperatures of the warm gas (see Figure 11)
are approximately one order of magnitude too cool to be diffuse group gas if $\beta_{spec}$ = 1 and if the group is virialized. The origin of this discrepancy 
is not known. It could be that the O~VI absorption is {\bf not} tracing the bulk of the intra-group gas but rather the gas at the temperature to which the O~VI
is most sensitive, T $\approx$ 10$^{5.5}$ K. In this case hotter gas likely 
is present in these groups but is not easily detected in the COS FUV absorption-line spectra. 

6. The observed, nearly-constant, cool cloud internal pressures \citep* [$\langle P/k \rangle \approx$ 10 cm$^{-3}$ K; ][] {stocke13, werk14} as a function of
impact parameter require a very large, diffuse intra-cloud gas to confine these clouds. While the warm absorbers could
be detections of this massive, diffuse gas reservoir, it is more plausible that these warm clouds are a tracer of even hotter, diffuse gas which confines the cool,
photoionized clouds. Three observables favor
this interpretation: (a) the warm absorber O~VI and H~I line widths are too narrow to be gas distributed over a large region of a bound or unbound galaxy
group; (b) the warm absorber temperatures are in a regime where diffuse gas is quite unstable to rapid cooling so that a large diffuse gas reservoir would be difficult
to maintain at that temperature; and (c) the warm absorber temperatures are
an order of magnitude too cool to be virialized gas in the groups we have found ($\sigma$=250-450 km s$^{-1}$; see Figure 11). 
In the context of a standard $\beta$-model with a core radius of $\sim$ 300 kpc and a full extent of $\sim$ 1 Mpc (based on the results described in the previous
Section) we find a total mass of $\sim$ 10$^{11}$ M$_{\odot}$ for this hotter gas based on pressure equilibrium with the cool, photoionized clouds. These values for 
the spiral intra-group medium suggests a similar extent to the gas in elliptical-dominated groups \citep {mulchaey00}, but less mass. This amount of gas also 
is quite similar to the amount of ``missing'' baryons required to bring spiral galaxy groups up to the cosmic mean ratio of baryons 
to dark matter. {\bf If} the conclusions summarized in this paragraph are correct, spiral galaxy groups could be ``closed boxes'' for cosmic evolution like their
more massive counterparts which contain dominant ellipticals. 

Regardless of their detailed interpretation, these warm absorbers are detections of local baryons not inventoried before. As such these absorbers will add to the
baryon census, but with an amount that can only be determined by making an independent measurement of their size and filling factor. New HST/COS
FUV spectroscopy of QSO sightlines passing through pre-defined low-redshift galaxy groups will be important in confirming or denying the
presence of warm gas in spiral-rich galaxy groups. These suggested observations could establish the cosmic importance of this warm gas and the hotter gas which it
probably traces.

\acknowledgements

This work was supported by NASA grants NNX08AC146 and NAS5-98043 to the University of Colorado 
at Boulder and the University of Wisconsin at Madison for the \hst/COS project.
JTS, BAK and HY are partially supported on NSF grant AST-1109117 for the galaxy redshift survey work.
TSK acknowledges current support from a European Research Council Starting Grant in cosmology and
the IGM under Grant Agreement (GA) 257670. ERW acknowledges support of Australian Research Council
Grant DP1095600. Dr. Steven Penton is thanked for help with the galaxy redshift
survey database. The authors thank an anonymous referee for her/his persistence in pointing out to us the line-width
problems created by models in which these absorber detections are a diffuse gas reservoir.

{\it Facilities:} \facility{HST (COS)}, \facility{FUSE}, \facility{WIYN (HYDRA)},
\facility {APO 3.5 m}, \facility {CTIO/Blanco (Hydra)}, \facility {AAT (AAOmega)}

\input ./biblist.tex
\begin{appendix}
\section{Detailed Reanalysis of a few ``Warm Absorber'' Candidates}

In this Appendix we present detailed line fits to a few BLA plus broad O~VI absorbers
which may contain T $>$ 10$^5$ K gas, but were not included as warm absorbers in Paper 1. 
These fits {\it do not} replace the line fitting done
in Paper 1 which was accomplished with no prior conditions and executed in a
conservative manner; e.g., minimum number of components, no requirement to have
Ly$\alpha$ align with O~VI necessarily, etc. (see Appendix in Paper 1). Here we take a
different tactic by adopting an aim to determine the maximum number of
plausible warm absorbers which might be present in this sample. We do this by first
analyzing the O~VI absorption which is less contaminated by lower temperature,
photo-ionized gas than Ly$\alpha$. If the O~VI absorption is smooth and symmetrical to
the limits of the S/N, then we fit this absorption with a single component and assume
that the observed $b$-value is mostly due to thermal motions. With the O~VI fit
in-mind we then fit the complex, multi-component Ly$\alpha$ line,
constraining the fit to include a component aligned with the broad O~VI absorption. If
this analysis shows that a BLA can be present with a velocity consistent with the O~VI 
and an observed $b$-value consistent with T $>$ 10$^5$ K gas, then this absorber
is added to the list of plausible warm absorbers in this sample.

See the Appendix and Table 5 of Paper 1 
for the Ly$\alpha$ and O~VI line fits of these 10 absorbers, which we now discuss one-by-one. For each
absorber we first describe the results of Paper 1 and then the new analysis is presented in the second
paragraph. We emphasize that these re-analyses represent alternative absorption-line component de-convolutions, not
necessarily superior ones, to those found in Paper 1. 

\subsection{PKS2155-304/0.05722}

Because the Ly$\alpha$ and O~VI absorptions do not align in velocity, Paper 1 did not
analyze this
system in detail. However, the O~VI absorption visible in a high-S/N  {\it FUSE}
spectrum is quite
symmetrical and has a velocity within the span of the Ly$\alpha$ profile. This system
has been studied
in detail by \citet* {shull98, shull03} including a reconciliation of the 
O~VI detection and possible
{\it Chandra} O~VIII detections. \citet {shull98} also identified a small foreground
group of
galaxies at the absorber redshift and provides accurate H~I emission velocities for
each group member
detected. Therefore, it is worthwhile to consider this absorber in the context of warm
gas.

Despite attempts to constrain a BLA absorption component to be coincident in velocity
with the broad
O~VI, the closest offset finds the O~VI centroid 46 km s$^{-1}$ higher in velocity
than the nearest
Ly$\alpha$ component. Because the Ly$\alpha$ absorption is found in both a COS and a
STIS spectrum and
the O~VI was detected by {\it FUSE}, a velocity offset is possible. However, the {\it
FUSE} LiF2a spectrum and
the COS spectrum detect the same Galactic Fe~II 1144.8\AA\ line with a velocity offset
of only -10 km s$^{-1}$.
Therefore, the O~VI absorption remains inconsistent ($\Delta$v $\geq$ 36 km s$^{-1}$)
with any
Ly$\alpha$ component and no warm absorber is present in agreement with the conclusion
of Paper 1.

\subsection{3C~273/0.09010}

Paper 1 reports that this system contains three Ly$\alpha$ absorption lines which are
nearly completely
disjoint in velocity while the O~VI absorber is displaced by $\approx$ 30 km s$^{-1}$
to higher
velocity. The best fit to the redward of two closely-spaced Ly$\alpha$ components (see
Figure 12)
is quite acceptable but leaves open the possibility for the presence of a broad
component that
could align with the observed, symmetric O~VI absorption.

In our reanalysis of this absorber we have fit the Ly$\alpha$ absorption with four
components,
constraining one of the components to align with the symmetrical O~VI absorption line.
Our best fit
for this aligned component finds $b_{H~I}=$ 49$\pm$3 km s$^{-1}$ and $b_{O~VI}\approx$
20 km s$^{-1}$ (see Table 9). The
combination of $b$-values results in a thermal width of $\approx$ 46 km s$^{-1}$ and
log T (K) $\approx$ 5.1.
Therefore, we judge that this absorption system likely, but not conclusively, contains
a warm
absorber. This new fit is shown in Figure 12.

\begin{figure}
  \epsscale{0.8}
  \plotone{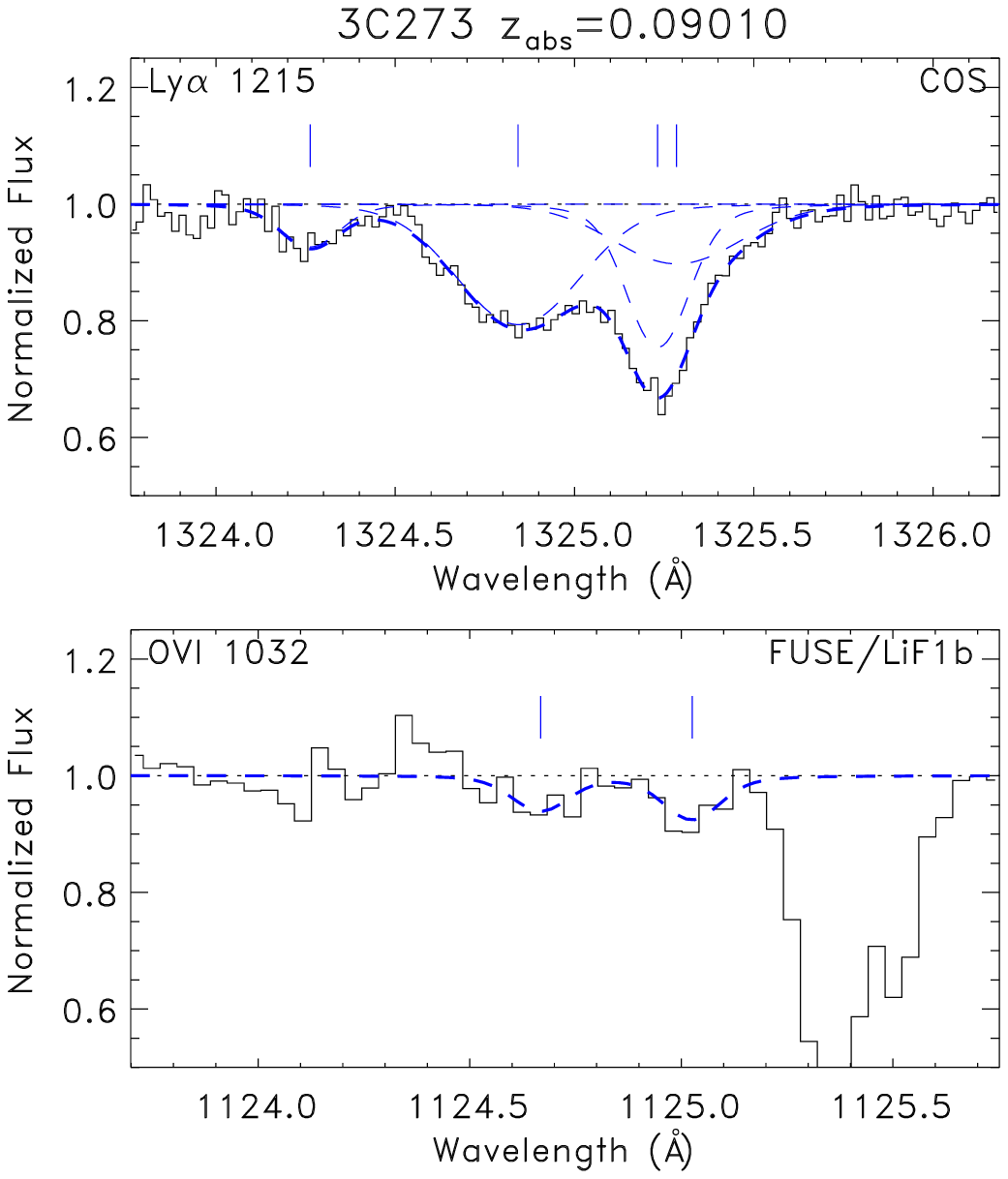}
  \caption{The Ly$\alpha$ and O~VI 1032 \AA\ absorption lines for the 3C\,273 $z_{\rm
abs}=0.09010$ absorber with
our revised models (see Table 9) in blue. The redward (longest wavelength) BLA was
constrained to match the redward of two
O~VI absorbers to $\Delta$v of $\pm$10 km s$^{-1}$}
\end{figure}

\begin{deluxetable}{lcccc}
\tabletypesize{\footnotesize}
\tablecolumns{5}
\tablewidth{0pt}
\tablecaption{Profile Fit Results: 3C\,273 $z_{\rm abs}=0.09010$}
\tablehead{\colhead{Species}    &
           \colhead{$\lambda_0$} &
           \colhead{$v$} &
           \colhead{$b$} &
           \colhead{$\log\,N$}     \\
           \colhead{}          &
           \colhead{(\AA)}     &
           \colhead{($\rm km~s^{-1}$)} &
           \colhead{($\rm km~s^{-1}$)} &
           \colhead{($\rm cm^{-2}$)} }
 \startdata
  Ly$\alpha$ & 1216 &$-226\pm3 $&$22\pm4$ &$12.47\pm0.05$\\
             &      &$ -83\pm2 $&$49\pm3$ &$13.23\pm0.02$\\
             &      &$ +12\pm2 $&$23\pm3$ &$13.06\pm0.09$\\
             &      &$ +26\pm10$&$52\pm11$&$12.92\pm0.10$\\
  O\,VI      & 1032 &$ -68\pm13$&$\sim20$ &$13.0\pm0.3$\\
             &      &$ +32\pm10$&$\sim20$ &$13.1\pm0.02$\\
  O\,VI      & 1038 &$ +35\pm10$&$\sim26$ &$13.5\pm0.02$\\
 \enddata
\end{deluxetable}

\subsection{ PG~1116+215/0.13850}

Despite a broad (b= 36 km s$^{-1}$), symmetric and aligned O~VI absorption line
well-measured with COS,
Paper 1 finds only minimal evidence for a BLA setting a temperature of log T (K)=4.72,
consistent with
very low density gas in photo-ionization equilibrium. The combination of the
Ly$\alpha$ and O~VI
$b$-values obtained by Paper 1, suggests that the line widths are dominated by
turbulence. This would
be an unusual situation given the striking symmetry and lack of structure in the
COS-observed O~VI
line.


\begin{figure}
  \epsscale{0.8}
  \plotone{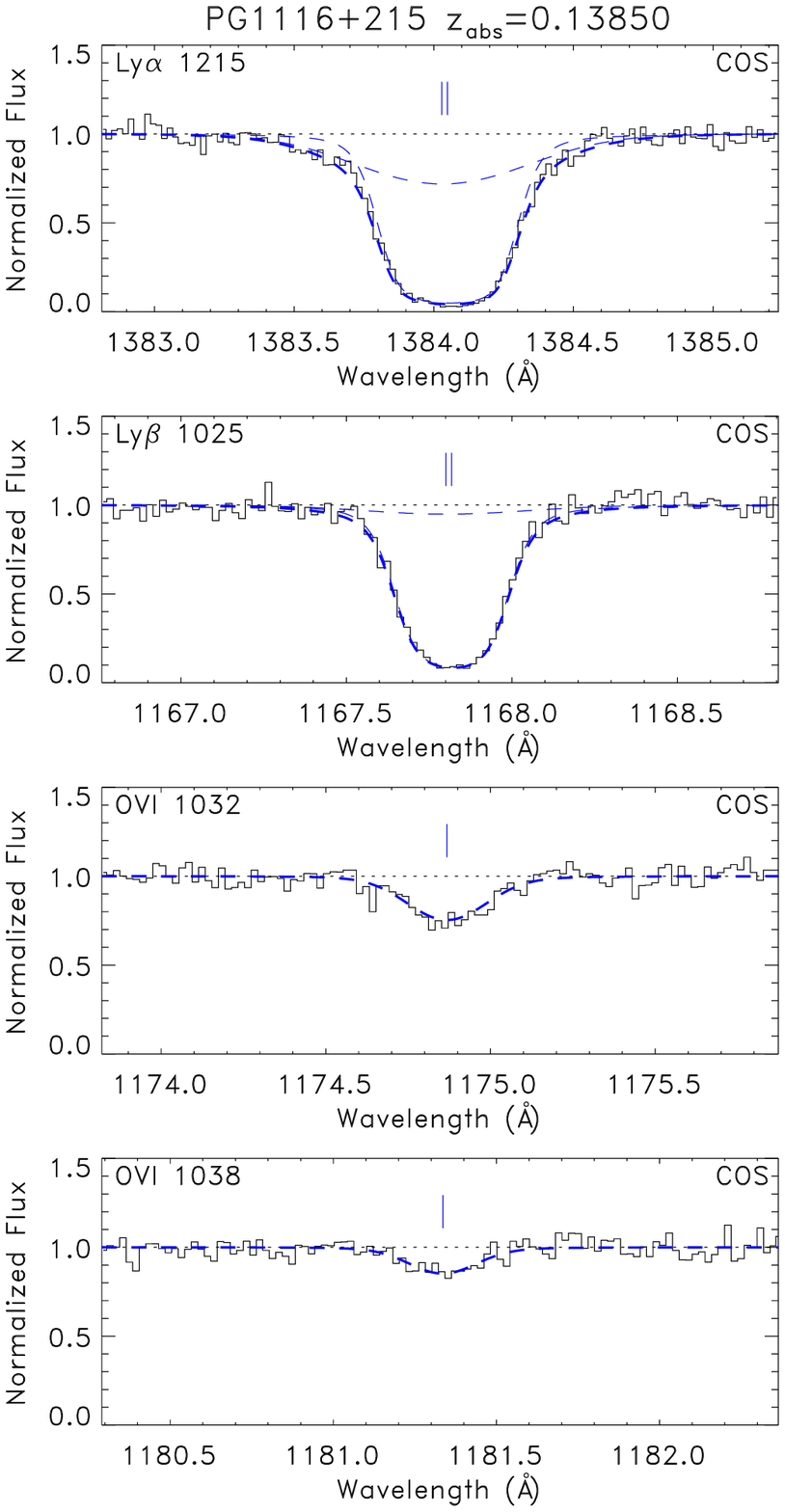}
  \caption{The Ly$\alpha$, Ly$\beta$ and O\,VI doublet absorption
  lines for the PG\,1116$+$215 $z_{\rm abs}=0.13850$ absorber with our
  revised models (see Table~10) over-plotted in blue.  The BLA
  component is displayed on the
  Ly$\beta$ profile for illustrative purposes only.}
\end{figure}

\begin{deluxetable}{lcccc}
\tabletypesize{\footnotesize}
\tablecolumns{5}
\tablewidth{0pt}
\tablecaption{Profile Fit Results: PG\,1116$+$215 $z_{\rm abs}=0.13850$}
\tablehead{\colhead{Species}    &
           \colhead{$\lambda_0$} &
           \colhead{$v$} &
           \colhead{$b$} &
           \colhead{$\log\,N$}     \\
           \colhead{}          &
           \colhead{(\AA)}     &
           \colhead{($\rm km~s^{-1}$)} &
           \colhead{($\rm km~s^{-1}$)} &
           \colhead{($\rm cm^{-2}$)} }
 \startdata
  Ly$\alpha$ & 1216 &$ +3\pm5$&$86\pm11$&$13.6\pm0.1$ \\
             &      &$ +8\pm1$&$31\pm2$ &$14.8\pm0.1$ \\
  Ly$\beta$  & 1026 &$+10\pm1$&$28\pm1$ &$15.3\pm0.1$ \\
  O\,VI      & 1032 &$ +6\pm2$&$37\pm3$ &$13.8\pm0.1$ \\
  O\,VI      & 1038 &$ +3\pm5$&$32\pm5$ &$13.8\pm0.1$ \\
  N\,V       & 1238 &$  0\pm2$&$14\pm3$ &$12.9\pm0.1$ \\
  Si\,IV     & 1394 &$ -7\pm2$&$\leq 5$ &$12.9\pm0.1$ \\
  C\,IV      & 1548 &$ -9\pm3$&$17\pm4$ &$13.3\pm0.1$ \\
             & 1550 &$ -1\pm6$&$26\pm5$ &$13.4\pm0.2$ \\
  Si\,III    & 1206 &$ -2\pm1$&$11\pm1$ &$12.8\pm0.1$ \\
  C\,II      & 1334 &$ -2\pm1$&$ 8\pm1$ &$14.0\pm0.1$ \\
             & 1036 &$ +2\pm1$&$14\pm2$ &$13.8\pm0.1$ \\
  Si\,II     & 1260 &$ -7\pm2$&$15\pm2$ &$12.5\pm0.1$ \\
             & 1193 &$ -7\pm1$&$13\pm2$ &$12.8\pm0.1$ \\
 \enddata
\end{deluxetable}

\begin{figure}
\epsscale{0.8}
\plottwo{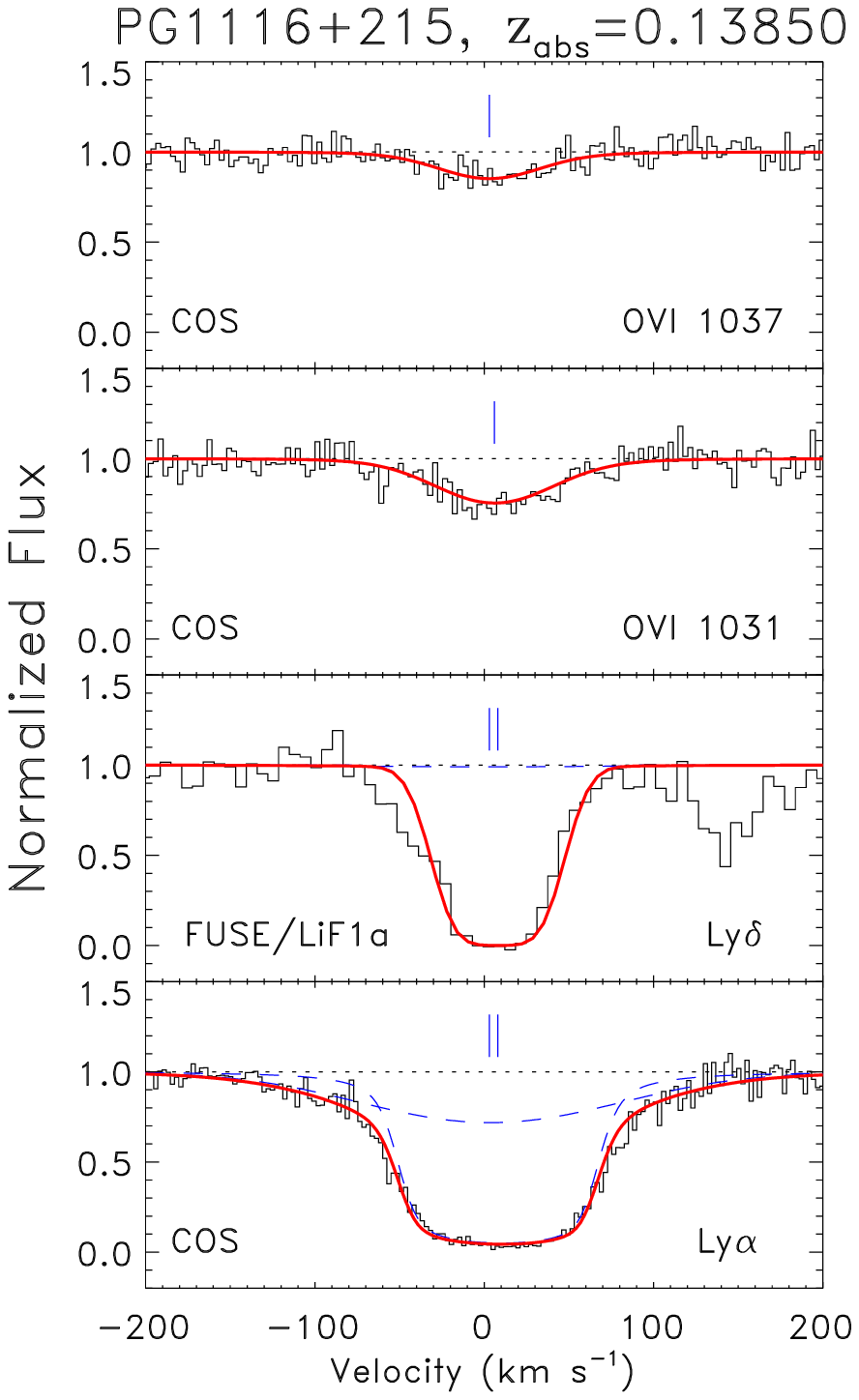}{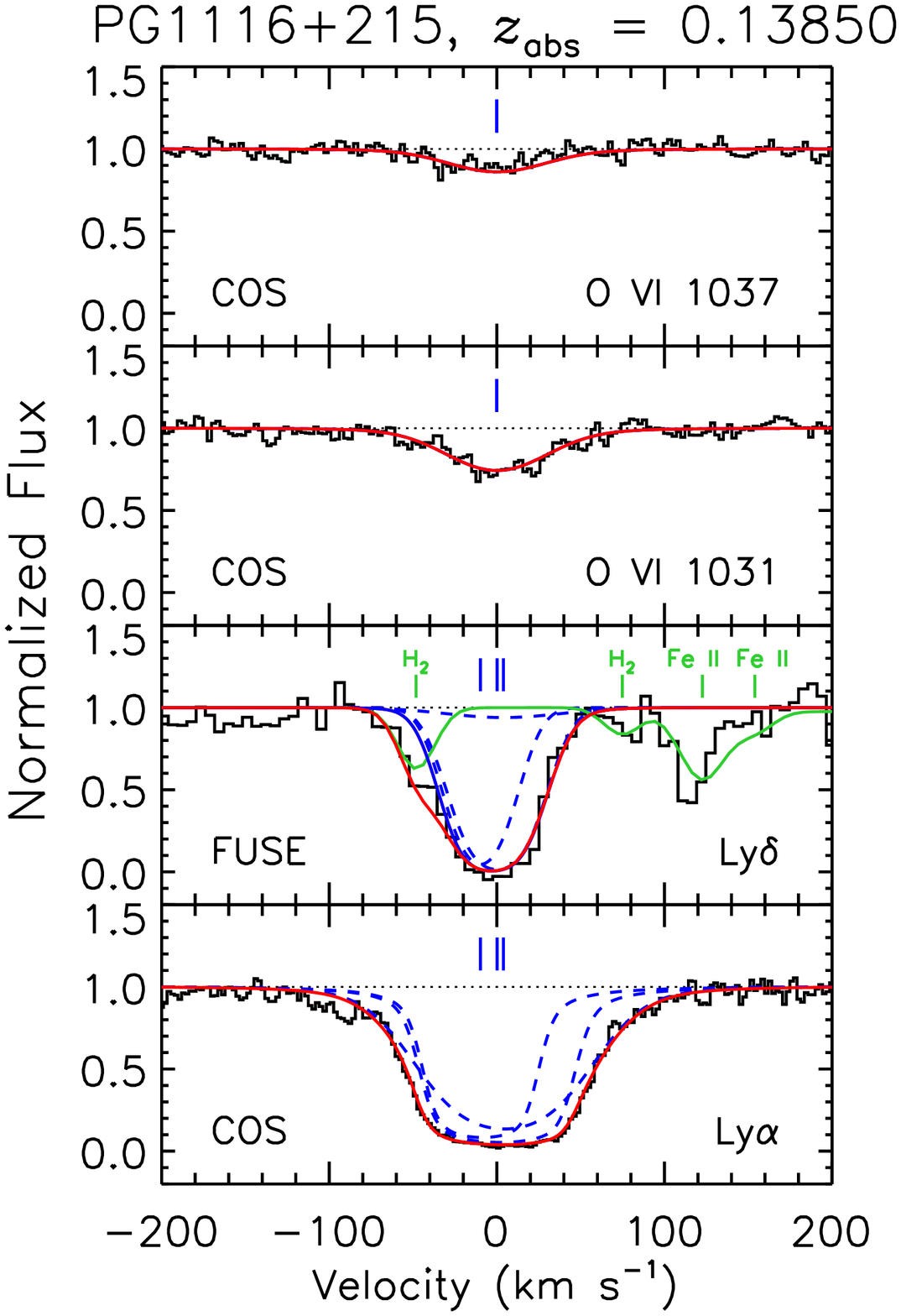}
\caption{The Ly$\alpha$ absorption lines as extracted and fitted by this analysis (at left) and by Paper 1 (at right). 
The Paper 1 extraction includes an attempt to remove fixed pattern noise (see Paper 1 for details), which this analysis does not employ.
In both cases a blue-wing is present which we fit as a BLA component using
the extraction at left (see Table 10 for detailed solutions).}
\end{figure}

An alternate extraction of the same data by Danforth et al. (2010) finds a slightly
stronger blue wing
to the Ly$\alpha$ line which is aligned with the O~VI (see Figure 12). A best-fit
$b(H~I)= 86\pm$ 11 km
s$^{-1}$ BLA is found, which is aligned to within the velocity errors with the O~VI
1031, 1038\AA\
$b=35\pm$5 km s$^{-1}$ absorber (see Table 10). This combination of observed b-values
yields $b_T =$ 81
$\pm$14 km s$^{-1}$, log T (K)= 5.6 (+0.1/-0.2)  and $b_{NT}\approx$30 km s$^{-1}$. A narrower
Ly$\alpha$ component
is well-matched ($b=31 \pm$ 2 km s$^{-1}$) to the widths of the low ion metal lines
(see Figure 13 and
Table 10). However, a curve-of-growth (COG) analysis using 3 Lyman-series lines finds
an order of
magnitude more column density than as shown for Ly$\alpha$ in Table 7 and a 10 km
s$^{-1}$ lower b-value
for the cool component. Also the N~V 1238 \AA\  and C~IV 1548, 1550 \AA\ absorptions
have  $b$-values
intermediate between the C~II and O~VI absorptions. Clearly the photo-ionized absorber
in this system
is more complex than a single homogenous cloud. Using the COG values for the cool
absorber does not change the
best-fit to the warm absorber component. 

For clarity in Figure 14 we show the Ly$\alpha$ absorption profiles for this absorber as
extracted and fitted by Paper 1 (at left) compared to the extraction and analysis reported herein
(at right). While a blue-wing is present in both cases, it is present more clearly in the 
extraction shown at left.
In the spirit of the reanalysis presented here, we adopt the line-fit shown at left in Figure 14
in which a BLA is found coincident in velocity with the broad O~VI (solution presented in Table 10).
We conclude that the PG~1116+215/0.13850 absorption complex very likely contains 
a warm absorber.

\subsection{PHL~1811/0.17651}

The best fit to the 1038 \AA\ O~VI absorption is broad (b = 20 km s$^{-1}$) and
symmetric; the 1032 \AA\
line is obscured by Galactic Ly$\alpha$.  The C~III 977 \AA\ line shows two components,
the red-ward
component quite narrow ($b \approx$ 8 km s$^{-1}$) and the blue-ward one unmatched
either in
Ly$\alpha$ or O~VI. The two-component fit to the Ly$\alpha$ line in Paper 1 does not
yield either a
broad or an aligned component. Therefore, no absorber temperature was derived and no BLA
obviously
detected. However, the O~VI velocity centroid is within the velocity spread of the
Ly$\alpha$
absorption, suggesting that a BLA at the O~VI velocity could be present.

However, in investigating this system further we find that for C~III the broader,
blue-ward component
(see Figure A4 in Paper 1) is identified as Ly$\beta$ at $z=$ 0.12051. Additionally,
Ly$\gamma$ in
this system is confused by the presence of an Fe~II HVC so that little useful detail can
be derived
from it. Refitting Ly$\alpha$ unconstrained by Ly$\gamma$ can produce a two component
fit with one
component aligned with the broad O~VI. However, the width of this line cannot be much
broader than the
fit in Paper 1. Therefore, our reanalysis of this system concludes that a warm absorber
is not present,
consistent with the conclusions of Paper 1.

\subsection{H~1821+643/0.17036}

In this case the O~VI absorption is fitted by Paper 1 as two components, one of which is
extremely
broad ($\approx$ 80 km s$^{-1}$). However, the continuum is poorly defined in this
region due to the
presence to the blue of Galactic Si~III and an intervening Lyman series line at a
different redshift.
The resulting O~VI absorption is quite shallow (see Figure A6 in Paper 1) and, while a
symmetrical
O~VI component could be present, the overall profile is not obviously Gaussian. The velocity
centroid of the O~VI is well within the velocity spread of the Ly$\alpha$ absorption
line complex,
which is fit by three overlapping components. The best-fit solution in Paper 1 does not
yield an
alignment between the very broad O~VI component and any of the three Ly$\alpha$
components. A second, weaker
O~VI line coincident in velocity with a BLA component yields log T (K)= 5.10, but the O~VI
line fitting is
uncertain due to an uncertain continuum.

In refitting this absorber we started by removing the higher-order Ly$\zeta$ line at
$z=$0.29686 using
a model for the line strength and width obtained using curve-of-growth fitted values.
Then, instead of
allowing the fit to proceed without constraint as in Paper 1, the two O~VI components
were constrained
to align with the two higher redshift components in Ly$\alpha$. In this case a poorer
fit to the data
was obtained compared to that of Paper 1 but two aligned components resulted (see Figure 15). 
For the Ly$\alpha$/O~VI
component at $\Delta$v $\approx$ -100 km s$^{-1}$ the fitted H~I and O~VI lines have
nearly equal
$b$-values of $\approx$ 50 km s$^{-1}$ implying that the line widths are dominated by NT
motions and the bulk temperature of the gas is quite low, consistent with
photo-ionization
equilibrium. In our minimally constrained model, the component at $\Delta v \approx$ +30
has $b_{OVI}
= 47\pm$8 km s$^{-1}$ and $b_{H~I} = 34\pm2$ km s$^{-1}$, an unphysical result. Adding a
fourth component specifically at the
broad O~VI velocity is allowed by the data but results in a $b_{H~I} \approx$ 50 km
s$^{-1}$, yielding
an absorber at log T (K) $<$ 5.0, with substantial NT motions. Thus, attempts to constrain
two
Ly$\alpha$ and O~VI absorbers to match in velocity result in a poorer fit than in Paper
1 and no
second warm absorber. Therefore, we adopt the solution of Paper 1 for this absorber as
its best
description.


\begin{figure}
  \epsscale{0.8}
  \plotone{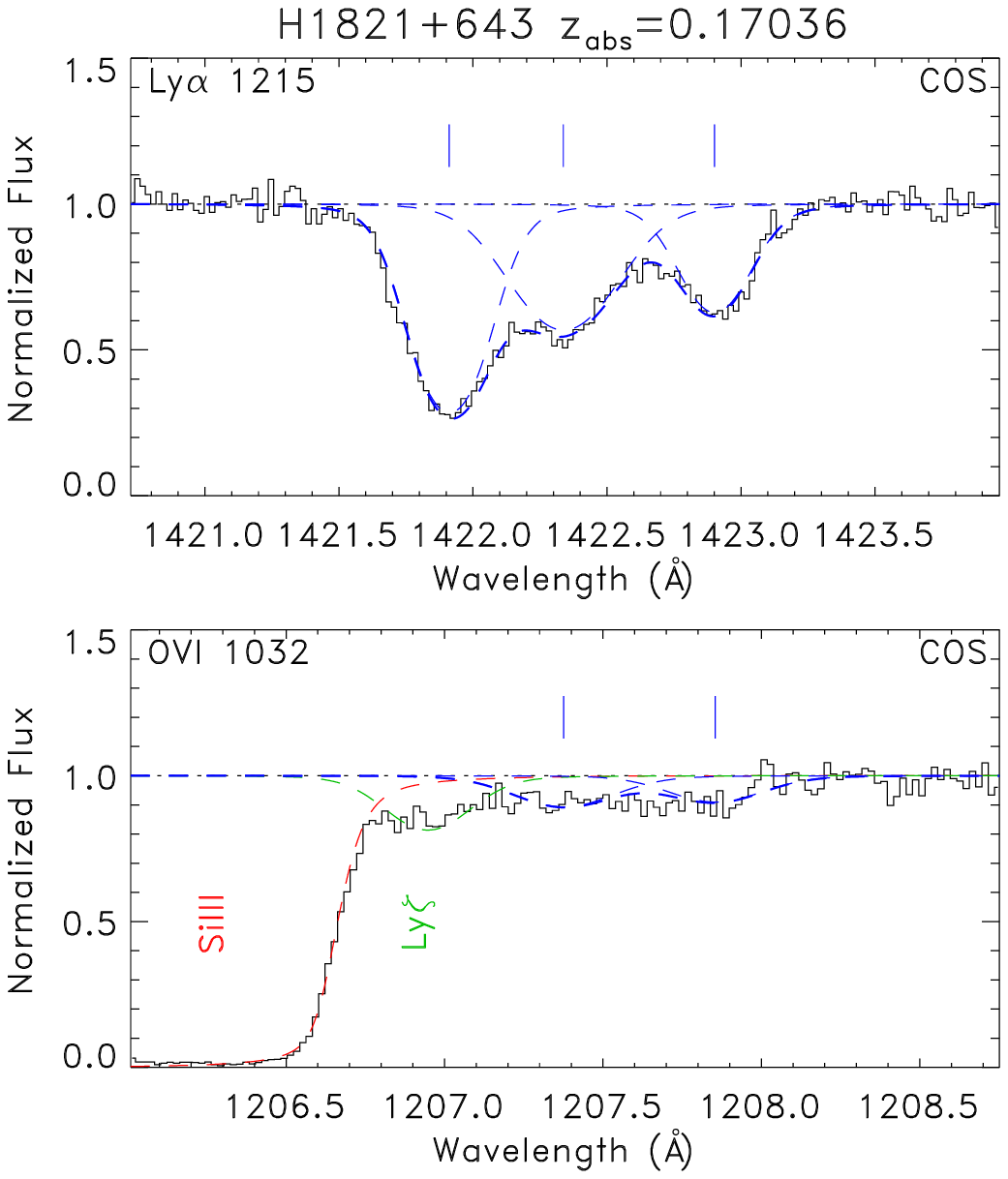}
  \caption{The Ly$\alpha$ and O\,VI doublet absorption lines for the
  H\,1821$+$643 $z_{\rm abs}=0.17036$ absorber with our revised models
  over-plotted in blue.  Careful fits to the Galactic Si\,III and
  intervening Ly$\zeta$ ($z_{\rm abs}=0..29675$) are shown in red and
  green, respectively.  In this case the new models are not so
  dissimilar from the Paper~1 models and no new warm absorber is claimed.}
\end{figure}

\subsection{Ton 236/0.19457}

The very broad ($b=$70 km s$^{-1}$) O~VI measured with COS is somewhat symmetrical (see
Figure A8 in
Paper 1), but could have excess absorption redward of line center. Paper 1 had already
fit a BLA
with $b=$82 km s$^{-1}$ which left most of the total $b$-value being due to NT motions
but
with a significant thermal component and a best-fit log T (K)=5.07. Thus, a warm absorber
was judged to be
present by Paper 1.

However, while neither the O~VI nor the Ly$\alpha$ profiles demand it (i.e., the fits
obtained in
Paper 1 are fully acceptable), we have refit both lines with two components. The second
component in
O~VI is $\sim$70 km s$^{-1}$ redward of the stronger component shown in Paper 1, Figure
A8.
Constraining Ly$\alpha$ to match the two O~VI components results in two BLAs with $b$=
45 and 90 km
s$^{-1}$.  These fits are shown in Figure 16 with parameters listed in Table 11. This
alternative fit
confirms that at least one warm absorber is present and that two broad components can
fit these data.
However, even attempting to constrain the lower redshift Ly$\alpha$ absorber does not
yield a velocity
alignment with the O~VI and further requires $b_{O~VI} \approx b_{H~I}$. Thus, this
component is
significantly misaligned and the O~VI and Ly$\alpha$ $b$-values cannot be used to
constrain physical
conditions. The higher redshift component has the Ly$\alpha$ and O~VI lines aligned to
within the
rather wide errors. Adopting $b_{O~VI} \approx 40$ km s$^{-1}$ for the higher redshift
absorber yields
a thermal $b_T \approx$ 83 (+16/-24) km s$^{-1}$ and log T (K)= 5.8 (+0.2/-0.3). This is
hotter than the
warm absorber derived in Paper 1. Although this two-component fit is not demanded by the
data, the
O~VI and Ly$\alpha$ profiles are suggestive of two absorbers, so we adopt the solution
in Table 11 for
the best description for this absorber. A single warm absorber is confirmed but at a
higher temperature than derived by Paper 1.

\begin{figure}
  \epsscale{0.8}
  \plotone{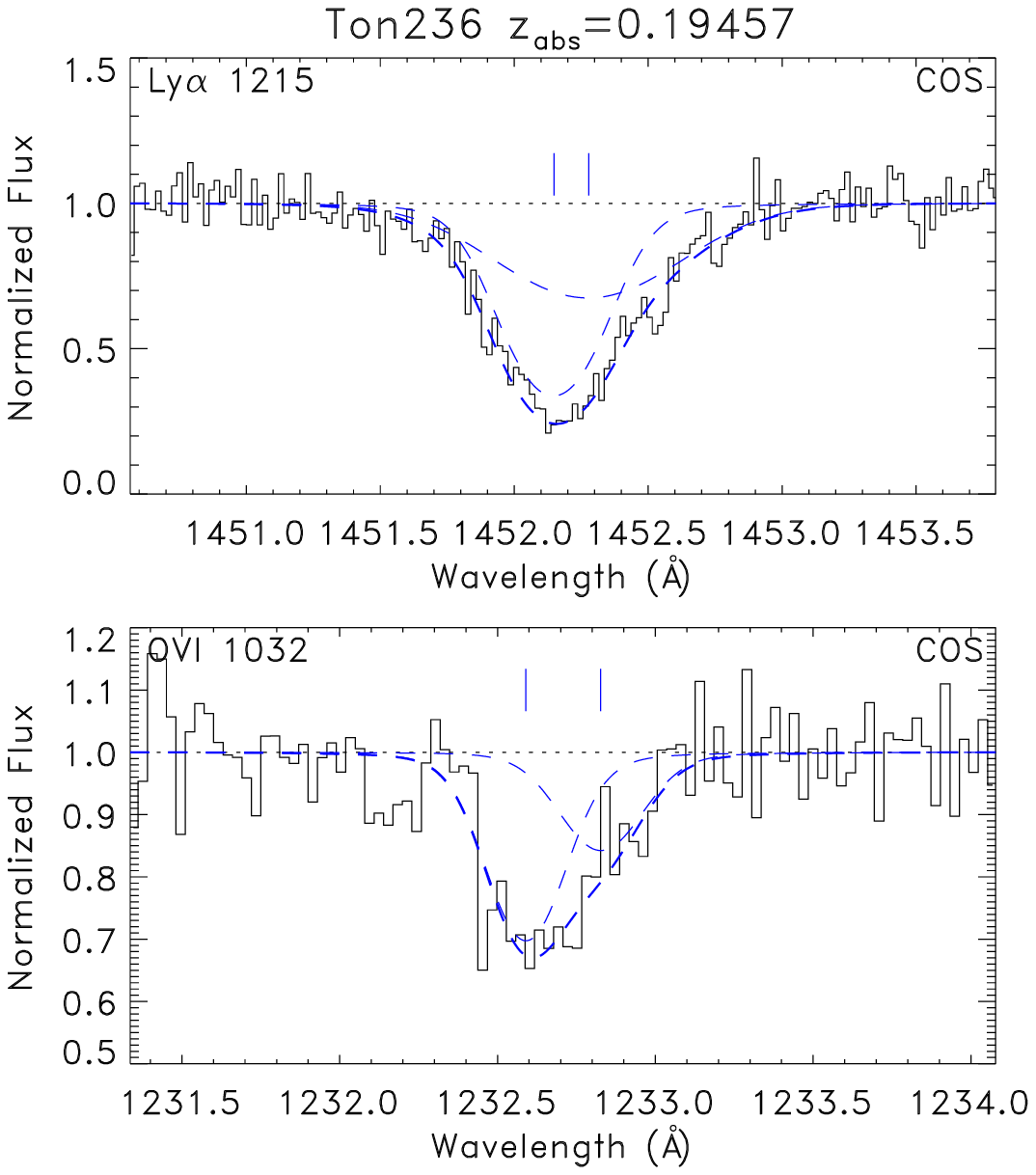}
  \caption{The Ly$\alpha$ and O\,VI 1032\AA\ absorption lines for the  Ton\,236 $z_{\rm
abs}=0.19547$ absorber with our revised models (see
  Table 11) over-plotted in blue.  The redder O\,VI component is
  aligned with the broader Ly$\alpha$ component to within the errors
  in the velocity calibration.}
\end{figure}

\begin{deluxetable}{lcccc}
\tabletypesize{\footnotesize}
\tablecolumns{5}
\tablewidth{0pt}
\tablecaption{Profile Fit Results: Ton\,236 $z_{\rm abs}=0.19457$}
\tablehead{\colhead{Species}    &
           \colhead{$\lambda_0$} &
           \colhead{$v$} &
           \colhead{$b$} &
           \colhead{$\log\,N$}     \\
           \colhead{}          &
           \colhead{(\AA)}     &
           \colhead{($\rm km~s^{-1}$)} &
           \colhead{($\rm km~s^{-1}$)} &
           \colhead{($\rm cm^{-2}$)} }
 \startdata  Ly$\alpha$ & 1216 &$-7\pm  3$&$45\pm 3$&$13.9\pm0.1$\\
             &      &$+25\pm10$&$90\pm 9$&$13.7\pm0.1$\\
  O\,VI      & 1032 &$-34\pm 6$&$37\pm 8$&$13.9\pm0.1$\\
             &      &$+35\pm10$&$40\pm17$&$13.6\pm0.2$\\
 \enddata
\end{deluxetable}

\subsection{HE~0153-4520/0.40052}

Both the O~VI and the H~I are broad and have a complex profile in this absorber. Because
the
Ly$\alpha$ components are comparably broad to the O~VI components aligned with them,
Paper 1 judged
most of the b-value to be NT and derived temperatures consistent with photo-ionization
in a
very low density gas.

A reanalysis of this system finds good two-component fits to O~VI, Ly$\alpha$ and
Ly$\beta$, which are
not inconsistent with the detailed fits in Paper 1. We conclude that, while a BLA is
present, the
$b$-values are consistent with photo-ionized gas in this system. Thus, we adopt the
component fit of
Paper 1 and conclude that no warm absorber is present.

\subsection{PKS~0405-123/0.09657}

Due to potentially different wavelength calibrations for the O~VI line in the {\it FUSE}
band and the Ly$\alpha$ line in the COS band, Paper 1 questioned whether the broad,
symmetric O~VI absorption at this redshift was associated with a narrow or a broad Ly$\alpha$ absorption, which differ
only by 11 km s$^{-1}$. In the end Paper 1 concluded that the broad O~VI was better aligned with the
narrow Ly$\alpha$ requiring log T (K) = 4.30 (+0.21/-01.02), but noted that the warm absorber association
appeared equally likely. 

In this absorber the Ly$\beta$ line is detected clearly in the {\it FUSE} spectrum
adjacent to the O~VI absorber at a velocity 17 km s$^{-1}$ lower than the O~VI absorption based on our analysis.  
The detected Ly$\beta$ must be associated almost exclusively with the narrow Ly$\alpha$ component, 
which is 4--5 times stronger than the BLA. This analysis provides a wavelength offset for these two spectra which 
places the O~VI absorption much closer in velocity to the BLA ($\Delta$ v = 6 km s$^{-1}$)
than the narrow H~I component. We conclude that the O~VI
is associated with a BLA in a warm absorber and retain the values suggested in Paper 1,
including log T(K)=5.48$\pm$0.05. However, because the PIE/CIE determination rests solely on a wavelength
solution which is uncertain at the level to which these two absorption systems differ, we conclude that this 
absorber's temperature can not be determined within a reasonable doubt. On this basis we have excluded this system from the
analysis herein and ascribe it neither as a warm nor a cool absorber.

\subsection{PKS~0405-123/0.29770}

In this case, the O~VI absorption is very broad and symmetrical. In order to fit
Ly$\alpha$ with the
minimum number of components (2), the width of the Ly$\alpha$ is only slightly broader
than its
associated O~VI. Paper 1 judged this O~VI absorber to have a width dominated by NT
motions
and a derived temperature of log T (K) =4.62 consistent with photo-ionization in a low
density gas.

The symmetrical O~VI absorption suggests that the dominant broadening mechanism in this
absorber is
thermal. In this case a third component was added at the velocity of the broad O~VI
absorber to the
two found in Paper 1. The resulting three Ly$\alpha$ components are shown in Figure 17
with their particulars
recorded in Table 12. While the $b$-value of the very broad component is not tightly
constrained, a
significant column density BLA with $b$= 86 (+14/-17) km s$^{-1}$ is suggested which
results in log
T (K) = 5.7(+0.1/-0.2), considerably hotter than the model adopted in Paper 1. We adopt this
new three
component fit which results in a new warm absorber in the sample.


\begin{figure}
  \epsscale{0.8}
  \plotone{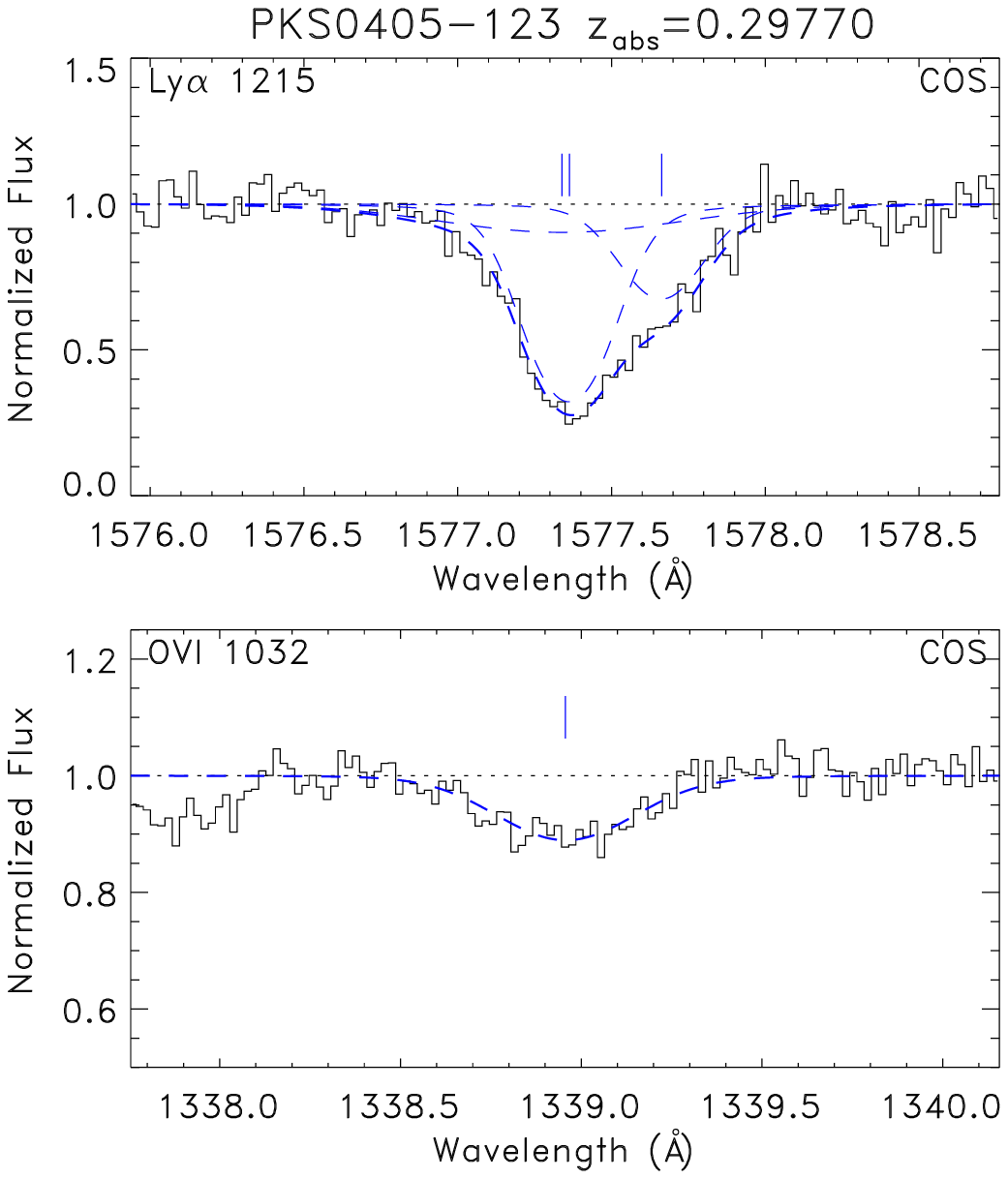}
  \caption{The Ly$\alpha$ and O\,VI 1032 \AA\ absorption lines for the
  PKS\,0405$-$123 $z_{\rm abs}=0.29770$ absorber with our revised
  models (see Table 12) over-plotted in blue.  The broad Ly$\alpha$
  component is aligned with the O\,VI profile.}
\end{figure}

\begin{deluxetable}{lcccc}
\tabletypesize{\footnotesize}
\tablecolumns{5}
\tablewidth{0pt}\tablecaption{Profile Fit Results: PKS\,0405$-$123 $z_{\rm
abs}=0.29770$}
\tablehead{\colhead{Species}    &
           \colhead{$\lambda_0$} &
           \colhead{$v$} &           \colhead{$b$} &
           \colhead{$\log\,N$}     \\
           \colhead{}          &
           \colhead{(\AA)}     &
           \colhead{($\rm km~s^{-1}$)} &
           \colhead{($\rm km~s^{-1}$)} &
           \colhead{($\rm cm^{-2}$)} }
 \startdata
  Ly$\alpha$& 1216 &$-51\pm30$&$100\pm10$&$13.15\pm0.10$\\
            &      &$-45\pm 3$&$ 30\pm 3$&$13.78\pm0.05$\\
            &      &$+29\pm10$&$ 31\pm 8$&$13.3\pm0.2$\\
  O\,VI     & 1032 &$-50\pm 4$&$ 56\pm 4$&$13.55\pm0.10$\\
 \enddata
\end{deluxetable}

\subsection{3C~263/0.11389}

This absorber has an O~VI aligned with a rather narrow ($b$= 20 km s$^{-1}$) H~I line
leading Paper 1
to conclude that this absorber is in photo-ionization equilibrium. In addition, there
are two BLAs
with no associated O~VI absorptions. Therefore, these absorbers are judged to be
misaligned and are
not interpreted in Paper 1.

Our analysis reproduces the detailed three component fits in Paper 1, including the
exact $b$-values
of each H~I component. In addition, we use the {\it FUSE} Ly$\beta$ absorption to make a
two-line
(Ly$\alpha$/Ly$\beta$) COG analysis which finds $b$= 48 km s$^{-1}$ for the lowest
velocity
component, consistent with the $b$=47$\pm$1 derived from the Ly$\alpha$ line alone. This
consistency
suggests, but does not demand, that the lowest velocity H~I component at $\Delta$v =
-219 km s$^{-1}$ is
a metal-deficient BLA and warm absorber. The $\Delta$v = -44 km s$^{-1}$ component is
confirmed by our reanalysis to have
a large $b$-value but the absence of detectable Ly$\beta$ for this component leaves us
unable to draw
any firm conclusion as to its physical details. Additionally, the $\Delta$v = -44 km
s$^{-1}$  portion of
the Ly$\alpha$ absorption complex (see Figure A13 in Paper 1) appears to have structure
in
the absorption making it an unlikely BLA. We conclude that the detailed fit to this
absorber obtained
by Paper 1 is the best description of this absorber but also that an H~I-only warm absorber is
{\bf possibly}
present with log T (K) $\approx$ 5.14.

\clearpage
\section{Basic Galaxy Data for Group Members Associated with Warm, Cool 
and Misaligned Absorbers}

In this Appendix we tabulate the group members for all warm, cool and misaligned
absorbers (in that order) with N$_{grp} \geq$ 8 members (see Tables 5,6 \& 7 in
the main text). The absorber 3C~263/0.11389 with N$_{grp}$ = 7 members has been
added because it may contain a warm absorber (see Appendix A and Figure 7). The galaxy
redshift data comes from a variety of sources including a combined redshift
catalog of nearby galaxy redshifts \citep [Catalog][] {stocke06}; the Sloan
Digital Sky Survey [SDSS]; and our own MOS galaxy redshift surveys in Australia
[AAT] or at WIYN or CTIO [HYDRA] \citep{keeney14}. A complete listing of all
galaxies surveyed in these latter redshift surveys will be presented at a later
time \citep{keeney14}. See Section 3 in the main text
for a description of these various sources which are listed in column (4) of
the following tables.

These tables include by column: (1) galaxy name or ``...'' if unnamed; (2) and (3)
galaxy RA and DEC in J2000.0 equinox coordinates; (4) the source for the redshift
as described above; (5) the heliocentric redshift ($\pm$ 30 km s$^{-1}$); (6) the
galaxy g-band luminosity in L$^*$ units; (7) the impact parameter from the
sightline to the galaxy in kpc; (8)the impact parameter from the 
sightline to the galaxy in units of the galaxy's virial radius \citep [see Figure 1
in][]{stocke13}; and (9) the velocity difference between the galaxy and the
absorber.
    
\LongTables
\input tabB1.tex
\input tabB2.tex
\clearpage
\input tabB3.tex
\input tabB4.tex
\input tabB5.tex
\input tabB6.tex
\input tabB7.tex
\input tabB8.tex
\input tabB9.tex
\clearpage
\input tabB10.tex
\input tabB11.tex
\input tabB12.tex
\input tabB13.tex

\end{appendix}

\end{document}

%% file: tab5.tex
\begin{turnpage}
\begin{deluxetable*}{lccccccccccccc}

\tabletypesize{\scriptsize}

\tablecolumns{14}
\tablewidth{0pt}

\tablecaption{Galaxy Groups Associated with Warm \OVI\ Absorbers
\label{tab:groups_warm}}

\tablehead{ \colhead{Sight Line} & 
            \colhead{$z_{\rm abs}$} & 
            \colhead{$L_{\rm ng}$} & 
            \colhead{$\rho_{\rm ng}$} &
            \colhead{$\rho_{\rm ng}/R_{\rm vir}$} & 
            \colhead{$\Delta v_{\rm ng}$} & 
            \colhead{$L_{\rm grp}$} & 
            \colhead{$N_{\rm grp}$} & 
            \colhead{$\rho_{\rm grp}$} &
            \colhead{$\rho_{\rm grp}/R_{\rm vir}$} & 
            \colhead{$\Delta v_{\rm grp}$} & 
            \colhead{$\sigma_{\rm grp}$} & 
            \colhead{$\sigma_{\rm vir}$} & 
            \colhead{$L_{\rm comp}$} \\
            & 
            & 
            \colhead{($L^*$)} & 
            \colhead{(kpc)} & 
            & 
            \colhead{(\kms)} & 
            \colhead{($L^*$)} & 
            & 
            \colhead{(kpc)} & 
            & 
            \colhead{(\kms)} & 
            \colhead{(\kms)} & 
            \colhead{(\kms)} & 
            \colhead{($L^*$)} }

\startdata
Mrk~876       & 0.00315                  &  0.25 &  188 & 1.65 &    \phn$-34$ &    0.30 &     ~~4 &     279 &    0.93 & \phn\phn$-4$ &  ~90                &     104 &               0.019 \\
3C~273        & 0.00336\tablenotemark{a} & 0.021 &  ~77 & 1.20 &       $-108$ &      32 &     161 &     525 &    0.37 &      \phs537 & $500^{+ 40}_{- 50}$ &     491 &            $<0.001$ \\
Mrk~290       & 0.01027                  &   1.7 &  434 & 2.00 &    \phn$-69$ &     6.0 &     ~24 &     814 &    1.00 &      \phs302 & $270^{+ 80}_{-110}$ &     282 &               0.011 \\
PHL~1811      & 0.07773\tablenotemark{b} &  0.94 &  310 & 1.75 &    \phn$-22$ &      55 &     126 &     410 &    0.24 &      \phs333 & $630^{+110}_{-120}$ &     591 &               0.060 \\
H~1821+643    & 0.12141                  &   2.3 &  158 & 0.65 &   \phs\phn34 & \nodata & \nodata & \nodata & \nodata & \nodata      & \nodata             & \nodata &   0.19\,(659) / 1.2 \\
PHL~1811      & 0.13280                  &   3.0 &  227 & 0.87 &    \phn$-48$ &      26 &     ~30 &    1114 &    0.85 &      \phs568 & $440^{+190}_{-200}$ &     458 &                0.19 \\
PG~1116+215   & 0.13850                  &   1.5 &  138 & 0.67 &    \phn$-55$ &     3.8 &     ~~4 &     568 &    0.81 & \phn\phn$-9$ &  ~50                &     242 &                0.33 \\
3C~263        & 0.14072                  &   1.7 &  620 & 2.88 &   \phs\phn39 & \nodata & \nodata & \nodata & \nodata & \nodata      & \nodata             & \nodata &                0.34 
\enddata

\tablenotetext{a}{The closest group galaxy to the QSO sight line is a $0.005\,L^*$ galaxy at $\rho = 46$~kpc ($0.89\,R_{\rm vir}$), but it has a large velocity separation of $\Delta v = 610$~\kms\ with respect to $z_{\rm abs}$ \citep{stocke04}.}
\tablenotetext{b}{The closest group galaxy to the QSO sight line is a $0.70\,L^*$  galaxy at $\rho = 35$~kpc ($0.22\,R_{\rm vir}$), but it has a large velocity separation of $\Delta v = 904$~\kms\ with respect to $z_{\rm abs}$ \citep{jenkins03}.}

\end{deluxetable*}
\end{turnpage}

%% file: tab6.tex
\begin{turnpage}
\begin{deluxetable*}{lccccccccccccc}

\tabletypesize{\scriptsize}

\tablecolumns{14}
\tablewidth{0pt}

\tablecaption{Galaxy Groups Associated with Cool \OVI\ Absorbers
\label{tab:groups_cool}}

\tablehead{ \colhead{Sight Line} & 
            \colhead{$z_{\rm abs}$} & 
            \colhead{$L_{\rm ng}$} & 
            \colhead{$\rho_{\rm ng}$} &
            \colhead{$\rho_{\rm ng}/R_{\rm vir}$} & 
            \colhead{$\Delta v_{\rm ng}$} & 
            \colhead{$L_{\rm grp}$} & 
            \colhead{$N_{\rm grp}$} & 
            \colhead{$\rho_{\rm grp}$} &
            \colhead{$\rho_{\rm grp}/R_{\rm vir}$} & 
            \colhead{$\Delta v_{\rm grp}$} & 
            \colhead{$\sigma_{\rm grp}$} & 
            \colhead{$\sigma_{\rm vir}$} & 
            \colhead{$L_{\rm comp}$} \\
            & 
            & 
            \colhead{($L^*$)} & 
            \colhead{(kpc)} & 
            & 
            \colhead{(\kms)} & 
            \colhead{($L^*$)} & 
            & 
            \colhead{(kpc)} & 
            & 
            \colhead{(\kms)} & 
            \colhead{(\kms)} & 
            \colhead{(\kms)} & 
            \colhead{($L^*$)} }

\startdata
PG~0953+414   & 0.00212 & 0.025 & ~227 & 3.44 &    \phn$-31$ &     1.5 &      13 &    ~722 &    1.40 &    \phn$-14$ & $130\pm80$          &     179 &                          $<0.001$ \\
HE~0226--4110 & 0.01750 &   2.8 & ~596 & 2.32 &   \phs\phn98 &      25 &      21 &    ~911 &    0.69 &       $-101$ & $480^{+120}_{-160}$ &     456 &               0.003\,(643) / 0.15 \\
H~1821+643    & 0.02443 &   1.7 & 1255 & 5.75 &   \phs\phn40 &     3.7 &      ~4 &    1260 &    1.82 &   \phs\phn72 &  240                &     241 & 0.008\,(148) / 0.048\,(891) / 1.2 \\
PG~1259+593   & 0.04625 &  0.50 & ~138 & 0.96 & \phn\phn$-3$ &     9.6 &      23 &    ~352 &    0.37 &   \phs\phn34 & $350^{+100}_{-170}$ &     329 &              0.032\,(1095) / 0.24 \\
PG~1116+215   & 0.05897 &  0.13 & ~131 & 1.40 &      \phs297 &     9.3 &      16 &    ~334 &    0.36 &      \phs340 & $230^{+ 50}_{- 80}$ &     326 &              0.053\,(1376) / 0.49 \\
PG~0953+414   & 0.06808 &  0.88 & ~607 & 3.50 &      \phs306 &     8.4 &      14 &    ~913 &    1.01 &      \phs318 & $290^{+ 70}_{-120}$ &     316 &              0.072\,(1572) / 0.63 \\
3C~273        & 0.12007 &   6.2 & 2119 & 6.36 &    \phn$-99$ &      11 &      ~4 &    2009 &    2.10 &   \phs\phn19 &  ~80                &     346 &                0.42\,(1306) / 2.0 \\
PG~0953+414   & 0.14231 &   4.7 & ~438 & 1.43 &      \phs128 &     9.5 &      ~6 &    ~331 &    0.35 &      \phs123 &  270                &     329 &                              0.35 \\
HE~0153--4520 & 0.14887 &   2.5 & 1081 & 4.37 &      \phs227 &     3.5 &      ~3 &    1084 &    1.60 &      \phs210 &  145                &     235 &                              0.24 
\enddata

\end{deluxetable*}
\end{turnpage}

%% file: tab7.tex
\begin{turnpage}
\begin{deluxetable*}{lccccccccccccc}

\tabletypesize{\scriptsize}

\tablecolumns{14}
\tablewidth{0pt}

\tablecaption{Galaxy Groups Associated with Misaligned \OVI\ Absorbers
\label{tab:groups_misaligned}}

\tablehead{ \colhead{Sight Line} & 
            \colhead{$z_{\rm abs}$} & 
            \colhead{$L_{\rm ng}$} & 
            \colhead{$\rho_{\rm ng}$} &
            \colhead{$\rho_{\rm ng}/R_{\rm vir}$} & 
            \colhead{$\Delta v_{\rm ng}$} & 
            \colhead{$L_{\rm grp}$} & 
            \colhead{$N_{\rm grp}$} & 
            \colhead{$\rho_{\rm grp}$} &
            \colhead{$\rho_{\rm grp}/R_{\rm vir}$} & 
            \colhead{$\Delta v_{\rm grp}$} & 
            \colhead{$\sigma_{\rm grp}$} & 
            \colhead{$\sigma_{\rm vir}$} & 
            \colhead{$L_{\rm comp}$} \\
            & 
            & 
            \colhead{($L^*$)} & 
            \colhead{(kpc)} & 
            & 
            \colhead{(\kms)} & 
            \colhead{($L^*$)} & 
            & 
            \colhead{(kpc)} & 
            & 
            \colhead{(\kms)} & 
            \colhead{(\kms)} & 
            \colhead{(\kms)} & 
            \colhead{($L^*$)} }

\startdata
Mrk~876       & 0.01169                  &  0.10 &  266 & 3.03 &   \phs\phn11 & \nodata & \nodata & \nodata & \nodata & \nodata      & \nodata              & \nodata &                 0.26 \\
PKS~2155--304 & 0.05422\tablenotemark{a} &   3.7 &  544 & 1.95 &    \phn$-91$ &      17 &      ~8 &     380 &    0.33 &      \phs648 & $340^{+ 70}_{-210}$  &     403 &  0.087\,(636) / 0.13 \\
PKS~2155--304 & 0.05722                  &   3.0 &  429 & 1.65 &       $-176$ &      17 &      ~8 &     380 &    0.33 &       $-205$ & $340^{+ 70}_{-210}$  &     403 &  0.097\,(669) / 0.14 \\
3C~263        & 0.06342                  &  0.21 &  ~63 & 0.58 &    \phn$-56$ & \nodata & \nodata & \nodata & \nodata & \nodata      & \nodata              & \nodata & 0.062\,(1472) / 0.30 \\
3C~273        & 0.09010                  &   1.9 &  478 & 2.14 &       $-110$ &      24 &      13 &    1833 &    1.43 &   \phn\phs89 & $560^{+ 60}_{-190}$  &     445 &   0.24\,(1014) / 1.1 \\
PKS~0405--123 & 0.09192                  & 0.020 &  ~72 & 1.14 &      \phs104 &    0.41 &      ~5 &     115 &    0.35 &       $-155$ &  210                 &     115 &  0.004\,(500) / 0.14 \\
3C~263        & 0.11389                  &   1.2 &  982 & 5.08 &      \phs318 &     4.8 &      ~7 &     958 &    1.27 &      \phs245 &  270                 &     262 &                 0.21 
\enddata

\tablenotetext{a}{This absorber and the absorber at $z=0.05722$ in the PKS~2155--304 sight line are associated with the same group of galaxies. The closest galaxy to the higher redshift absorber is the galaxy with the smallest impact parameter from this absorber, but it has a large velocity difference of $\Delta v = 677$~\kms.}

\end{deluxetable*}
\end{turnpage}

%% file: tabB1.tex
\begin{deluxetable*}{lllccccc}

\tabletypesize{\scriptsize}

\tablecolumns{8}
\tablewidth{0pt}

\tablecaption{Galaxy Group Members Near 3C~273 ($z_{\rm abs} = 0.00336$)}
\label{tab:3c273_z00336}

\tablehead{ \colhead{Name} & 
            \colhead{Position} & 
            \colhead{Source} & 
            \colhead{$z_{\rm gal}$} & 
            \colhead{$L_{\rm gal}$} & 
            \colhead{$\rho$} & 
            \colhead{$\rho/R_{\rm vir}$} & 
            \colhead{$\Delta v$} \\ 
            & 
            \colhead{(J2000)} & 
            & 
            & 
            \colhead{($L^*$)} & 
            \colhead{(kpc)} & 
            & 
            \colhead{(\kms)} } 

\startdata
             McLin77 &  12:29:50.50, +02:02:08.8 & Catalog & 0.00519 &  0.009 &   46 &  0.81 & $ 547$ \\
          TGN388Z111 &  12:29:50.57, +02:01:54.2 & Catalog & 0.00540 &  0.012 &   46 &  0.79 & $ 610$ \\
          TGN387Z059 &  12:28:15.95, +01:49:43.7 & Catalog & 0.00300 &  0.021 &   77 &  1.20 & $-108$ \\
          A1225+0152 &  12:27:46.50, +01:36:03.0 & Catalog & 0.00433 &  0.021 &  141 &  2.19 & $ 290$ \\
          12285+0157 &  12:31:03.80, +01:40:34.0 & Catalog & 0.00369 &  0.043 &  155 &  2.12 & $  97$ \\
          12277+0254 &  12:30:14.20, +02:37:32.0 & Catalog & 0.00545 &  0.057 &  160 &  2.08 & $ 626$ \\
          12265+0300 &  12:29:02.80, +02:43:19.0 & Catalog & 0.00525 &  0.061 &  168 &  2.14 & $ 565$ \\
          A1225+0258 &  12:28:15.90, +02:42:02.0 & Catalog & 0.00777 &  0.021 &  171 &  2.66 & $1317$ \\
\enddata

\end{deluxetable*}

%% file: tabB2.tex
\begin{deluxetable*}{lllccccc}

\tabletypesize{\scriptsize}

\tablecolumns{8}
\tablewidth{0pt}

\tablecaption{Galaxy Group Members Near Mrk~290 ($z_{\rm abs} = 0.01027$)}
\label{tab:mrk290_z01027}

\tablehead{ \colhead{Name} & 
            \colhead{Position} & 
            \colhead{Source} & 
            \colhead{$z_{\rm gal}$} & 
            \colhead{$L_{\rm gal}$} & 
            \colhead{$\rho$} & 
            \colhead{$\rho/R_{\rm vir}$} & 
            \colhead{$\Delta v$} \\ 
            & 
            \colhead{(J2000)} & 
            & 
            & 
            \colhead{($L^*$)} & 
            \colhead{(kpc)} & 
            & 
            \colhead{(\kms)} } 

\startdata
             \nodata &  15:35:14.26, +57:30:53.2 &    SDSS & 0.01033 &  0.073 &  302 &  3.72 & $  17$ \\
             \nodata &  15:38:02.75, +57:30:18.3 &    SDSS & 0.01180 &  0.021 &  375 &  5.89 & $ 454$ \\
            NGC 5987 &  15:39:57.38, +58:04:46.3 & Catalog & 0.01004 &  1.700 &  434 &  2.00 & $ -69$ \\
           NGC 5965A &  15:34:13.80, +57:17:06.0 & Catalog & 0.01104 &  0.169 &  499 &  4.92 & $ 229$ \\
             \nodata &  15:30:40.83, +57:53:01.1 &    SDSS & 0.00969 &  0.001 &  526 & 12.30 & $-173$ \\
          A1540+5741 &  15:41:09.60, +57:31:51.0 & Catalog & 0.01210 &  0.023 &  608 &  9.35 & $ 543$ \\
             \nodata &  15:37:42.04, +57:05:06.5 &    SDSS & 0.01158 &  0.027 &  650 &  9.73 & $ 390$ \\
             \nodata &  15:33:57.97, +56:50:50.4 &    SDSS & 0.01081 &  0.059 &  827 & 10.63 & $ 160$ \\
\enddata

\end{deluxetable*}

%% file: tabB3.tex
\begin{deluxetable*}{lllccccc}

\tabletypesize{\scriptsize}

\tablecolumns{8}
\tablewidth{0pt}

\tablecaption{Galaxy Group Members Near PHL~1811 ($z_{\rm abs} = 0.07773$)}
\label{tab:phl1811_z07773}

\tablehead{ \colhead{Name} & 
            \colhead{Position} & 
            \colhead{Source} & 
            \colhead{$z_{\rm gal}$} & 
            \colhead{$L_{\rm gal}$} & 
            \colhead{$\rho$} & 
            \colhead{$\rho/R_{\rm vir}$} & 
            \colhead{$\Delta v$} \\ 
            & 
            \colhead{(J2000)} & 
            & 
            & 
            \colhead{($L^*$)} & 
            \colhead{(kpc)} & 
            & 
            \colhead{(\kms)} } 

\startdata
 PROCHASKA\_22\_1734 & 21:54:59.90, --09:22:24.0 & Catalog & 0.08098 &  0.723 &   35 &  0.21 & $  904$ \\
 PROCHASKA\_22\_2091 & 21:54:50.80, --09:22:33.0 & Catalog & 0.07880 &  0.324 &  235 &  1.88 & $  298$ \\
 PROCHASKA\_22\_2033 & 21:54:52.20, --09:24:37.0 & Catalog & 0.08080 &  0.089 &  283 &  3.31 & $  854$ \\
 PROCHASKA\_22\_2226 & 21:54:47.50, --09:22:54.0 & Catalog & 0.07765 &  0.944 &  310 &  1.75 & $  -22$ \\
 PROCHASKA\_22\_2103 & 21:54:50.90, --09:20:03.0 & Catalog & 0.07942 &  0.318 &  312 &  2.52 & $  470$ \\
 PROCHASKA\_22\_2225 & 21:54:47.30, --09:23:05.0 & Catalog & 0.07791 &  0.103 &  317 &  3.59 & $   50$ \\
 PROCHASKA\_22\_1202 & 21:55:14.10, --09:24:34.0 & Catalog & 0.08157 &  0.179 &  336 &  3.27 & $ 1068$ \\
 PROCHASKA\_22\_1109 & 21:55:16.40, --09:24:07.0 & Catalog & 0.07376 &  0.838 &  360 &  2.11 & $-1104$ \\
\enddata

\end{deluxetable*}

%% file: tabB4.tex
\begin{deluxetable*}{lllccccc}

\tabletypesize{\scriptsize}

\tablecolumns{8}
\tablewidth{0pt}

\tablecaption{Galaxy Group Members Near PHL~1811 ($z_{\rm abs} = 0.13280$)}
\label{tab:phl1811_z13280}

\tablehead{ \colhead{Name} & 
            \colhead{Position} & 
            \colhead{Source} & 
            \colhead{$z_{\rm gal}$} & 
            \colhead{$L_{\rm gal}$} & 
            \colhead{$\rho$} & 
            \colhead{$\rho/R_{\rm vir}$} & 
            \colhead{$\Delta v$} \\ 
            & 
            \colhead{(J2000)} & 
            & 
            & 
            \colhead{($L^*$)} & 
            \colhead{(kpc)} & 
            & 
            \colhead{(\kms)} } 

\startdata
             \nodata & 21:55:06.29, --09:23:20.1 &     AAT & 0.13743 &  0.230 &  214 &  1.92 & $1225$ \\
 PROCHASKA\_22\_1474 & 21:55:06.50, --09:23:25.0 & Catalog & 0.13262 &  2.950 &  227 &  0.87 & $ -48$ \\
 PROCHASKA\_22\_1802 & 21:54:58.40, --09:18:56.0 & Catalog & 0.13541 &  0.616 &  507 &  3.28 & $ 691$ \\
             \nodata & 21:55:09.52, --09:19:24.4 &     AAT & 0.13207 &  0.230 &  512 &  4.61 & $-193$ \\
 PROCHASKA\_22\_2425 & 21:54:42.30, --09:21:54.0 & Catalog & 0.13248 &  0.557 &  679 &  4.53 & $ -85$ \\
 PROCHASKA\_22\_2426 & 21:54:42.50, --09:21:12.0 & Catalog & 0.13220 &  0.583 &  690 &  4.53 & $-159$ \\
 PROCHASKA\_22\_1848 & 21:54:56.40, --09:28:28.0 & Catalog & 0.13463 &  0.450 &  882 &  6.35 & $ 484$ \\
             \nodata & 21:55:29.92, --09:22:35.3 &     AAT & 0.13980 &  0.276 &  999 &  8.44 & $1853$ \\
\enddata

\end{deluxetable*}

%% file: tabB5.tex
\begin{deluxetable*}{lllccccc}

\tabletypesize{\scriptsize}

\tablecolumns{8}
\tablewidth{0pt}

\tablecaption{Galaxy Group Members Near PG~0953+414 ($z_{\rm abs} = 0.00212$)}
\label{tab:pg0953_z00212}

\tablehead{ \colhead{Name} & 
            \colhead{Position} & 
            \colhead{Source} & 
            \colhead{$z_{\rm gal}$} & 
            \colhead{$L_{\rm gal}$} & 
            \colhead{$\rho$} & 
            \colhead{$\rho/R_{\rm vir}$} & 
            \colhead{$\Delta v$} \\ 
            & 
            \colhead{(J2000)} & 
            & 
            & 
            \colhead{($L^*$)} & 
            \colhead{(kpc)} & 
            & 
            \colhead{(\kms)} } 

\startdata
            NGC 3104 &  10:03:57.80, +40:45:28.0 & Catalog & 0.00201 &  0.025 &  227 &  3.44 & $ -31$ \\
          10043+3913 &  10:07:22.90, +38:58:08.0 & Catalog & 0.00196 &  0.008 &  483 &  8.85 & $ -48$ \\
          10133+4125 &  10:16:15.60, +41:09:57.0 & Catalog & 0.00177 &  0.009 &  579 & 10.39 & $-104$ \\
            NGC 3184 &  10:18:17.20, +41:25:26.0 & Catalog & 0.00197 &  0.817 &  639 &  3.77 & $ -43$ \\
             \nodata &  10:17:32.98, +43:08:41.4 &    SDSS & 0.00378 &  0.002 &  678 & 14.97 & $ 496$ \\
             \nodata &  10:25:50.79, +43:55:58.2 &    SDSS & 0.00219 &  0.001 &  947 & 22.77 & $  20$ \\
            NGC 3198 &  10:19:54.80, +45:32:58.0 & Catalog & 0.00221 &  0.432 &  951 &  6.91 & $  26$ \\
             \nodata &  10:07:18.96, +47:00:23.4 &    SDSS & 0.00211 &  0.018 &  960 & 15.39 & $  -4$ \\
\enddata

\end{deluxetable*}

%% file: tabB6.tex
\begin{deluxetable*}{lllccccc}

\tabletypesize{\scriptsize}

\tablecolumns{8}
\tablewidth{0pt}

\tablecaption{Galaxy Group Members Near HE~0226--4110 ($z_{\rm abs} = 0.01750$)}
\label{tab:he0226_z01750}

\tablehead{ \colhead{Name} & 
            \colhead{Position} & 
            \colhead{Source} & 
            \colhead{$z_{\rm gal}$} & 
            \colhead{$L_{\rm gal}$} & 
            \colhead{$\rho$} & 
            \colhead{$\rho/R_{\rm vir}$} & 
            \colhead{$\Delta v$} \\ 
            & 
            \colhead{(J2000)} & 
            & 
            & 
            \colhead{($L^*$)} & 
            \colhead{(kpc)} & 
            & 
            \colhead{(\kms)} } 

\startdata
             \nodata & 02:27:46.33, --40:51:17.1 &   HYDRA & 0.01490 &  0.005 &  173 &  3.39 & $-766$ \\
             \nodata & 02:27:44.79, --41:05:23.6 &   HYDRA & 0.01524 &  0.012 &  214 &  3.63 & $-666$ \\
             \nodata & 02:29:20.48, --41:02:30.2 &   HYDRA & 0.01947 &  0.005 &  287 &  5.62 & $ 580$ \\
             \nodata & 02:27:41.67, --41:09:48.3 &   HYDRA & 0.01511 &  0.005 &  302 &  5.92 & $-704$ \\
             \nodata & 02:26:38.02, --40:39:17.6 &     AAT & 0.01792 &  0.004 &  551 & 11.12 & $ 124$ \\
               N 954 & 02:28:52.20, --41:24:09.0 & Catalog & 0.01783 &  2.793 &  596 &  2.32 & $  98$ \\
          A0223-4039 & 02:25:14.62, --40:25:47.9 & Catalog & 0.01790 &  2.281 &  996 &  4.16 & $ 118$ \\
          A0222-4136 & 02:24:36.00, --41:23:06.0 & Catalog & 0.01759 &  0.499 & 1044 &  7.24 & $  25$ \\
\enddata

\end{deluxetable*}

%% file: tabB7.tex
\begin{deluxetable*}{lllccccc}

\tabletypesize{\scriptsize}

\tablecolumns{8}
\tablewidth{0pt}

\tablecaption{Galaxy Group Members Near PG~1259+593 ($z_{\rm abs} = 0.04625$)}
\label{tab:pg1259_z04625}

\tablehead{ \colhead{Name} & 
            \colhead{Position} & 
            \colhead{Source} & 
            \colhead{$z_{\rm gal}$} & 
            \colhead{$L_{\rm gal}$} & 
            \colhead{$\rho$} & 
            \colhead{$\rho/R_{\rm vir}$} & 
            \colhead{$\Delta v$} \\ 
            & 
            \colhead{(J2000)} & 
            & 
            & 
            \colhead{($L^*$)} & 
            \colhead{(kpc)} & 
            & 
            \colhead{(\kms)} } 

\startdata
             \nodata &  13:01:14.75, +59:03:43.3 &   HYDRA & 0.04608 &  0.068 &   89 &  1.11 & $ -49$ \\
             \nodata &  13:01:01.04, +59:00:06.8 &    SDSS & 0.04624 &  0.504 &  138 &  0.96 & $  -3$ \\
             \nodata &  13:01:00.55, +58:58:04.4 &    SDSS & 0.04621 &  0.752 &  238 &  1.44 & $ -13$ \\
             \nodata &  13:00:33.95, +58:58:56.9 &    SDSS & 0.04601 &  2.378 &  325 &  1.34 & $ -69$ \\
             \nodata &  13:00:25.54, +59:02:18.6 &    SDSS & 0.04658 &  0.188 &  334 &  3.20 & $  94$ \\
             \nodata &  13:00:50.54, +58:56:28.8 &   HYDRA & 0.04627 &  0.033 &  347 &  5.00 & $   6$ \\
             \nodata &  13:01:40.19, +59:07:29.4 &   HYDRA & 0.04700 &  0.067 &  351 &  4.38 & $ 215$ \\
             \nodata &  13:00:22.13, +59:01:26.8 &    SDSS & 0.04645 &  0.630 &  360 &  2.31 & $  56$ \\
\enddata

\end{deluxetable*}

%% file: tabB8.tex
\begin{deluxetable*}{lllccccc}

\tabletypesize{\scriptsize}

\tablecolumns{8}
\tablewidth{0pt}

\tablecaption{Galaxy Group Members Near PG~1116+215 ($z_{\rm abs} = 0.05897$)}
\label{tab:pg1116_z05897}

\tablehead{ \colhead{Name} & 
            \colhead{Position} & 
            \colhead{Source} & 
            \colhead{$z_{\rm gal}$} & 
            \colhead{$L_{\rm gal}$} & 
            \colhead{$\rho$} & 
            \colhead{$\rho/R_{\rm vir}$} & 
            \colhead{$\Delta v$} \\ 
            & 
            \colhead{(J2000)} & 
            & 
            & 
            \colhead{($L^*$)} & 
            \colhead{(kpc)} & 
            & 
            \colhead{(\kms)} } 

\startdata
 PROCHASKA\_12\_2071 &  11:19:05.50, +21:17:33.0 & Catalog & 0.06002 &  0.129 &  131 &  1.40 & $ 297$ \\
 PROCHASKA\_12\_2154 &  11:19:03.10, +21:16:23.0 & Catalog & 0.05939 &  0.047 &  220 &  2.97 & $ 119$ \\
 PROCHASKA\_12\_2048 &  11:19:05.40, +21:15:38.0 & Catalog & 0.05873 &  0.824 &  258 &  1.51 & $ -68$ \\
 PROCHASKA\_12\_1571 &  11:19:19.80, +21:22:28.0 & Catalog & 0.06128 &  0.189 &  281 &  2.68 & $ 654$ \\
  PROCHASKA\_12\_808 &  11:19:38.40, +21:21:42.0 & Catalog & 0.06063 &  0.246 &  504 &  4.43 & $ 470$ \\
 PROCHASKA\_12\_1312 &  11:19:24.30, +21:10:30.0 & Catalog & 0.05899 &  1.458 &  655 &  3.19 & $   6$ \\
          A1115+2129 &  11:18:33.70, +21:13:00.0 & Catalog & 0.05963 &  1.431 &  708 &  3.44 & $ 186$ \\
  PROCHASKA\_12\_674 &  11:19:42.00, +21:26:10.0 & Catalog & 0.06138 &  0.125 &  713 &  7.64 & $ 682$ \\
\enddata

\end{deluxetable*}

%% file: tabB9.tex
\begin{deluxetable*}{lllccccc}

\tabletypesize{\scriptsize}

\tablecolumns{8}
\tablewidth{0pt}

\tablecaption{Galaxy Group Members Near PG~0953+414 ($z_{\rm abs} = 0.06808$)}
\label{tab:pg0953_z06808}

\tablehead{ \colhead{Name} & 
            \colhead{Position} & 
            \colhead{Source} & 
            \colhead{$z_{\rm gal}$} & 
            \colhead{$L_{\rm gal}$} & 
            \colhead{$\rho$} & 
            \colhead{$\rho/R_{\rm vir}$} & 
            \colhead{$\Delta v$} \\ 
            & 
            \colhead{(J2000)} & 
            & 
            & 
            \colhead{($L^*$)} & 
            \colhead{(kpc)} & 
            & 
            \colhead{(\kms)} } 

\startdata
             \nodata &  09:56:12.27, +41:14:39.9 &   HYDRA & 0.07018 &  0.137 &  595 &  6.23 & $ 589$ \\
             \nodata &  09:56:11.37, +41:14:50.4 &    SDSS & 0.06917 &  0.877 &  607 &  3.50 & $ 306$ \\
             \nodata &  09:56:03.77, +41:20:03.3 &   HYDRA & 0.06970 &  0.084 &  807 &  9.59 & $ 455$ \\
             \nodata &  09:56:47.27, +41:03:27.8 &   HYDRA & 0.07128 &  0.151 &  939 &  9.56 & $ 898$ \\
             \nodata &  09:56:36.33, +41:03:04.0 &    SDSS & 0.06875 &  0.615 &  996 &  6.45 & $ 189$ \\
             \nodata &  09:56:19.93, +41:03:13.7 &    SDSS & 0.06869 &  1.009 & 1068 &  5.88 & $ 170$ \\
             \nodata &  09:56:27.65, +41:01:53.6 &   HYDRA & 0.06899 &  0.241 & 1121 &  9.90 & $ 255$ \\
             \nodata &  09:57:55.36, +41:06:25.9 &   HYDRA & 0.06854 &  0.321 & 1166 &  9.35 & $ 129$ \\
\enddata

\end{deluxetable*}

%% file: tabB10.tex
\begin{deluxetable*}{lllccccc}

\tabletypesize{\scriptsize}

\tablecolumns{8}
\tablewidth{0pt}

\tablecaption{Galaxy Group Members Near PKS~2155--304 ($z_{\rm abs} = 0.05422$)}
\label{tab:pks2155_z05422}

\tablehead{ \colhead{Name} & 
            \colhead{Position} & 
            \colhead{Source} & 
            \colhead{$z_{\rm gal}$} & 
            \colhead{$L_{\rm gal}$} & 
            \colhead{$\rho$} & 
            \colhead{$\rho/R_{\rm vir}$} & 
            \colhead{$\Delta v$} \\ 
            & 
            \colhead{(J2000)} & 
            & 
            & 
            \colhead{($L^*$)} & 
            \colhead{(kpc)} & 
            & 
            \colhead{(\kms)} } 

\startdata
         A2155-3033B & 21:58:41.60, --30:19:19.0 & Catalog & 0.05702 &  2.045 &  395 &  1.71 & $ 795$ \\
          TGS329Z002 & 21:58:40.87, --30:19:28.8 & Catalog & 0.05660 &  2.959 &  408 &  1.57 & $ 677$ \\
          TGS330Z126 & 21:58:23.84, --30:19:32.4 & Catalog & 0.05390 &  3.671 &  544 &  1.95 & $ -91$ \\
          TGS330Z102 & 21:58:57.17, --30:02:45.9 & Catalog & 0.05710 &  2.495 &  689 &  2.79 & $ 819$ \\
          TGS331Z156 & 21:59:49.75, --30:16:19.2 & Catalog & 0.05630 &  3.264 &  812 &  3.02 & $ 592$ \\
          TGS329Z018 & 21:57:52.80, --30:19:32.6 & Catalog & 0.05680 &  0.657 &  899 &  5.69 & $ 734$ \\
          TGS330Z061 & 22:00:02.15, --30:02:03.4 & Catalog & 0.05460 &  0.511 & 1209 &  8.31 & $ 108$ \\
          TGS330Z063 & 22:00:01.89, --30:38:05.1 & Catalog & 0.05580 &  1.878 & 1832 &  8.20 & $ 449$
\enddata

\end{deluxetable*}

%% file: tabB11.tex
\begin{deluxetable*}{lllccccc}

\tabletypesize{\scriptsize}

\tablecolumns{8}
\tablewidth{0pt}

\tablecaption{Galaxy Group Members Near PKS~2155--304 ($z_{\rm abs} = 0.05722$)}
\label{tab:pks2155_z05722}

\tablehead{ \colhead{Name} & 
            \colhead{Position} & 
            \colhead{Source} & 
            \colhead{$z_{\rm gal}$} & 
            \colhead{$L_{\rm gal}$} & 
            \colhead{$\rho$} & 
            \colhead{$\rho/R_{\rm vir}$} & 
            \colhead{$\Delta v$} \\ 
            & 
            \colhead{(J2000)} & 
            & 
            & 
            \colhead{($L^*$)} & 
            \colhead{(kpc)} & 
            & 
            \colhead{(\kms)} } 

\startdata
         A2155-3033B & 21:58:41.60, --30:19:19.0 & Catalog & 0.05702 &  2.045 &  415 &  1.80 & $ -58$ \\
          TGS329Z002 & 21:58:40.87, --30:19:28.8 & Catalog & 0.05660 &  2.959 &  429 &  1.65 & $-176$ \\
          TGS330Z126 & 21:58:23.84, --30:19:32.4 & Catalog & 0.05390 &  3.672 &  572 &  2.05 & $-941$ \\
          TGS330Z102 & 21:58:57.17, --30:02:45.9 & Catalog & 0.05710 &  2.494 &  724 &  2.93 & $ -34$ \\
          TGS331Z156 & 21:59:49.75, --30:16:19.2 & Catalog & 0.05630 &  3.263 &  854 &  3.18 & $-261$ \\
          TGS329Z018 & 21:57:52.80, --30:19:32.6 & Catalog & 0.05680 &  0.657 &  946 &  5.98 & $-119$ \\
          TGS330Z061 & 22:00:02.15, --30:02:03.4 & Catalog & 0.05460 &  0.511 & 1272 &  8.74 & $-743$ \\
          TGS330Z063 & 22:00:01.89, --30:38:05.1 & Catalog & 0.05580 &  1.879 & 1926 &  8.62 & $-403$
\enddata

\end{deluxetable*}

%% file: tabB12.tex
\begin{deluxetable*}{lllccccc}

\tabletypesize{\scriptsize}

\tablecolumns{8}
\tablewidth{0pt}

\tablecaption{Galaxy Group Members Near 3C~273 ($z_{\rm abs} = 0.09010$)}
\label{tab:3c273_z09010}

\tablehead{ \colhead{Name} & 
            \colhead{Position} & 
            \colhead{Source} & 
            \colhead{$z_{\rm gal}$} & 
            \colhead{$L_{\rm gal}$} & 
            \colhead{$\rho$} & 
            \colhead{$\rho/R_{\rm vir}$} & 
            \colhead{$\Delta v$} \\ 
            & 
            \colhead{(J2000)} & 
            & 
            & 
            \colhead{($L^*$)} & 
            \colhead{(kpc)} & 
            & 
            \colhead{(\kms)} } 

\startdata
          TGN387Z035 &  12:28:51.87, +02:06:03.2 & Catalog & 0.08970 &  1.867 &  478 &  2.14 & $-110$ \\
             McLin10 &  12:28:30.32, +02:10:43.8 & Catalog & 0.08760 &  0.112 & 1201 & 13.29 & $-687$ \\
          A1225+0227 &  12:28:28.50, +02:11:08.0 & Catalog & 0.08798 &  0.912 & 1261 &  7.16 & $-584$ \\
          TGN387Z067 &  12:28:07.65, +02:02:52.0 & Catalog & 0.09030 &  1.642 & 1497 &  6.95 & $  55$ \\
          TGN387Z068 &  12:28:06.93, +01:59:59.6 & Catalog & 0.08970 &  4.632 & 1548 &  5.09 & $-110$ \\
          TGN387Z042 &  12:28:34.52, +01:48:15.7 & Catalog & 0.09210 &  2.311 & 1715 &  7.17 & $ 550$ \\
          TGN387Z088 &  12:27:46.03, +01:56:15.6 & Catalog & 0.09180 &  1.265 & 2160 & 11.00 & $ 468$ \\
          A1225+0212 &  12:27:45.00, +01:56:15.0 & Catalog & 0.09251 &  0.759 & 2161 & 13.05 & $ 662$ \\
\enddata

\end{deluxetable*}

%% file: tabB13.tex
\begin{deluxetable*}{lllccccc}

\tabletypesize{\scriptsize}

\tablecolumns{8}
\tablewidth{0pt}

\tablecaption{Galaxy Group Members Near 3C~263 ($z_{\rm abs} = 0.11389$)}
\label{tab:3c263_z11389}

\tablehead{ \colhead{Name} & 
            \colhead{Position} & 
            \colhead{Source} & 
            \colhead{$z_{\rm gal}$} & 
            \colhead{$L_{\rm gal}$} & 
            \colhead{$\rho$} & 
            \colhead{$\rho/R_{\rm vir}$} & 
            \colhead{$\Delta v$} \\ 
            & 
            \colhead{(J2000)} & 
            & 
            & 
            \colhead{($L^*$)} & 
            \colhead{(kpc)} & 
            & 
            \colhead{(\kms)} } 

\startdata
             \nodata &  11:40:38.59, +65:51:21.8 &   HYDRA & 0.11369 &  0.375 &  691 &  5.29 & $ -54$ \\
             \nodata &  11:39:41.36, +65:53:33.3 &   HYDRA & 0.11581 &  1.713 &  743 &  3.43 & $ 517$ \\
             \nodata &  11:39:48.94, +65:55:33.2 &   HYDRA & 0.11560 &  0.455 &  970 &  6.93 & $ 460$ \\
             \nodata &  11:39:44.02, +65:55:34.5 &    SDSS & 0.11507 &  1.191 &  982 &  5.08 & $ 318$ \\
             \nodata &  11:38:35.42, +65:53:05.2 &   HYDRA & 0.11313 &  0.247 & 1232 & 10.83 & $-205$ \\
             \nodata &  11:38:55.34, +65:58:00.6 &   HYDRA & 0.11489 &  0.557 & 1495 &  9.97 & $ 269$ \\
             \nodata &  11:39:57.80, +66:01:19.9 &   HYDRA & 0.11485 &  0.277 & 1686 & 14.25 & $ 258$
\enddata

\end{deluxetable*}

%% file: ms.bbl
\begin{thebibliography}



\bibitem[Anderson \& Bregman(2010)]{anderson10} Anderson, M. E. \& Bregman, J. N. 2010, \apj, 714, 320

\bibitem[Anderson \& Bregman(2011)]{anderson11} Anderson, M. E. \& Bregman, J. N. 2011, \apj, 737, 22



\bibitem[Bahcall \& Spitzer(1969)]{bahcall69} Bahcall, J.N. \& Spitzer, L., 1969, \apjl, 156, L63

\bibitem[Bahcall et~al.(1991)]{bahcall91} Bahcall, J. N., Jannuzi, B. T., Schneider, D. P., et~al. 1991, \apjl, 377, L5



\bibitem[Beers, Flynn \& Gebhardt (1990)]{beers90} Beers, T. C., Flynn, K. \& Gebhardt, K. 1990, \aj, 100, 32


\bibitem[Berlind et al.(2006)]{berlind06} Berlind, A.A. et~al. 2006, \apjs, 167, 1

\bibitem[Berlind et al.(2008)]{berlind08} Berlind, A.A. et~al. 2008, yCat, 21670001

\bibitem[Bickel \& Sakov (2008)]{bickel08} Bickel, P.J. \& Sakov, A. 2008, Statistica Sinica, 18, 967

\bibitem[Binney \& Tremaine(1987)]{binney87} Binney, J. \& Tremaine, S. 1987, Galactic Dynamics (Princeton: Princeton Univ. Press), pp. 567-575









\bibitem[Bregman(2007)]{bregman07} Bregman, J. N. 2007, \araa, 45, 221

\bibitem[Bryan \& Norman (1998)]{bryan98} Bryan, G. \& Norman, M. 1998, \apj, 495, 80





\bibitem[Cen \& Ostriker(1999)]{cen99} Cen, R. \& Ostriker, J. P. 1999, \apj, 514, 1

\bibitem[Cen(2013)]{cen13} Cen, R. 2013, \apj, 770, 139



\bibitem[Chen \& Mulchaey(2009)]{chen09} Chen, H.-W. \& Mulchaey, J. S. 2009, \apj, 701, 1219



\bibitem[Chomiuk \& Povich(2011)]{chomiuk11} Chomiuk, L. \& Povich, M. S. 2011, \aj, 142, 197

\bibitem[Churchill et~al.(2000)]{churchill00} Churchill, C. W., Mellon, R. R., Charlton, J. C., et~al. 2000, \apj, 543, 577













\bibitem[Danforth \& Shull(2008)]{danforth08} Danforth, C. W. \& Shull, J. M. 2008, \apj, 679, 194 


\bibitem[Danforth et~al.(2010)Danforth, Stocke \& Shull]{danforth10} Danforth, C. W., Stocke, J. T. \& Shull, J. M. 2010, \apj, 710, 613 

\bibitem[Danforth et~al.(2014)Danforth, Shull, \& Stocke]{danforth14} Danforth, C. W., Shull, J. M., Stocke et al. 2014, in preparation

\bibitem[Dav\'e et~al.(1999)]{dave99} Dav\'e, R., Hernquist, L., Katz, N., \& Weinberg, D. H. 1999, \apj, 511, 521

\bibitem[Dav\'e \& Oppenheimer(2007)]{dave07} Dav\'e, R. \& Oppenheimer, B. D. 2007, \mnras, 374, 427








\bibitem[Fang et~al.(2007)Fang, Canizares, \& Yao]{fang07} Fang, T., Canizares, C. R., \& Yao, Y. 2007, \apj, 670, 992

\bibitem[Fang et~al.(2013)Fang, Bullock, \& Boylan-Kolchin]{fang13} Fang, T., Bullock J. S., \& Boylan-Kolchin, M. 2013, \apj, 762, 20






\bibitem[Fukugita et~al.(1998)Fukugita, Hogan, \& Peebles]{fukugita98} Fukugita, M., Hogan, C. J., \& Peebles, P. J. E. 1998, \apj, 503, 518



\bibitem[Girardi \& Giuricin(2000)]{girardi00} Girardi, M. \& Giuricin, G.  2000, \apj, 540, 45 


\bibitem[Green et~al.(2012)]{green12} Green, J. C., Froning, C. S., Osterman, S., et~al. 2012, \apj, 744, 60


\bibitem[Gupta et~al.(2012)]{gupta12} Gupta, A., Mathur, S., Krongold, Y., Nicastro, F., \& Galeazzi, M. 2012, \apjl, 756, L8






\bibitem[Helsdon, Ponman \& Mulchaey(2005)]{helsdon05} Helsdon, S.F., Ponman, T.J. \& Mulchaey, J.S. 2005, \apj, 618, 679 

\bibitem[Hinshaw et~al. (2013)]{hinshaw13} Hinshaw, G., Larson, D., Komatsu, E., et al. 2013, ApJS, 208, 19




\bibitem[Jenkins et~al.(2003)]{jenkins03} Jenkins, E. B., Bowen, D. V., Tripp, T. M., et~al. 2003, \aj, 125, 2824

\bibitem[Johnson, Chen \& Mulchaey(2013)]{johnson13} Johnson, S.D., Chen, H-W. \& Mulchaey, J.S. 2013, \mnras, 434, 1765 



\bibitem[Kacprzak et~al.(2010)]{kacprzak10} Kacprzak, G. G., Churchill, C. W., Ceverino, D., et~al. 2010, \apj, 711, 533

\bibitem[Kacprzak et~al.(2011)]{kacprzak11} Kacprzak, G. G., Churchill, C. W., Barton, E. J., \& Cooke, J. 2011, \apj, 733, 105


\bibitem[Karachentseva, Karachentsev \& Sharina(2010)]{karachentseva10} Karachentseva, V.E., Karachentsev, I.D. \& Sharina, M.E. 2010, Astrophysics, 53, 462





\bibitem[Keeney et~al.(2013)]{keeney13} Keeney, B. A., et~al. 2013, \apj, 765, 27

\bibitem[Keeney et~al.(2014)]{keeney14} Keeney, B. A., et~al. 2014, in preparation



\bibitem[Kere\u{s} \& Hernquist(2009)]{keres09} Kere\u{s}, D. \& Hernquist, L. 2009a, \apjl, 700, L1


\bibitem[Klypin et~al.(2001)]{klypin01} Klypin, A., Kravtsov, A. V., Bullock, J. S. \& Primack, J. R. 2001, \apj, 554, 903




\bibitem[Larson(1972)]{larson72} Larson, R. B. 1972, \mnras, 157, 121



\bibitem[Lehner et~al.(2007)]{lehner07} Lehner, N., Savage, B. D., Richter, P., et~al. 2007, \apj, 658, 680


\bibitem[Lehner \& Howk(2011)]{lehner11} Lehner, N. \& Howk, J. C. 2011, Science, 334, 955

\bibitem[Lehner et~al.(2013)]{lehner13} Lehner, N., Howk, J. C., Tripp, T. M., et~al. 2013, \apj, 770, 138







\bibitem[McGaugh et~al.(2000)]{mcgaugh00} McGaugh, S. S., Schombert, J. M., Bothun, G. D., \& de Blok, W. J. G. 2000, \apjl, 533, L99







\bibitem[Miles et~al.(2004)]{miles04} Miles, T.A. et~al. 2004, \mnras, 355, 785

\bibitem[Montero-Dorta \& Prada(2009)]{montero09} Montero-Dorta, A. D. \& Prada, F. 2009, \mnras, 399, 1106

\bibitem[Morris et~al.(1991)]{morris91} Morris, S. L., Weymann, R. J., Savage, B. D., \& Gilliland, R. L. 1991, \apjl, 377, L21

\bibitem[Morris et~al.(1993)]{morris93} Morris, S. L., Weymann, R. J., Dressler, A., et~al. 1993, \apj, 419, 524



\bibitem[Moster et~al.(2013)Moster, Naab, \& White]{moster13} Moster, B. P., Naab, T., \& White, S. D. M., 2013, \mnras, 428, 1321

\bibitem[Mulchaey et~al.(1996)]{mulchaey96} Mulchaey, J. S., Mushotzky, R. F., Burstein, D., \& Davis, D. S. 1996, \apjl, 456, L5 (M96)

\bibitem[Mulchaey \& Zabludoff(1998)]{mulchaey98} Mulchaey, J.S. \& Zabludoff, A. I. 1998, \apj, 496, 73

\bibitem[Mulchaey(2000)]{mulchaey00} Mulchaey, J. S. 2000, \araa, 38, 289

\bibitem[Narayanan, Wakker \& Savage(2009)]{narayanan09} Narayanan, A.P., Wakker, B.P \& Savage, B.D.  2009, \apj, 703, 74

\bibitem[Narayanan  et~al.(2010)]{narayanan10} Narayanan, A.P., Wakker, B.P, Savage, B.D., et~al. 2010, \apj, 721, 960

\bibitem[Osmond \& Ponman(2004)]{osmond04} Osmond, J. P. F \& Ponman, T.J. 2004, \mnras, 350, 1511

\bibitem[Osterman et~al.(2011)]{osterman11} Osterman, S., Green, J., Froning, C., et~al. 2011, \apss, 335, 257


\bibitem[Pagel (2008)]{pagel08} Pagel, B.E.J. 2008, in ``Pathways through
an Eclectic Universe", ed. J. H. Knapen, T. J. Mahoney, and A. Vazdekis (San
Francisco: Astronomical Society of the Pacific), ASP Conference \#390, 483
 






\bibitem[Pisani, Ramella \& Geller(2003)]{pisani03} Pisani, A., Ramella, M \& Geller, M.J. 2003, \aj, 126, 1677


\bibitem[Prochaska et~al.(2011b)]{prochaska11b} Prochaska, J. S., Weiner, B., Chan, H.-W., Cooksey, K., \& Mulchaey, J. 2011, \apjs, 193, 28

\bibitem[Prochaska et~al.(2011a)]{prochaska11a} Prochaska, J. X., Weiner, B., Chen, H.-W., Mulchaey, J., \& Cooksey, K. 2011a, \apj, 740, 91






\bibitem[Richter et~al.(2004)]{richter04} Richter, P., Savage, B. D., Tripp, T. M. \& Sembach, K. R. 2004, \apjs, 153, 165

 



\bibitem[Rosenberg et~al.(2003)]{rosenberg03} Rosenberg, J. L., Ganguly, R., Giroux, M. L., \& Stocke, J. T. 2003, \apj, 597, 677


\bibitem[Sarazin(1988)]{sarazin88} Sarazin, C. L. 1988, X-ray Emission from Clusters of Galaxies (Cambridge: Cambridge University Press)


\bibitem[Savage et~al.(2010)]{savage10} Savage, B. D., Narayanan, A., Wakker, B. P., et~al. 2010, \apj, 721, 960

\bibitem[Savage et~al.(2011a)]{savage11a} Savage, B. D., Narayanan, A., Lehner, N., \& Wakker, B. P. 2011, \apj, 731, 14

\bibitem[Savage et~al.(2011b)Savage, Lehner, \& Narayanan]{savage11b} Savage, B. D., Lehner, N., \& Narayanan, A. 2011, \apj, 743, 180

\bibitem[Savage et~al.(2014)]{savage14} Savage, B.D. et~al. 2014, \apjs, in press; arXiv:1403.7542 (Paper 1)










\bibitem[Shull et~al.(1998)]{shull98} Shull, J. M., Penton, S. V., Stocke, J.T. et~al. 1998, \aj, 116, 2094


\bibitem[Shull et~al.(2003)]{shull03} Shull, J. M., Tumlinson, J.T. \& Giroux, M.L. 2003, \apjl, 594, L107



\bibitem[Shull et~al.(2012)] {shull12} Shull, J. M., Smith, B. D., \& Danforth, C. W. 2012, \apj, 759, 23

\bibitem[Simionescu et al. (2011)] {simion11} Simionescu. A. et~al. 2011, Science, 331, 1576

\bibitem[Smith et~al.(2011)]{smith11} Smith, B.D., Hallman, E.J., Shull,
J.M. \& O'Shea, B.W. 2011, \apj, 731, 6


\bibitem[Spitzer(1956)]{spitzer56} Spitzer, L. 1956, \apj, 124, 20



\bibitem[Steidel(1995)]{steidel95} Steidel, C. C. 1995, in QSO Absorption Lines, ed. G. Meylan (Garching: Springer), 139





\bibitem[Stocke et~al.(2004)]{stocke04} Stocke, J. T., Keeney, B. A., McLin, K. M., et~al. 2004, \apj, 609, 94

\bibitem[Stocke et~al.(2006)]{stocke06} Stocke, J. T., Penton, S. V., Danforth, C. W., et~al. 2006, \apj, 641, 217



\bibitem[Stocke et~al.(2013)]{stocke13} Stocke, J. T., Keeney, B. A., \& Danforth, C.W. et~al. 2013, \apj, 763, 148






\bibitem[Thom et~al.(2012)]{thom12} Thom, C., Tumlinson, J., Werk, J. K., et~al. 2012, \apjl, 758, L41

\bibitem[Tilton et~al.(2012)]{tilton12} Tilton, E. M., Danforth, C. W., Shull, J. M., \& Ross, T. L. 2012, \apj, 759, 112


\bibitem[Tripp et~al.(1998)Tripp, Lu, \& Savage]{tripp98} Tripp, T. M., Lu, L., \& Savage, B. D. 1998, \apj, 508, 200

\bibitem[Tripp et~al.(2002)]{tripp02} Tripp, T. M., Jenkins, E. B., Williger, G. M., et~al. 2002, \apj, 575, 697

\bibitem[Tripp et~al.(2003)]{tripp03} Tripp, T. M., Wakker, B. P., Jenkins, E. B., et~al. 2003, \aj, 125, 3122

\bibitem[Tripp et~al.(2008)]{tripp08} Tripp, T. M., Sembach, K. R., Bowen, D. V., et~al. 2008, \apjs, 177, 39


\bibitem[Tully et~al.(2009)]{tully09} Tully, B. R., Rizzi, L., Shaya, E. J., et~al. 2009, \aj, 138, 323

\bibitem[Tumlinson \& Fang(2005)]{tumfang05} Tumlinson, J. \& Fang, T. 2005, \apjl, 623, L97


\bibitem[Tumlinson et~al.(2011)]{tumlinson11} Tumlinson, J., Thom, C., Werk, J. K., et~al. 2011, Science, 334, 948

\bibitem[Tumlinson et~al.(2013)]{tumlinson13} Tumlinson, J., Thom, C., Werk, J. K., et~al. 2013, \apj, 777, 59


\bibitem[Voit \& Bryan(2001)]{voit01} Voit, G.M \& Bryan, G.L. 2001, Nature, 414, 425


\bibitem[Wakker(2001)]{wakker01} Wakker, B. P. 2001, \apjs, 136, 463

\bibitem[Wakker et~al.(2007)]{wakker07} Wakker, B. P., York, D. G., Howk, J. C., et~al. 2007, \apjl, 670, L113

\bibitem[Wakker et~al.(2008)]{wakker08} Wakker, B. P., York, D. G., Wilhelm, R., et~al. 2008, \apj, 672, 298

\bibitem[Wakker \& Savage(2009)]{wakker09} Wakker, B. P. \& Savage, B. D. 2009, \apjs, 182, 378


\bibitem[Werk et~al.(2013)]{werk13} Werk, J., Prochaska, J. X., Thom, C., et~al.  2013, \apjs, 204, 17

\bibitem[Werk et~al.(2014)]{werk14} Werk, J., Prochaska, J.X., Tumlinson, J.T., et~al. 2014, \apj, in press (arXiv: 1403.0947).


\bibitem[White et~al.(1993)]{white93} White, S. D. M., Navarro, J. F., Evrard, A. E., \& Frenk, C. S. 1993, \nat, 366, 429

\bibitem[White(2000)]{white00} White, D.A. 2000, \mnras, 312, 663
 


\bibitem[Wu, Xue \& Fang(1999)]{wu99} Wu X-P, Xue Y-J, Fang L-Z. 1999, \apj, 524, 22


\bibitem[Yao et~al.(2012)]{yao12} Yao, Y., Shull, J. M., Wang, Q. D., \& Cash, W. 2012, \apj, 746, 166

\bibitem[Yoon et~al. (2012)]{yoon12} Yoon, J. H.  Putman, M. E., Thom, C. Chen, H-W and
Bryan, G. L. 201, 2ApJ, 754, 84

\bibitem[Zabludoff \& Mulchaey(1998)]{zabludoff98} Zabludoff, A.I. \& Mulchaey, J.S. 1998, \apj, 496, 39

\end{thebibliography}
